\documentclass[12pt]{article}
\usepackage{latexsym}
\usepackage{exscale}
\usepackage{psfig}
\newcommand{\mbb}[1]{\mbox{\boldmath $#1$}}

\begin{document}
\renewcommand{\thefootnote}{\fnsymbol{footnote}}

\begin{titlepage}
\begin{center}
\LARGE \bf
QED in Dispersing and Absorbing Media\footnote{In
   {\sl Coherence and Statistics of Photons and Atoms},
   edited by J. Pe\v{r}ina (Wiley, New York, 2001), p. 1.}
\\[4ex]
\Large
Ludwig Kn\"oll, Stefan~Scheel, Dirk-Gunnar~Welsch\\[2ex]
\large 
Theoretisch-Physikalisches Institut\\ 
Friedrich-Schiller-Universit\"{a}t Jena\\
Max-Wien-Platz 1, D-07743 Jena, Germany\\[4ex] 
\end{center}
\begin{abstract}
After giving an outline of the quantization scheme based on the
microscopic Hopfield model of a dielectric bulk material,
we show how the classical phenomenological Maxwell equations of
the electromagnetic field in the presence of dielectric matter
of given space- and frequency-dependent complex permittivity
can be transferred to quantum theory. Including in the
theory the interaction of the medium-assisted field with atomic
systems, we present both the minimal-coupling Hamiltonian
and the multipolar-coupling Hamiltonian in the Coulomb gauge.
To illustrate the concept, we discuss 
the input--output relations of radiation
and the transformation of radiation-field quantum states
at absorbing four-port devices, and the spontaneous decay of an
excited atom near the surface of an absorbing body
and in a spherical micro-cavity with intrinsic
material losses. Finally, we give an extension of the
quantization scheme to other media such as amplifying media,
magnetic media, and nonlinear media.  
\end{abstract}
\end{titlepage}
\thispagestyle{empty}

\setcounter{tocdepth}{2}
\tableofcontents


\newpage
\renewcommand{\thefootnote}{\arabic{footnote}}
\setcounter{footnote}{0}
\renewcommand{\theequation}{\arabic{section}.\arabic{equation}}
\setcounter{equation}{0}
\section{Introduction}

As is already known from classical optics, the use of instruments
in optical experiments needs careful examination with regard to their
action on the light under study. Fibres, beam splitters, cavities,
spectral filters etc. are familiar examples of optical instruments,
which are typically built up by dielectric bodies. 
In quantum optics an important aspect  is the influence of
such material bodies on the quantum features of light
including its interaction with (microscopic) atomic systems. For
example, let us consider a $50\%/50\%$ beam splitter oriented at
$45^\circ$ to an incident light beam. In classical optics the beam
splitter simply divides the incoming beam into two (apart from a phase
shift) equal outgoing parts propagating perpendicular to each other,
and with the same scaling factor the classical noise of the incident
field is transferred to the two fields in the output channels of the
beam splitter. It is intuitively clear that in quantum optics the
noise of the vacuum in the unused input port of the beam splitter
introduces additional noise in the two output beams and thus the
quantum statistics of the output field may differ significantly from
that of the input field provided that the input field is prepared in a 
nonclassical state.

Moreover, when two light beams that are prepared
in non-classical states are superimposed by the beam splitter, then
the outgoing field is prepared in a nonclassically correlated
bipartite state, also called an entangled state. Entanglement as a
typical quantum coherence feature plays an important role in
quantum communication. Let us assume that two light beams have been
prepared in an entangled state. In order to use them, e.g.,
in quantum teleportation, the beams should be transmitted through
optical channels such as fibres. Here the crucial point is to
what extent the entanglement can be preserved during the propagation
of the beams, because in any real fibre the absorption gives
necessarily rise to an entanglement degradation.    
 
{F}rom a more theoretical point of view, a very interesting question
is that of the Casimir force between material  bodies.
A crude physical explanation of the Casimir force is that the
vacuum energies of two regions spatially separated by the bodies
differ due to the presence of the matter, which in a
rough approximation can be described simply by a boundary
condition on the quantization volume.
In fact, this simple model does not take account of the
dispersive and absorptive properties of the matter and 
fails in the high-frequency limit where the matter becomes
transparent. 

These few examples show that it is necessary to include in the theory
the presence of material bodies when considering the quantized
radiation field and its interaction with atoms.
In principle, material bodies could be included as a part 
of the matter to which the radiation field is coupled and treated 
microscopically. However, there is a class of material bodies whose
action can be included in the quantum theory exactly,
namely dielectric bodies that respond linearly to the electromagnetic
field, the response being described in terms  of a phenomenologically
introduced space-dependent dielectric
permittivity.
Such a concept has -- similar to classical optics -- the benefit
of being universally valid, because it uses only general physical 
properties, without the need of involved 
{\em ab initio} calculations. 

The quantum theory of radiation in the presence of dielectric matter
has been studied over a long period.   
Quantization of the electromagnetic field in dielectric media with assumed
real and frequency-independent permittivity has been treated extensively
\cite{Jauch48,Shen67,Carniglia71,Birula72,Abram87,Birula87,Knoll87,%
Kennedy88,Khosravi91,Glauber91,Knoll92a,Vogel,Dalton96,Bordag98,Dalton99}.
In the same context, dispersive dielectrics have been considered
\cite{Watson49,Agarwal75,Huttner91,Drummond91,Milonni95,Santos95,%
Hradil96,Drummond99} and attempts have been made to extend the concepts to
nonlinear media \cite{Hillery84,Abram91,Joneckis93,Duan97}.
However, it is well known that the permittivity is a complex function of 
frequency which has to satisfy the Kramers--Kronig relations, which state 
that the real part of the permittivity (responsible for dispersion) and 
the imaginary part (responsible for absorption) are necessarily
connected with each other. A consequence of the existence
of the imaginary part of the permittivity is that the commonly used
mode expansion of the (macroscopic) electromagnetic field fails,
at least in frequency intervals where the absorption cannot be
disregarded. Obviously, an expansion of the field in terms of
damped (non-orthogonal) waves would not be complete. {F}rom 
statistical mechanics it is clear that dissipation is unavoidably
connected with the appearance of a random force which gives rise to
an additional noise source of the electromagnetic field.
Hence, any quantum theory that is based
on the assumption of a real permittivity can only be valid
for narrow-bandwidth fields far from medium resonances 
where absorption can safely be disregarded.

For the last years there has been an increasing number of
articles dealing with the problem of the formulation of quantum
electrodynamics in dielectric media of given complex permittivity
satisfying the Kramers--Kronig relations \cite{Fleischhauer91,%
Knoll92,Huttner92,Ho93,Matloob95,Gruner96a,Gruner96c,Barnett96a,Matloob96,%
Tip97,Tip98,Ho98,Scheel98,Anglin96,Bechler99,Matloob99,DiStefano99,%
DiStefano00}. 
A systematic and quantum-theoretically consistent approach 
to the problem of the quantization of the radiation field
in absorbing bulk dielectrics is given in \cite{Huttner92}
on the basis of the microscopic Hopfield model of a dielectric
\cite{Hopfield58}. It is based on an explicit Fano-type
diagonalization \cite{Fano56} of a Hamiltonian consisting of the
electromagnetic field, a (harmonic-oscillator)
polarization field representing the dielectric matter,
and a continuous set of (harmonic-oscillator) reservoir variables
accounting for absorption. The resulting expression for the
vector potential can be written in terms of the Green tensor of the
classical scattering problem, which, in fact, makes it possible to
perform the quantization of the electromagnetic field in the presence
of arbitrary dielectric bodies of (phenomenologically) given
permittivities \cite{Gruner96a,Gruner96c,Gruner95}, without referring
to specific microscopic models of the bodies, which were hard to
establish for general systems. The concept is based on a source-quantity 
representation of the electromagnetic field, in which the 
electromagnetic-field operators are expressed in terms of a continuous
set of fundamental bosonic fields via the Green tensor of the classical
problem. 

Let us give a brief guide to the topics covered.
After giving an outline of the quantization scheme based on the
microscopic Hopfield model of a dielectric bulk material
(Sec.~\ref{sec:hopfield}),
we show how the classical phenomenological Maxwell equations of
the electromagnetic field in the presence of dielectric matter
of given space- and frequency-dependent permittivity
can be transferred to quantum theory (Sec.~\ref{sec:quantization}).
For this purpose we first summarize the basic properties of
the classical Maxwell equations and express the electromagnetic field
in terms of the Green tensor and a continuous set
of appropriately chosen dynamical field variables
(Sec.~\ref{section1.2.1}). We then perform the quantization by
identifying the dynamical field variables with bosonic fields
associated with the elementary excitations of the composed system
(Sec.~\ref{sec:maxquant}).     
Having quantized the electromagnetic field, the question arises
of how to include in the theory the interaction of the
medium-assisted field with atomic systems
(Sec.~\ref{sec:source}). In order to answer it, we present
both the minimal-coupling Hamiltonian (Sec.~\ref{section2.4.3.1})
and the multipolar-coupling Hamiltonian (Sec.~\ref{section2.4.3.2})
in the Coulomb gauge. To illustrate the basic theoretical concept,
we discuss a number of applications
such as the input--output relations of radiation
(Sec.~\ref{sec:inout}) and the transformation
of radiation-field quantum states (Sec.~\ref{sec:state})
at absorbing four-port devices, and the spontaneous decay of an
excited atom near the (planar) surface of an absorbing body
(Sec.~\ref{sec:planar}) and in a (spherical) micro-cavity with intrinsic
material losses (Secs.~\ref{sec:realcavity} and \ref{sec:beyond}).  
For the sake of transparency, we perform the calculations for
isotropic media and outline the extension to
other media in a separate section at the end
(Sec.~\ref{sec:extensions}).

\setcounter{equation}{0}
\section{Hopfield model and Fano diagonalization}
\label{sec:hopfield}

Following the quantization scheme in \cite{Huttner92}, we consider a
Hopfield model \cite{Hopfield58} of a bulk dielectric in which $N$ harmonic
oscillator fields describing the polarization of the dielectric medium
are linearly coupled to a continuum of harmonic oscillators standing for 
a reservoir. Such a model leads to an energy flow
essentially only in one direction, namely from the medium to the
reservoir where it ``disappears'', hence it is absorbed. The overall
system that consists of the radiation, the dielectric-medium
polarization, the reservoir, and couplings between them may be
regarded as being a Hamiltonian system whose Lagrangian reads
\begin{equation}
\label{hb1.1}
L = \int {\rm d}^3{\bf r}\,{\cal L}
=  \int {\rm d}^3{\bf r}\,({\cal L}_{\rm rad}+{\cal L}_{\rm mat}
+ {\cal L}_{\rm int}),
\end{equation}
where
\begin{equation}
\label{hb1.2a}
{\cal L}_{\rm rad} = {\textstyle\frac{1}{2}} \varepsilon_0
\big[  (\dot{\bf A} +\mbb{\nabla}U)^2
-c^2 (\mbb{\nabla}\times{\bf A})^2
\big]
\end{equation}
($U$, scalar potential; ${\bf A}$, vector potential;
$c^{-2}$ $\!=$ $\!\varepsilon_0 \mu_0$), and
\begin{equation}
\label{hb1.2b}
{\cal L}_{\rm mat} = \sum\limits_{i=1}^N
{\textstyle\frac{1}{2}}\mu \big( \dot{\bf X}_i^2
-\omega_i^2 {\bf X}_i^2 \big)
+\int\limits_0^\infty \!{\rm d}\omega \,
{\textstyle\frac{1}{2}}\mu 
\big[ \dot{\bf X}^2(\omega) -\omega^2 {\bf X}^2(\omega) \big] ,
\end{equation}
\begin{equation}
\label{hb1.2d}
{\cal L}_{\rm int} =
-\sum\limits_{i=1}^N \alpha_i \big( {\bf A}\dot{\bf X}_i
+U\mbb{\nabla}{\bf X}_i \big)
-\int\limits_0^\infty \!{\rm d}\omega \, \dot{\bf X}(\omega)
\sum\limits_{i=1}^N v_i(\omega) {\bf X}_i .
\end{equation}
Here ${\cal L}_{\rm rad}$ and ${\cal L}_{\rm mat}$
are respectively the free Lagrangian densities of the
radiation field\footnote{Note that the vector potential fulfills the
   Coulomb-gauge condition $\mbb{\nabla}{\bf A}$ $\!=$ $\!0$.}
and the matter fields [i.e., the medium oscillator fields ${\bf X}_i$
and the reservoir oscillator fields ${\bf X}(\omega)$ with density $\mu$], 
and ${\cal L}_{\rm int}$ is the interacting part, where
the medium--field coupling constants $\alpha_i$ play the role of
(electric) polarizabilities,
and the medium--reservoir coupling constants $v_i(\omega)$
are assumed to be square-integrable functions of $\omega$.
Introducing the canonical momenta
\begin{equation}
\label{hb1.2e}
{\bf \Pi} = \frac{\partial {\cal L}}{\partial \dot{\bf A}}
= \varepsilon_0 \dot{\bf A} ,
\end{equation}
\begin{equation}
\label{hb1.2f}
{\bf Q}_i = \frac{\partial {\cal L}}{\partial \dot{\bf X}_i} =
\mu \dot{\bf X}_i -\alpha_i {\bf A} ,
\end{equation}
\begin{equation}
\label{hb1.2g}
{\bf Q}(\omega) =
\frac{\partial {\cal L}}{\partial \dot{\bf X}(\omega)} =
\mu \dot{\bf X}(\omega) -\sum\limits_{i=1}^N v_i(\omega) {\bf X}_i ,
\end{equation}
it is not difficult to perform the Legendre transformation and
construct the Hamiltonian $H$ $\!=$ $\!H_{\rm rad}$ $\!+$ $\!H_{\rm mat}$
$\!+$ $\!H_{\rm int}$ of the overall system.

Since the dielectric medium is assumed to be infinitely extended,
it is convenient to go to the reciprocal space\footnote{The spatial
   Fourier transform $\tilde{F}({\bf k})$ of a function $F({\bf r})$
   is defined according to the relation \mbox{$F({\bf r})$ $\!=$
   $\!(2\pi)^{-3/2}\int {\rm d}^3{\bf k}\,\tilde{F}({\bf k})
   {\rm e}^{i{\bf k}{\bf r}}$}.},  
\begin{equation}
\label{hb1.4b}
{\bf A}({\bf r}) \to \tilde{\bf A}({\bf k})
= \sum\limits_{\lambda=1}^2 {A}_\lambda({\bf k}) {\bf e}_\lambda({\bf k}),  
\end{equation}
\begin{equation}
\label{hb1.4a}
{\bf X}_i({\bf r}) \to \tilde{\bf X}_i({\bf k})
= X_{i\|}({\bf k})\mbb{\kappa}
+\sum\limits_{\lambda=1}^2 X_{i\lambda}({\bf k}) 
{\bf e}_\lambda({\bf k})
\end{equation}
[$\mbb{\kappa}$ $\!=$ $\!{\bf k}/|{\bf k}|$,
${\bf e}_\lambda({\bf k})\perp{\bf k}$],
and $\tilde{\bf \Pi}({\bf k})$, $\tilde{\bf Q}_i({\bf k})$,
$\tilde{\bf X}({\bf k},\omega)$, $\tilde{\bf Q}({\bf k},\omega)$
accordingly, and to introduce, with regard to quantization, the
new variables  
\begin{eqnarray}
\label{hb1.4c}
a_\lambda({\bf k}) &\hspace{-1ex}=&\hspace{-1ex}
\sqrt{\frac{\varepsilon_0}{2\hbar\tilde{k} c}}
 \left[ \tilde{k} cA_\lambda({\bf k})
+\frac{i}{\varepsilon_0} \Pi_\lambda({\bf k}) \right] ,
\\
\label{hb1.4d}
b_{i\lambda}({\bf k}) &\hspace{-1ex}=&\hspace{-1ex}
\sqrt{\frac{\mu}{2\hbar\tilde{\omega_i}}}
\left[ \tilde{\omega}_i X_{i\lambda}({\bf k})
+\frac{i}{\mu}\, Q_{i\lambda}({\bf k}) \right] ,
\\
\label{hb1.4e}
b_{\lambda}({\bf k},\omega) &\hspace{-1ex}=&\hspace{-1ex}
\sqrt{\frac{\mu}{2\hbar\omega}}
\left[ -i\omega X_{\lambda}({\bf k},\omega)
+\frac{1}{\mu}\, Q_{\lambda}({\bf k},\omega) \right], 
\end{eqnarray}
with
\begin{equation}
\label{hb1.4e1}
\tilde{k}^2 = k^2+\sum\limits_{i=1}^N \frac{\alpha_i^2}{\mu
\varepsilon_0 c^2} \,, \qquad
\tilde{\omega_i}^2 = \omega_i^2+ \int_0^\infty{\rm d}\omega\,
\frac{v_i^2(\omega)}{\mu^2} \,,
\end{equation}
and
\begin{eqnarray}
\label{hb1.4e2}
b_{i\|}({\bf k}) &\hspace{-1ex}=&\hspace{-1ex}
\sqrt{\frac{\mu}{2\hbar{\tilde{\omega}_{i\|}}}}
\left[\tilde{\omega}_{i\|} X_{i\|}({\bf k})
+\frac{i}{\mu}\, Q_{i\|}({\bf k}) \right] ,
\\
\label{hb1.4e3}
b_{\|}({\bf k},\omega) &\hspace{-1ex}=&\hspace{-1ex}
\sqrt{\frac{\mu}{2\hbar\omega}}
\left[ -i\omega X_{\|}({\bf k},\omega)
+\frac{1}{\mu}\, Q_{\|}({\bf k},\omega) \right]
\end{eqnarray}
[$\tilde{\omega}^2_{i\|}$ $\!=$
$\!\tilde{\omega}^2_i+\alpha_i^2/(\mu\epsilon_0)$].

The fields are now quantized in the familiar way by conversion
of the complex amplitudes
$a_\lambda({\bf k})$, $b_{i\lambda(\|)}({\bf k})$, and
$b_{\lambda(\|)}({\bf k},\omega)$ into bosonic annihilation operators
$\hat{a}_\lambda({\bf k})$, $\hat{b}_{i\lambda(\|)}({\bf k})$, and
$\hat{b}_{\lambda(\|)}({\bf k},\omega)$,
and the Hamiltonian can be expressed in terms of the annihilation
and creation operators as follows:
\begin{equation}
\label{hb1.4j}
\hat{H} = \hat{H}^\perp + \hat{H}^\|_{\rm mat},
\quad
\hat{H}^\perp = \hat{H}_{\rm rad}+ \hat{H}^\perp_{\rm mat}
+ \hat{H}^\perp_{\rm int},
\end{equation}
\begin{eqnarray}
\label{hb1.4i}
\lefteqn{
\hspace*{-4ex}
\hat{H}^\|_{\rm mat} = \int\!{\rm d}^3{\bf k} \,\Bigg\{
\int\limits_0^\infty \!{\rm d}\omega \,\hbar\omega\,
\hat{b}_{\|}^\dagger({\bf k},\omega) \hat{b}_{\|}({\bf k},\omega)
+\sum\limits_{i=1}^N \hbar \tilde{\omega}_{i\|}\,
\hat{b}_{i\|}^\dagger({\bf k}) \hat{b}_{i\|}({\bf k})
}
\nonumber \\ &&\hspace*{-6ex} 
+\int\limits_0^\infty \!{\rm d}\omega \sum\limits_{i=1}^N \bigg[
{\textstyle\frac{1}{2}}\hbar V_i(\omega)
\left( \hat{b}_{i\|}^\dagger({\bf k}) \!+\!\hat{b}_{i\|}(-{\bf k}) \right) 
\left( \hat{b}_{\|}^\dagger(-{\bf k},\omega)
\!+\!\hat{b}_{\|}({\bf k},\omega) \right) \bigg] \Bigg\} ,
\end{eqnarray}
\begin{equation}
\label{hb1.4f}
\hat{H}^\perp_{\rm rad}
= \sum\limits_{\lambda=1}^2
\int\!{\rm d}^3{\bf k} \, \hbar \tilde{k}c\, 
\hat{a}_\lambda^\dagger({\bf k}) \hat{a}_\lambda({\bf k}) ,
\end{equation}
\begin{eqnarray}
\label{hb1.4g}
\lefteqn{
\hat{H}^\perp_{\rm mat} = \sum\limits_{\lambda=1}^2
\int\!{\rm d}^3{\bf k} \Bigg\{
\int\limits_0^\infty \!{\rm d}\omega \,\hbar\omega\,
\hat{b}_{\lambda}^\dagger({\bf k},\omega) \hat{b}_{\lambda}({\bf k},\omega)
+\sum\limits_{i=1}^N  \hbar \tilde{\omega}_i\,
\hat{b}_{i\lambda}^\dagger({\bf k}) \hat{b}_{i\lambda}({\bf k})
}
\nonumber \\ && 
+  \int\limits_0^\infty \!{\rm d}\omega  \sum\limits_{i=1}^N  \bigg[
{\textstyle\frac{1}{2}}\hbar V_i(\omega)
\left( \hat{b}_{i\lambda}^\dagger({\bf k})
\!+\!\hat{b}_{i\lambda}(-{\bf k}) \right) 
\left( \hat{b}_{\lambda}^\dagger(-{\bf k},\omega)
\!+\!\hat{b}_{\lambda}({\bf k},\omega) \right)
\qquad
\nonumber \\ && 
+\!\sum\limits_{j\ne i}
{\textstyle\frac{1}{2}}\hbar V_i(\omega) V_j(\omega)
\left( \hat{b}_{i\lambda}^\dagger({\bf k})
\!+\!\hat{b}_{i\lambda}(-{\bf k}) \right)
\left( \hat{b}_{j\lambda}^\dagger(-{\bf k})
\!+\!\hat{b}_{j\lambda}({\bf k}) \right) 
\bigg] \Bigg\} , 
\end{eqnarray}
\begin{equation}
\label{hb1.4h}
\hat{H}^\perp_{\rm int} = \sum\limits_{\lambda=1}^2
\sum\limits_{i=1}^N \int\!{\rm d}^3{\bf k} \,
{\textstyle\frac{1}{2}}i\hbar \Lambda_i(k)
\left[ \hat{a}_\lambda^\dagger(-{\bf k}) +\hat{a}_\lambda({\bf k}) \right] 
\left[ \hat{b}_{i\lambda}^\dagger({\bf k})
+\hat{b}_{i\lambda}(-{\bf k}) \right] ,
\end{equation} 
where $V_i(\omega)$ $\!=$
$\!(v_i(\omega)/\mu)(\omega/\tilde{\omega}_i)^{1/2}$
and $\Lambda_i(k)$ $\!=$
$\![(\tilde{\omega}_i \alpha_i^2)/(\mu c\varepsilon_0 \tilde{k})]^{1/2}$. 
The bilinear Hamiltonian can be diagonalized (separately for the
transverse and longitudinal parts) by applying a Fano-type
technique \cite{Fano56} such that\footnote{A different technique for
   diagonalization is utilized in \cite{Anglin96,Bechler99}, where
   path-integral quantization of the Lagrangian (\ref{hb1.1}) is
   performed.}
\begin{equation}
\label{hb1.5}
\hat{H}^\|_{\rm mat} = \int {\rm d}^3{\bf k}
\int\limits_0^\infty {\rm d}\omega \, \hbar\omega \,
\hat{B}_\|^\dagger({\bf k},\omega) \hat{B}_\|({\bf k},\omega),
\end{equation}
\begin{equation}
\hat{H}^\perp
=\sum\limits_{\lambda=1}^2 \int {\rm d}^3{\bf k}
\int\limits_0^\infty {\rm d}\omega 
\, \hbar\omega \,\hat{C}_\lambda^\dagger({\bf k},\omega) 
\hat{C}_\lambda({\bf k},\omega).
\end{equation} 
Since the derivation of the formulas for expressing the old
bosonic operators
$\hat{b}_{i\|}({\bf k})$,
$\!\hat{b}_{i\|}^\dagger({\bf k})$,
$\!\hat{b}_\|({\bf k},\omega)$,
$\!\hat{b}_\|^\dagger({\bf k},\omega)$ and
$\hat{a}_\lambda({\bf k},\omega)$,
$\!\hat{a}_\lambda^\dagger({\bf k},\omega)$,
$\!\hat{b}_{i\lambda}({\bf k})$,
$\!\hat{b}_{i\lambda}^\dagger({\bf k})$,
$\!\hat{b}_\lambda({\bf k},\omega)$,
$\!\hat{b}_\lambda^\dagger({\bf k},\omega)$
in terms of the new bosonic operators
$\hat{B}_\|({\bf k},\omega)$, $\hat{B}_\|^\dagger({\bf k},\omega)$
and $\hat{C}_\lambda({\bf k},\omega)$,
$\hat{C}_\lambda^\dagger({\bf k},\omega)$, respectively, 
is rather lengthy, we renounce it here and refer the reader to
\cite{Huttner92,Schmidt97}.

The vector potential and the transverse part of the medium
polarization can then be expressed in terms of the
(polariton-like) operators $\hat{C}_\lambda({\bf k},\omega)$,
$\hat{C}_\lambda^\dagger({\bf k},\omega)$
as \cite{Huttner92,Gruner96c,Gruner95}
\begin{equation}
\label{hb1.8}
\hat{\tilde{\bf A}}({\bf k}) = - \sum\limits_{\lambda=1}^2
{\bf e}_\lambda({\bf k}) \sqrt{\frac{\hbar}{\pi\varepsilon_0}}
\int\limits_0^\infty\!{\rm d}\omega \,
\omega\sqrt{\varepsilon_{\rm I}(\omega)}\,\tilde G({\bf k},\omega) 
\hat{C}_\lambda({\bf k},\omega) +\mbox{H.c.} 
\end{equation}
and
\begin{samepage}
\begin{eqnarray}
\label{hb1.7}
\lefteqn{
\hat{\tilde{\bf P}}({\bf k})
= \sum\limits_{\lambda=1}^2 \sum\limits_{i=1}^N \alpha_i
{\hat{\bf X}}_{i\lambda}({\bf k})
=-i\sum\limits_{\lambda=1}^2 {\bf e}_\lambda({\bf k})
\sqrt{\frac{\hbar\varepsilon_0}{\pi}}
}
\nonumber \\[.5ex] && \times\,
\left[
\int\limits_0^\infty{\rm d}\omega\,
\omega^2\left[\varepsilon(\omega)\!-\!1\right]
\sqrt{\varepsilon_{\rm I}(\omega)}\,\tilde G({\bf k},\omega)
\hat{C}_\lambda({\bf k},\omega)
\!+\! {\rm H.c.}\right] + \hat{\tilde{\bf P}}_{\rm N}({\bf k}),
\qquad
\end{eqnarray}
\end{samepage}
\begin{equation}
\label{hb1.7a}
\hat{\tilde{\bf P}}_{\rm N}({\bf k}) =
\sum\limits_{\lambda=1}^2 {\bf e}_\lambda({\bf k})
\sqrt{\frac{\hbar\varepsilon_0}{\pi}}
\int\limits_0^\infty\!{\rm d}\omega\,
i\sqrt{\varepsilon_{\rm I}(\omega)}
\hat{C}_\lambda({\bf k},\omega) + {\rm H.c.}.
\end{equation}
Here, $\varepsilon(\omega)$
$\!=$ $\!\varepsilon_{\rm R}(\omega)$ $\!+$
$\!i\varepsilon_{\rm I}(\omega)$ is the complex (model) permittivity
of the medium, and 
\begin{equation}
\label{hb1.7b}
\tilde G({\bf k},\omega)
= - \frac{c^2}{\omega^2\varepsilon(\omega)-k^2c^2}
\end{equation}
is the Green function of the classical Maxwell equations with
that permittivity. {F}rom Eq.~(\ref{hb1.7}) [together with
Eq.~(\ref{hb1.7a})] it is seen
that the (transverse) polarization of the medium consists of two
qualitatively different terms. Obviously, the first term is the induced
polarization, whose frequency components are given by the frequency
components of the electric field strength of the radiation multiplied by
$\varepsilon_0[\varepsilon(\omega)$ $\!-$ $\!1]$, and the second
term is the (noise) polarization, i.e., the fluctuating component of
the polarization that is associated with absorption.   

Recalling that the longitudinal part of the electric field operator is 
given by $\epsilon_0\hat{\tilde{\bf E}}_\|({\bf k})$ $\!=$
$\!-\sum_i\alpha_i\hat{\tilde{X}}_{i\|}({\bf k})\mbb{\kappa}$ and
recalling the transversality of the displacement operator
$\hat{\tilde{\bf D}}({\bf k})$, a relation between
$\hat{\tilde{\bf E}}_\|$ and
a longitudinal (noise) polarization
defined analogously to Eq.~(\ref{hb1.7a})
[by replacing $\hat{C}_\lambda({\bf k},\omega)$ with
$\hat{B}_\|({\bf k},\omega)$] can be derived.

In summary, the quantized electromagnetic field can be
expressed, via a source-quantity representation with the classical
Green function, in terms of the permittivity and a continuum
of harmonic oscillators. It is worth noting that in this formulation
there is no explicit hint at the underlying microscopic model. Thus,
it seems quite natural to generalize the theory to arbitrary
dielectric matter of given space- and frequency-dependent permittivity
by transferring the classical source-quantity representation
of the electromagnetic field directly to quantum theory.

\setcounter{equation}{0}
\section{The medium-assisted Maxwell field}  
\label{sec:quantization}

{F}rom now we will not refer to one or the other microscopic
model of the dielectric media. Instead we will start from the familiar
phenomenological Maxwell equations, assuming that the permittivity
is known, e.g., from measurements. In order to allow for an arbitrary
formation of different (non-moving) dielectric bodies in space, we
will assume that the permittivity varies with 
space. For the sake of transparency we will disregard the tensor
character of the permittivity, restricting our attention to
isotropic media (for the extension to anisotropic media,
see Sec.~\ref{sec:extensions2}).

\subsection{Classical basic equations}
\label{section1.2.1}

Let us first briefly outline
the classical theory and bring it in a form suitable for
quantization.  
The phenomenological Maxwell equations of the electromagnetic field 
in the presence of dielectric bodies but without additional charge 
and current densities read   
\begin{equation}
\label{2.01}
\mbb{\nabla}{\bf B}({\bf r}) = 0,
\end{equation}
\begin{equation}
\label{2.02}
\mbb{\nabla}\times{\bf E}({\bf r}) + \dot {\bf B}({\bf r}) =0, 
\end{equation}
\begin{equation}
\label{2.03}
\mbb{\nabla}{\bf D}({\bf r}) = 0,
\end{equation}
\begin{equation}
\label{2.04}
\mbb{\nabla}\times{\bf H}({\bf r}) - \dot {\bf D}({\bf r}) =0, 
\end{equation}
where the displacement field ${\bf D}$ is related to the
electric field ${\bf E}$ and the polarization field ${\bf P}$ 
according to
\begin{equation}
\label{2.05}
{\bf D}({\bf r}) = \varepsilon_0 {\bf E}({\bf r}) 
+ {\bf P}({\bf r}), 
\end{equation}
and for nonmagnetic matter it may be assumed that
\begin{equation}
\label{2.06}
{\bf H}({\bf r}) = \frac{1}{\mu_0}\, {\bf B}({\bf r})
\end{equation}
(for the extension to magnetic matter, see Sec.~\ref{sec:extensions3}).

Let us consider arbitrarily inhomogeneous (isotropic) media 
and assume that the polarization responds linearly and locally
to the electric field. In this case, the most general
relation between the polarization and the electric field
which is in agreement with the causality principle and the
dissipation-fluctuation theorem is
\begin{equation}
\label{2.07}
{\bf P}({\bf r},t) 
= \varepsilon_0 \int_0^\infty\!{\rm d}\tau \, \chi({\bf r},\tau)
{\bf E}({\bf r},t-\tau) + {\bf P}_{\rm N}({\bf r},t),
\label{2.4.6}
\end{equation} 
where $\chi({\bf r},\tau)$ is the dielectric susceptibility
as a function of space and time, and ${\bf P}_{\rm N}$ is the
(noise) polarization associated with absorption. 

Substitution of this expression into 
Eq.~(\ref{2.05}) together with Fourier transformation converts 
this equation to
\begin{equation}
\label{2.08}
\underline{\bf D}({\bf r},\omega) 
= \varepsilon_0 \varepsilon({\bf r},\omega)
\underline{\bf E}({\bf r},\omega) 
+ \underline{\bf P}_{\rm N}({\bf r},\omega), 
\label{2.4.7}
\end{equation}
and thus
\begin{equation}
\label{2.4.7a}
\underline{\bf P}({\bf r},\omega) = \varepsilon_0
\left[ \varepsilon({\bf r},\omega) -1 \right]
\underline{\bf E}({\bf r},\omega) +
\underline{\bf P}_{\rm N}({\bf r},\omega),
\end{equation}
where
\begin{equation}
\label{2.09}
\varepsilon({\bf r},\omega)
= 1 + \int_0^\infty\! {\rm d}\tau \, \chi({\bf r},\tau)
{\rm e}^{i\omega\tau}
\end{equation}
is the (relative) permittivity,
and the Maxwell equations (\ref{2.01}) -- (\ref{2.04}) read in 
the Fourier domain as\footnote{Here and in the following
   the Fourier transform $\underline{F}(\omega)$ of a real function
   $F(t)$ is defined according to the relation $F(t)$ $\!=$
   $\!F^{(+)}(t)$ $\!+$ $\!F^{(-)}(t)$, where $F^{(+)}(t)$
   $\!=$ $\!\int_0^\infty\!{\rm d}\omega\, \underline{F}(\omega)
   {\rm e}^{-i\omega t}$ and $F^{(-)}(t)$ $\!=$ $\![F^{(+)}(t)]^*$.}
\begin{equation}
\label{2.13}
\mbb{\nabla}\underline{\bf B}({\bf r},\omega)=0, 
\end{equation}
\begin{equation}
\label{2.14}
\mbb{\nabla}\times\underline{\bf E}({\bf r},\omega)
- i\omega \underline{\bf B}({\bf r},\omega) =0,
\end{equation}
\begin{equation}
\label{2.15}
\varepsilon_0 \mbb{\nabla}
\varepsilon({\bf r},\omega) \underline{\bf E}({\bf r},\omega)
= \underline{\rho}_{\rm N}({\bf r},\omega),  
\end{equation}
\begin{equation}
\label{2.16}
\mbb{\nabla}\times\underline{\bf B}({\bf r},\omega)
+i\frac{\omega}{c^2} \varepsilon({\bf r},\omega) 
\underline{\bf E}({\bf r},\omega)
= \mu_0 \underline{\bf j}_{\rm N}({\bf r},\omega) .
\end{equation}
Here we have introduced the (noise) charge density
\begin{equation}
\label{2.17}
\underline{\rho}_{\rm N}({\bf r},\omega)
= - \mbb{\nabla} \underline{\bf P}_{\rm N}({\bf r},\omega)
\end{equation}
and the (noise) current density
\begin{equation}
\label{2.18}
\underline{\bf j}_{\rm N}({\bf r},\omega)
= - i\omega \underline{\bf P}_{\rm N}({\bf r},\omega),
\end{equation}
which obey the continuity equation 
\begin{equation}
\label{2.19}
\mbb{\nabla} \underline{\bf j}_{\rm N}({\bf r},\omega) - 
i\omega \underline{\rho}_{\rm N}({\bf r},\omega) = 0.
\end{equation}

According to Eq.~(\ref{2.09}), the permittivity
$\varepsilon({\bf r},\omega)$ is a complex function of frequency,
\begin{equation}
\label{2.19a}
\varepsilon({\bf r},\omega) 
= \varepsilon_{\rm R}({\bf r},\omega)
+ i\,\varepsilon_{\rm I}({\bf r},\omega).
\end{equation}
The real and imaginary parts, which are responsible
for dispersion and absorption, respectively, are uniquely
related to each other through the Kramers--Kronig relations
\begin{equation}
\label{2.11}
\varepsilon_{\rm R}({\bf r},\omega) - 1 =
\frac{{\cal P}}{\pi} \int\! {\rm d}\omega' \,
\frac{\varepsilon_{\rm I}({\bf r},\omega')}{\omega'-\omega}\,, 
\label{2.4.9}
\end{equation}
\begin{equation}
\label{2.12}
\varepsilon_{\rm I}({\bf r},\omega)
= - \frac{{\cal P}}{\pi} 
\int\! {\rm d}\omega' \,
\frac{\varepsilon_{\rm R}({\bf r},\omega')-1}{\omega'-\omega} 
\end{equation}
(${\cal P}$, principal value). Further,
$\varepsilon({\bf r},\omega)$ as a function of complex $\omega$
satisfies the relation 
\begin{equation}
\label{2.12a}
\varepsilon({\bf r},-\omega^\ast)
= \varepsilon^\ast({\bf r},\omega) 
\end{equation}
and is holomorphic in the upper complex half-plane without 
zeros. In particular, it approaches 
unity in the high-frequency limit, i.e., 
$\varepsilon({\bf r},\omega)$ $\!\to$ $\!1$ if
\mbox{$|\omega|$ $\!\to$ $\!\infty$} \cite{Altarelli72,LandauVIII}.

The Maxwell equations (\ref{2.14}) and 
(\ref{2.16}) imply that  
$\underline{\bf E}({\bf r},\omega)$ 
obeys the partial differential equation 
\begin{equation}
\label{2.20}
\mbb{\nabla}\times  
\mbb{\nabla}\times
\underline{\bf E}({\bf r},\omega) 
-\frac{\omega^2}{c^2} \varepsilon({\bf r},\omega) 
\underline{\bf E}({\bf r},\omega) \!
= \! i\omega \mu_0 \underline{\bf j}_{\rm N}({\bf r},\omega),
\end{equation}
the solution of which can be represented in the form
\begin{equation}
\label{2.21}
\underline{\bf E}({\bf r},\omega)
= i\omega\mu_0 \int\! {\rm d}^3{\bf r}' \,
\mbb{G}({\bf r},{\bf r}',\omega)
\underline{\bf j}_{\rm N}({\bf r}',\omega),
\end{equation}
where the Green tensor  
$\mbb{G}({\bf r},{\bf r}',\omega)$
has to be determined from the equation
\begin{equation}
\label{2.22}
\mbb{\nabla}\times  
\mbb{\nabla}\times
\mbb{G}({\bf r},{\bf r}',\omega) 
-\frac{\omega^2}{c^2} \varepsilon({\bf r},\omega) 
\mbb{G}({\bf r},{\bf r}',\omega)
= \mbb{\delta}({\bf r}-{\bf r}')
\end{equation}
together with the boundary condition at infinity.
In Cartesian coordinates, Eq.(\ref{2.22}) reads
\begin{equation}
\label{2.22a}
\Big[ \left(\partial^r_i \partial^r_k - \delta_{ik} 
\Delta^r\right)- \delta_{ik}
{\omega^2\over c^2}\varepsilon({\bf r},\omega) \Big] 
G_{kj}({\bf r}, {\bf r}',\omega) 
= \delta_{ij} \delta({\bf r}-{\bf r}')  
\end{equation}
($\partial^r_i$ $\!=$ $\!\partial/\partial x_i$), where over repeated
vector-component indices is summed.  The Green tensor has the
properties that
\begin{equation}
\label{2.22b}
G_{ij}^\ast({\bf r},{\bf r}',\omega) =
G_{ij}({\bf r},{\bf r}',-\omega^\ast),
\end{equation}
\begin{equation}
\label{2.22c}
G_{ji}({\bf r}',{\bf r},\omega) =  G_{ij}({\bf r},{\bf r}',\omega),
\end{equation}
and
\begin{equation}
\label{2.22d}
\int\! {\rm d}^3{\bf s} \, \frac{\omega^2}{c^2} 
\,\varepsilon_{\rm I}({\bf s},\omega)\, G_{ik}({\bf r},{\bf s},\omega)     
G^\ast_{jk}({\bf r'},{\bf s},\omega) = {\rm Im} \, 
G_{ij}({\bf r},{\bf r'},\omega). 
\end{equation} 
The property (\ref{2.22b}) is a direct consequence of the
corresponding relation (\ref{2.12a}) for the permittivity.
The reciprocity relation (\ref{2.22c}) and the integral
relation (\ref{2.22d}) are proven in Appendix \ref{sectionA}. 

The Fourier components of the magnetic induction, 
$\underline{\bf B}({\bf r},\omega)$, and the  displacement field,
$\underline{\bf D}({\bf r},\omega)$, are directly related to the
Fourier components of the
electric field, $\underline{\bf E}({\bf r},\omega)$,   
\begin{equation}
\label{2.23}
\underline{\bf B}({\bf r},\omega)
= (i\omega)^{-1}\mbb{\nabla}\times
\underline{\bf E}({\bf r},\omega), 
\end{equation}
\begin{equation}
\label{2.24}
\underline{\bf D}({\bf r},\omega)
= (\mu_0\omega^2)^{-1} \mbb{\nabla}\times
\mbb{\nabla} \times \underline{\bf E}({\bf r},\omega)
\end{equation}
[see Eqs.~(\ref{2.14}), (\ref{2.08}), (\ref{2.18}), and (\ref{2.20})], 
and $\underline{\bf E}({\bf r},\omega)$ is determined, according to
Eq.~(\ref{2.21}), by $\underline{\bf j}_{\rm N}({\bf r},\omega)$.
The continuous set of (complex) fields  
$\underline{\bf j}_{\rm N}({\bf r},\omega)$ [or equivalently,
$\underline{\bf P}_{\rm N}({\bf r},\omega)$] can therefore be regarded
as playing the role of the set of dynamical variables of the overall
system composed of the
electromagnetic field and the medium (including the dissipative system). 
For the following it is convenient to split off some factor
from $\underline{\bf P}_{\rm N}({\bf r},\omega)$ and to define
the fundamental dynamical variables ${\bf f}({\bf r},\omega)$ 
according to
\begin{equation}
\label{2.25}
\underline{\bf P}_{\rm N}({\bf r},\omega) =
i\sqrt{\frac{\hbar\varepsilon_0}{\pi}\,\varepsilon_{\rm I}({\bf r},\omega)}
\,  {\bf f}({\bf r},\omega).
\end{equation}

\subsection{Field quantization}
\label{sec:maxquant}

The transition from classical to quantum theory now consists
in the replacement of the classical fields 
${\bf f}({\bf r},\omega)$ and ${\bf f}^\ast({\bf r},\omega)$
by the operator-valued bosonic fields 
$\hat{\bf f}({\bf r},\omega)$ and $\hat{\bf f}^\dagger({\bf r},\omega)$,
respectively, which are associated with the
elementary excitations of the composed system
within the framework of linear light--matter interaction.
Thus the commutation relations are
\begin{equation}
\label{2.26}
\big[\hat{f}_k({\bf r},\omega),\hat{f}_{k'}^\dagger({\bf r}',\omega')\big]
= \delta_{kk'} \delta({\bf r}\!-\!{\bf r}')\delta(\omega\!-\!\omega'), 
\end{equation}
\begin{equation}
\label{2.27}
\big[\hat{f}_k({\bf r},\omega),\hat{f}_{k'}({\bf r}',\omega')\big] = 0, 
\end{equation}
and the Hamiltonian of the composed system is
\begin{equation}
\label{2.28}
\hat{H} = \int\!{\rm d}^3{\bf r} \int_0^\infty\!{\rm d}\omega
\,\hbar \omega\, 
\hat{\bf f}^\dagger({\bf r},\omega)\hat{\bf f}({\bf r},\omega).
\end{equation}

Replacing  
$\underline{\bf E}({\bf r},\omega)$ [Eq.~(\ref{2.21})], 
$\underline{\bf B}({\bf r},\omega)$ [Eq.~(\ref{2.23})], and 
$\underline{\bf D}({\bf r},\omega)$ [Eqs.~(\ref{2.4.7}), (\ref{2.24})]
by the quantum-mechanical operators, we find,
on recalling Eqs.~(\ref{2.18}) and (\ref{2.25}), that
\begin{equation}
\label{2.29}
\underline{\hat{\bf E}}({\bf r},\omega)
= i \sqrt{\frac{\hbar}{\pi\varepsilon_0}}\,\frac{\omega^2}{c^2} 
\int\! {\rm d}^3{\bf r}' \,
\sqrt{\varepsilon_{\rm I}({\bf r}',\omega) }\,
\mbb{G}({\bf r},{\bf r}',\omega)
 \hat{\bf f}({\bf r}',\omega),
\end{equation}
\begin{equation}
\label{2.30}
\underline{\hat{\bf B}}({\bf r},\omega)
= (i\omega)^{-1} \mbb{\nabla}\times
\underline{\hat{\bf E}}({\bf r},\omega), 
\end{equation}
and
\begin{eqnarray}
\label{2.31}
\lefteqn{
\underline{\hat{\bf D}}({\bf r},\omega)
=\varepsilon_0\varepsilon({\bf r},\omega)\underline{\hat{\bf E}}({\bf
r},\omega)
+ \underline{\hat{\bf P}}_{\rm N}({\bf
r},\omega)
}
\nonumber\\&&\hspace{3.5ex}
= (\mu_0\omega^2)^{-1} \mbb{\nabla}\times
\mbb{\nabla} \times 
\underline{\hat{\bf E}}({\bf r},\omega),  
\end{eqnarray}
from which the electromagnetic field operators in the Schr\"{o}dinger
picture are obtained by integration over $\omega$:
\begin{equation}
\label{2.32}
\hat{\bf E}({\bf r})
= \int_0^\infty\! {\rm d}\omega \, \underline{\hat{\bf E}}({\bf
r},\omega) + \mbox{H.c.},
\end{equation}
\begin{equation}
\label{2.33}
\hat{\bf B}({\bf r})
= \int_0^\infty\! {\rm d}\omega \,
\underline{\hat{\bf B}}({\bf r},\omega) + \mbox{H.c.},
\end{equation}
and\footnote{Here the longitudinal (${\bf F}^\|$) and transverse 
   (${\bf F}^\perp$) parts of a vector field ${\bf F}$ are defined by
   ${\bf F}^{\|(\perp)}({\bf r})$ $\!=$ $\!\int\! {\rm d}^3{\bf r}'$
   $\!\mbb{\delta}^{\|(\perp)}({\bf r}$ $\!-$ $\!{\bf r}')
   {\bf F}({\bf r}')$,  
   with $\mbb{\delta}^{(\|)}({\bf r})$ and 
   $\mbb{\delta}^{(\perp)}({\bf r})$ being respectively
   the longitudinal and transverse tensor-valued $\delta$-functions
   [Eqs.~(\ref{deltalong}) and (\ref{deltatrans})].
   Note that for bulk material the transverse polarization
   field $\hat{\bf P}^\perp$ $\!=$ $\!{\hat{\bf D}}$ $\!-$
   $\!\varepsilon_0\hat{\bf E}^\perp$, with ${\hat{\bf E}}$ 
   and $\hat{\bf D}$ being respectively given by
   Eqs.~(\ref{2.32}) and (\ref{2.34}) together with Eq.~(\ref{2.29}) 
   and (\ref{2.31}) exactly corresponds to Eq.~(\ref{hb1.7}) together
   with Eq.~(\ref{hb1.7a}).}
\begin{equation}
\label{2.34}
\hat{\bf D}({\bf r})
= \hat{\bf D}^\perp({\bf r}) 
= \int_0^\infty\! {\rm d}\omega \,
\underline{\hat{\bf D}}({\bf r},\omega) + \mbox{H.c.}.
\end{equation}
In this way, the electromagnetic field is expressed in terms of
the classical Green tensor $\mbb{G}({\bf r},{\bf r}',\omega)$
satisfying the generalized Helmholtz equation (\ref{2.22})
and the continuum of the fundamental bosonic field variables 
$\hat{\bf f}({\bf r},\omega)$ [and $\hat{\bf f}^\dagger({\bf r},\omega)$]. 
All the information about the dielectric matter (such as its formation
in space and its dispersive and absorptive properties) 
is contained [via the permittivity $\varepsilon({\bf r},\omega)$] 
in the Green tensor of the classical problem.
Eqs.~(\ref{2.32}) -- (\ref{2.34}), together with
Eqs.~(\ref{2.29}) -- (\ref{2.31}),
can be considered as the generalization of the familiar mode decomposition.

A similar formalism which also starts
from a causal relation between the polarization and the
electric field strength
is developed in \cite{Tip97,Tip98}. The auxiliary fields that are
introduced there in order to construct a unitary time
evolution in an enlarged Hilbert space can be shown to lead essentially
to the field variables $\hat{\bf f}({\bf r},\omega)$ and
$\hat{\bf f}^\dagger({\bf r},\omega)$ considered here.
Thus, the representation of the electromagnetic field
in \cite{Tip97,Tip98} corresponds
to the Green function representation in Eqs.~(\ref{2.29}) --
(\ref{2.34}).

The quantization scheme meets all the basic requirements of quantum
electrodynamics. So it can be shown
by using very general properties of the permittivity and the Green tensor
that the electric and magnetic fields satisfy the correct (equal-time)
commutation relations (see Appendix \ref{sectionB}) 
\begin{equation}
\label{2.35}
\big[\hat{E}_k({\bf r}),\hat{E}_l({\bf r}')\big] = 0 =
\big[\hat{B}_k({\bf r}),\hat{B}_l({\bf r}')\big], 
\end{equation}
\begin{equation}
\label{2.36}
\big[\varepsilon_0\hat{E}_k({\bf r}),\hat{B}_l({\bf r}')\big] = 
-i \hbar \,\epsilon_{klm} \,\partial^r_m \delta({\bf r}-{\bf r}'). 
\end{equation}
Obviously, the electromagnetic field operators in the Heisenberg picture 
satisfy the Maxwell equations (\ref{2.01}) -- (\ref{2.04}), 
with the time derivative of any operator $\hat{Q}$ being given by 
\begin{equation}
\label{2.37}
\dot{\hat{Q}} = (i \hbar)^{-1} \big[\hat{Q},\hat{H}\big],
\end{equation}
where $\hat{H}$ is the Hamiltonian (\ref{2.28}).

Let us briefly comment on the (zero-temperature) statistical
implications of the quantization scheme. The vacuum expectation value
of $\underline{\hat{\bf E}}({\bf r},\omega)$ is obviously zero whereas the
fluctuation of $\underline{\hat{\bf E}}({\bf r},\omega)$ is not. From
Eq.~(\ref{2.29}) together with the commutation relations (\ref{2.26})
and (\ref{2.27}) we derive, on using the integral
relation (\ref{2.22d}),
\begin{equation}
\label{dft}
\langle 0| \underline{\hat{E}}_k({\bf r},\omega)
\underline{\hat{E}}_l^\dagger({\bf r}',\omega') | 0 \rangle
= \frac{\hbar\omega^2}{\pi\epsilon_0c^2}
\,{\rm Im}\,G_{kl}({\bf r},{\bf r}',\omega) \,\delta(\omega-\omega').
\end{equation}
Equation (\ref{dft}) reveals that the fluctuation of the
electromagnetic field is determined by the imaginary part of the
Green tensor -- a result that is consistent with the
dissipation--fluctuation theorem\footnote{Note that
   the Green tensor plays the role of the response function
   of the electromagnetic field to an external perturbation.}
\cite{Abrikosov}.
Thus, the quantization scheme
respects both the basic requirements of quantum theory (in terms of
the correct commutation relations) and statistical physics (in terms
of the dissipation--fluctuation theorem).

So far we have considered the electromagnetic field strengths.
Instead, scalar ($\hat{\varphi}$) and vector ($\hat{\bf A}$) 
potentials can be introduced and expressed in terms
of the fundamental bosonic fields $\hat{\bf f}({\bf r},\omega)$
and $\hat{\bf f}^\dagger({\bf r},\omega)$. In particular, the
potentials in the Coulomb gauge are defined by
\begin{equation}
\label{2.39}
-\mbb{\nabla}\hat{\varphi}({\bf r}) 
= \hat{\bf E}^\parallel({\bf r}) , 
\end{equation}
\begin{equation}
\label{2.38}
\hat{\bf A}({\bf r}) = 
\int_0^\infty\! {\rm d} \omega  
 \underline{\hat{\bf A}}({\bf r},\omega)
 + \mbox{H.c.} \ ,
\end{equation}
where\footnote{Note that for bulk material Eqs.~(\ref{2.38})
   and (\ref{2.38a}) together with Eq.~(\ref{2.29}) exactly
   correspond to Eq.~(\ref{hb1.8}).}
\begin{equation}
\label{2.38a}
\underline{\hat{\bf A}}({\bf r},\omega) 
= (i\omega)^{-1} \underline{\hat{\bf E}}^\perp({\bf r},\omega) .
\end{equation}
  
The canonically conjugated momentum field with respect to 
$\hat{\bf A}({\bf r})$ is  
\begin{equation}
\label{2.41}
\hat{\bf \Pi}({\bf r})
= -i\varepsilon_0 \int_0^\infty\! {\rm d}\omega \, 
\omega \underline{\hat{\bf A}}({\bf r},\omega)
+ \mbox{H.c.}\,,
\end{equation}
and it is not difficult to verify that $\hat{\bf \Pi}$
$\!=$ $\!-\varepsilon_0\hat{\bf E}^\perp$,
$\mbb{\nabla}\times\hat{\bf A}$ 
$\!=$ $\!\hat{\bf B}$, and
$-\dot{\hat{\bf A}}$ $\!-$ 
$\!\mbb{\nabla}\hat{\varphi}$ $\!=$ $\!\hat{\bf E}$.
In addition, $\hat{\bf A}$ and $\hat{\bf \Pi}$
satisfy the well-known commutation
relations  (Appendix \ref{sectionB})
\begin{equation}
\label{2.42}
\big[\hat{A}_k({\bf r}),\hat{A}_{k'}({\bf r}')\big] = 0 =
\big[\hat{\Pi}_k({\bf r}),\hat{\Pi}_{k'}({\bf r}')\big], 
\end{equation}
\begin{equation}
\label{2.43}
\big[\hat{A}_{k}({\bf r}),\hat{\Pi}_{k'}({\bf r}')\big] = 
i \hbar \,\delta^{\perp}_{kk'}({\bf r}-{\bf r}'). 
\end{equation}

\subsubsection{One-dimensional systems}
\label{sec:1ded}

Let us illustrate the main features of the concept for
linearly polarized radiation propagating in the $x$\,direction,
which effectively reduces the system to one spatial dimension
[$\hat{\bf A}({\bf r})$ $\!\to$ $\!\hat{A}_y(x)$
$\!\equiv$ $\!\hat{A}(x)$,
$\hat{\bf \Pi}({\bf r})$ $\!\to$ $\!\hat{\Pi}_y(x)$
$\!\equiv$ $\!\hat{\Pi}(x)$,
$\hat{\bf f}({\bf r},\omega)$ $\!\to$ $\!\hat{f}(x,\omega)$].
According to Eqs.~(\ref{2.38}) and (\ref{2.38a}) together
with Eq.~(\ref{2.29}), the operator of the vector potential is 
\begin{equation}
\label{2.44}
\hat{A}(x)
= \sqrt{\frac{\hbar}{\pi\varepsilon_0{\cal A}}}
\int_0^\infty\! {\rm d}\omega \int\! {\rm d}x'\,
\frac{\omega}{c^2}\sqrt{\varepsilon_{\rm I}(x',\omega)}
\,G(x,x',\omega)  \hat{f}(x',\omega)+ \mbox{H.c.} \,, 
\end{equation}
and $\hat{\Pi}$ accordingly (${\cal A}$, normalization area perpendicular 
to the $x$\,direction).
Here the Green function $G(x,x',\omega)$ satisfies the equation
\begin{equation}
\label{2.45}
-\frac{\partial^2}{\partial x^2}G(x,x',\omega)
- \frac{\omega^2}{c^2}\varepsilon(x,\omega)G(x,x',\omega)
= \delta(x-x').
\end{equation}
In the simplest case when the spatial variation of the
permittivity can be disregarded, then the solution
of Eq.~(\ref{2.45}) that satisfies the boundary
conditions at $|x|,|x'|$ $\!\to$ $\!\infty$ is 
\begin{equation}
\label{2.46}
G(x,x',\omega)
= - \left[2i\frac{\omega}{c}n(\omega)\right]^{-1}
\exp\!\left[i\frac{\omega}{c}n(\omega)|x-x'|
\right],
\end{equation}
with $n(\omega)$ $\!=$ $\!\sqrt{\varepsilon(\omega)}$
$\!=$ $\!n_{\rm R}(\omega)$ $\!+$ $\!i\,n_{\rm I}(\omega)$ being the
complex refractive index of the medium.

Substituting the Green function (\ref{2.46}) into 
Eq.~(\ref{2.44}), we may rewrite the $x'$-integral to obtain
\begin{samepage}
\begin{eqnarray}
\label{2.47}
\lefteqn{
\hat{A}(x)
= \int_0^\infty {\rm d}\omega\,
\Bigg\{
\sqrt{\frac{\hbar}{4\pi\varepsilon_0 c \omega n_{\rm R}(\omega) {\cal A}}}
\frac{n_{\rm R}(\omega)}{n(\omega)}
}
\nonumber \\[.5ex] &&\hspace{2ex} \times\,
\left[{\rm e}^{i n_{\rm R}(\omega)\omega x/c} \hat{a}_+(x,\omega) +
{\rm e}^{-i n_{\rm R}(\omega)\omega x/c} \hat{a}_-(x,\omega) \right]
+ \mbox{H.c.}
\Bigg\},
\end{eqnarray}
\end{samepage}
where 
\begin{equation}
\label{2.48}
\hat{a}_\pm(x,\omega)
= i \sqrt{2n_{\rm I}(\omega)\omega/c}\;
{\rm e}^{\mp n_{\rm I}(\omega)\omega x/c}
\int_{-\infty}^{\pm x}\! {\rm d}x' \,
{\rm e}^{-i n(\omega)\omega x'/c}\hat{f}(\pm x',\omega), 
\end{equation}
and from 
Eq.~(\ref{2.26}) it follows that
\begin{equation}
\label{2.49}
\big[\hat{a}_\pm(x,\omega),\hat{a}_\pm^\dagger(x',\omega')\big] 
= {\rm e}^{- n_{\rm I}(\omega)\omega|x-x'|/c} \,\delta(\omega-\omega').
\end{equation}
Obviously, the space-dependent operators $\hat{a}_\pm(x,\omega)$ 
describe the amplitudes of the damped monochromatic waves propagating 
to the right (subscript $+$) and left (subscript $-$), and
from Eq.~(\ref{2.48}) it follows that $\hat{a}_\pm(x,\omega)$
and $\hat{a}_\pm(x',\omega)$ are related by (spatial) 
quantum Langevin equations \cite{Huttner92,Gruner96a},
\begin{equation}
\label{2.49a}
\frac{\partial}{\partial x}\hat{a}_\pm(x,\omega)
= \mp\, n_{\rm I}(\omega)\frac{\omega}{c}\,\hat{a}_\pm(x,\omega)
+ \hat{F}_\pm(x,\omega),
\end{equation}
with
\begin{equation}
\label{2.49b}
\hat{F}_\pm(x,\omega) = \pm i
\sqrt{2n_{\rm I}(\omega)\omega/c}\;
{\rm e}^{\mp i n_{\rm R}(\omega)\omega x/c}\hat{f}(x,\omega)
\end{equation}
being the operator Langevin noise sources.
In particular, when $\langle\hat{f}(x'',\omega)\rangle$ $\!=$ $\!0$
for $|x''$ $\!-$ $\!x|$ $\!<$ $\!|x$ $\!-$ $\!x'|$
and $|x''$ $\!-$ $\!x'|$ $\!<$ $\!|x$ $\!-$ $\!x'|$, then 
\begin{equation}
\label{2.50}
\langle\hat{a}_\pm(x,\omega)\rangle =
\langle\hat{a}_\pm(x',\omega)\rangle \exp[-n_{\rm I}(\omega)\omega|x-x'|/c],
\qquad
\pm x \mp x' \ge 0
\end{equation}
for arbitrary $\langle\hat{a}_\pm(x',\omega)\rangle$.

Equation (\ref{2.47}) is the extension of the familiar
mode decomposition to absorbing media. 
Let us assume that in a frequency interval $\Delta\omega$ the absorption
is sufficiently small, so that for a chosen (finite) propagation
interval $|x$ $\!-$ $\!x'|$ the condition \mbox{$n_{\rm I}(\omega)\omega
|x-x'|/c$ $\!\ll$ $\!1$} holds. Then the amplitude 
operators $\hat{a}_\pm(x,\omega)$ can be regarded as being independent 
of $x$ for that propagation distance and satisfying the
ordinary Bose commutation relations, as is seen from Eqs.~(\ref{2.48})
and (\ref{2.49}). In the chosen frequency and space
intervals, Eq.~(\ref{2.47}) exactly reduces to the familiar expression
obtained by mode expansion.

\setcounter{equation}{0}
\section{Atom--Field Interaction}
\label{sec:source}

The interaction of the quantized electromagnetic field with atoms placed
inside a dielectric-matter configuration or near dielectric bodies can
strongly be influenced by the dielectric medium. A well-known example is the
dependence of the rate of spontaneous decay of an excited atom on the
properties of a  dielectric environment (Sec.~\ref{sec:decay}).
In order to study such and related phenomena, the Hamiltonian
(\ref{2.28}) must be supplemented with the Hamiltonian of additional
charged particles and their interaction energy with the
medium-assisted electromagnetic field. 
 
\subsection{The minimal-coupling Hamiltonian}
\label{section2.4.3.1}

Applying the minimal-coupling scheme, we may write the
total Hamiltonian in the form\footnote{Here and in the following the
   subscripts A and M are introduced in order to distinguish between
   atom- and medium-assisted quantities.}
\begin{eqnarray}
\label{2.51}
\lefteqn{
\hspace{-4ex}
\hat{H} = \int\! {\rm d}^3{\bf r} \int_0^\infty\! {\rm d}\omega\,
\hbar\omega\,\hat{\bf f}^\dagger({\bf r},\omega) 
\hat{\bf f}({\bf r},\omega) + \sum_\alpha {1\over 2m_\alpha}
\left[ \hat{\bf p}_\alpha - q_\alpha 
\hat{\bf A}(\hat{\bf r}_\alpha) \right]^2    
}
\nonumber\\&&\hspace{8ex}
+{\textstyle\frac{1}{2}} \int\! {\rm d}^3{\bf r}\, 
\hat{\rho}_{\rm A}({\bf r}) \hat{\varphi}_{\rm A}({\bf r})
+ \int\! {\rm d}^3{\bf r}\, 
\hat{\rho}_{\rm A}({\bf r}) \hat{\varphi}_{\rm M}({\bf r}) ,
\end{eqnarray}
where $\hat{\bf r}_\alpha$ is the position operator and
$\hat{\bf p}_\alpha$ is the canonical momentum operator of the
$\alpha$th (nonrelativistic) particle of charge $q_\alpha$ and mass
$m_\alpha$. 
The Hamiltonian (\ref{2.51}) consists of four terms. The first term is the
energy of the electromagnetic field and the medium (including the
dissipative system), as introduced in Eq.~(\ref{2.28}).
The second term is the kinetic energy of the charged particles, and the third
term is their Coulomb energy, where the corresponding scalar potential
$\hat{\varphi}_{\rm A}$ is given by 
\begin{equation}
\label{2.52}
\hat{\varphi}_{\rm A}({\bf r}) = 
\int\!{\rm d}^3{\bf r}' \frac {\hat{\rho}_{\rm A}({\bf r}')}
{4\pi\varepsilon_0|{\bf r}-{\bf r}'|} \,,
\end{equation}
with
\begin{equation}
\label{2.53}
\hat{\rho}_{\rm A}({\bf r}) = 
\sum_\alpha q_\alpha 
\delta({\bf r}-\hat{\bf r}_\alpha) 
\end{equation}
being the charge density of the particles.
The last term is the Coulomb energy of interaction
of the particles with the medium.
{F}rom Eq.~(\ref{2.51}) it follows that 
the interaction Hamiltonian reads  
\begin{equation}
\label{2.63}
\hat{H}_{\rm int} = 
- \sum_\alpha {q_\alpha\over m_\alpha} \left[ \hat{\bf p}_\alpha 
- {\textstyle{1\over 2}}q_\alpha  \hat{\bf A}(\hat{\bf r}_\alpha)
\right] \hat{\bf A}(\hat{\bf r}_\alpha)
+ \int\! {\rm d}^3{\bf r}\, 
\hat{\rho}_{\rm A}({\bf r}) \hat{\varphi}_{\rm M}({\bf r}).
\end{equation}
Note that in Eqs.~(\ref{2.51}) and (\ref{2.63}), the vector potential
$\hat{\bf A}$ and the scalar potential $\hat{\varphi}_{\rm M}$,
respectively, must be thought of as being expressed, on using 
Eq.~(\ref{2.38}) [together with Eqs.~(\ref{2.38a}) and
(\ref{2.29})] and Eq.~(\ref{2.39}) [together with Eqs.~(\ref{2.32}) and
(\ref{2.29})], in terms of the fundamental fields
$\hat{\bf f}({\bf r},\omega)$
[and $\hat{\bf f}^\dagger({\bf r},\omega)$].

In a straightforward but somewhat lengthy calculation
(for an example, see Appendix \ref{sectionC})
it can be shown (by means of
the commutation relations derived in
Appendix \ref{sectionB}) that both the operator-valued
Maxwell equations 
\begin{equation}
\label{2.54}
\mbb{\nabla}\hat{{\bf B}}({\bf r}) = 0,
\end{equation}
\begin{equation}
\label{2.55}
\mbb{\nabla}\times\hat{{\bf E}}({\bf r})
+ \dot{\hat{{\bf B}}}({\bf r}) =0,             
\end{equation}
\begin{equation}
\label{2.56}
\mbb{\nabla}\hat{\bf D}({\bf r}) = 
\hat{\rho}_{\rm A}({\bf r}),
\end{equation}
\begin{equation}
\label{2.57}
\mbb{\nabla}\times\hat{\bf H}({\bf r})
- \dot{\hat{\bf D}}({\bf r}) = \hat{\bf j}_{\rm A}({\bf r}),             
\end{equation}
where
\begin{equation}
\label{2.62}
\hat{\bf j}_{\rm A}({\bf r}) = 
{\textstyle\frac{1}{2}} 
\sum_\alpha q_\alpha 
\left[\dot{\hat{\bf r}}_\alpha \delta({\bf r}-\hat{\bf r}_\alpha)
+\delta({\bf r}-\hat{\bf r}_\alpha)\,\dot{\hat{\bf r}}_\alpha
\right],  
\end{equation}
and the operator-valued Newtonian equations of motion 
\begin{equation}
\label{2.58}
\dot{\hat{\bf r}}_\alpha = 
{1\over m_\alpha} \left[ \hat{\bf p}_\alpha 
- q_\alpha \hat{\bf A}(\hat{\bf r}_\alpha) \right],
\end{equation}
\begin{equation}
\label{2.59}
m_\alpha\ddot{\hat{\bf r}}_\alpha = 
q_\alpha
\left\{
\hat{\bf E}(\hat{\bf r}_\alpha) 
+{\textstyle\frac{1}{2}}
\left[
\dot{\hat{\bf r}}_\alpha \times \hat{\bf B}(\hat{\bf r}_\alpha) - 
\hat{\bf B}(\hat{\bf r}_\alpha) \times \dot{\hat{\bf r}}_\alpha
\right]\right\}
\end{equation}
are fulfilled. In Eqs.~(\ref{2.55}) -- (\ref{2.57}),
the (longitudinal part of the) electric field and the displacement
field now contain [compared with Eqs.~(\ref{2.32}) and (\ref{2.34})]
additional longitudinal components that result from the
charge distribution $\hat{\rho}_{\rm A}({\bf r})$, i.e.,
\begin{equation}
\label{2.60}
\hat{\bf E}({\bf r})  =
\hat{\bf E}_{\rm M}({\bf r}) - 
\mbb{\nabla}\hat{\varphi}_{\rm A}({\bf r})
= \left[ \int_0^\infty\! {\rm d}\omega \, 
\underline{\hat{\bf E}}({\bf r},\omega) + \mbox{H.c.} \right]
-  \mbb{\nabla}\hat{\varphi}_{\rm A}({\bf r}),
\end{equation}
\begin{equation}
\label{2.61}
\hat{\bf D}({\bf r}) = 
\hat{\bf D}_{\rm M}({\bf r}) - 
\varepsilon_0\mbb{\nabla}\hat{\varphi}_{\rm A}({\bf r})
= \left[ \int_0^\infty\! {\rm d}\omega \, 
\underline{\hat{\bf D}}({\bf r},\omega) + \mbox{H.c.} \right]
-  \varepsilon_0\mbb{\nabla}
\hat{\varphi}_{\rm A}({\bf r}),
\end{equation}
with $\underline{\hat{\bf E}}({\bf r},\omega)$ and
$\underline{\hat{\bf D}}({\bf r},\omega)$ being
defined by Eqs.~(\ref{2.29}) and (\ref{2.31}).
The Maxwell equations (\ref{2.54}) and (\ref{2.56}),
respectively, simply follow from the definition of
$\hat{{\bf B}}({\bf r})$ [Eq.~(\ref{2.33})
together with Eq.~(\ref{2.30})] and
$\hat{{\bf D}}({\bf r})$ [Eq.~(\ref{2.61}) together
with Eqs.~(\ref{2.31}) and (\ref{2.52})].
The Maxwell equations (\ref{2.55}) and (\ref{2.57})
are respectively the Heisenberg equations of motion of
$\hat{\bf B}({\bf r})$ and $\hat{\bf D}({\bf r})$
according to Eq.~(\ref{2.37}), with the Hamiltonian being
given by Eq.~(\ref{2.51}), and the Newtonian
equations of motion (\ref{2.58}) and (\ref{2.59}) are
respectively obtained from the Heisenberg equations of motion of
$\hat{\bf r}_\alpha$ and $\hat{\bf p}_\alpha$.


\subsection{The multipolar-coupling Hamiltonian}
\label{section2.4.3.2}

In the interaction Hamiltonian (\ref{2.63}) used in the
minimal-coupling scheme the electromagnetic field is expressed 
in terms of the potentials. 
With regard to the interaction of the electromagnetic
field with (localized) atomic systems (atoms,
molecules etc.) the interaction energy is commonly
desired to be treated in terms of the field strengths and the
atomic polarization and magnetization. This can be achieved 
by means of a unitary transformation.

Let us consider an atomic system localized at position ${\bf r}_{\rm A}$
and introduce the atomic polarization
\begin{equation}
\label{2.65}
\hat{{\bf P}}_{\rm A}({\bf r}) = 
\sum_\alpha q_\alpha \left(\hat{{\bf r}}_{\alpha}- {\bf r}_{\rm A}\right) 
\int_0^1\! {\rm d}\lambda\, 
\delta\!\left[{\bf r}\!-\!{\bf r}_{\rm A}\!
-\!\lambda\left(\hat{{\bf r}}_\alpha\!
-\!{\bf r}_{\rm A}\right)\right], 
\end{equation}
so that the charge density (\ref{2.53}) can be rewritten as
\begin{equation}
\label{2.64}
\hat{\rho}_{\rm A}({\bf r}) = 
q_{\rm A} 
\delta({\bf r}-\hat{\bf r}_{\rm A}) 
- \mbb{\nabla}\hat{{\bf P}}_{\rm A}({\bf r}),
\end{equation}
with
\begin{equation}
\label{2.64a}
q_{\rm A} = \sum_\alpha q_\alpha
\end{equation}
being the total charge of the atomic system.
In order to perform the transition from the minimal-coupling scheme 
to the multipolar-coupling scheme, we apply to the variables the unitary 
operator 
\begin{equation}
\label{2.66}
\hat{U} = 
\exp\!\left[\frac{i}{\hbar}\int\! {\rm d}^3{\bf r}\,
\hat{{\bf P}}_{\rm A}({\bf r})\hat{\bf A}({\bf r})\right]
\end{equation}
which is known as the Power--Zienau transformation
\cite{Power59,Craig,Ackerhalt84,Cohen}.
It is not difficult to prove that the following transformation
rules are valid:
\begin{equation}
\label{2.67}
\hat{\bf r}_\alpha' 
= \hat{U} \hat{\bf r}_\alpha \hat{U}^\dagger = \hat{\bf r}_\alpha,
\end{equation}
\begin{samepage}
\begin{eqnarray}
\label{2.68}
\lefteqn{
\hat{\bf p}_\alpha' = \hat{U} \hat{\bf p}_\alpha \hat{U}^\dagger
}
\nonumber\\&&\hspace{-1ex} 
= \hat{\bf p}_\alpha  - q_\alpha \hat{\bf A}(\hat{\bf r}_\alpha)
- q_\alpha \int_0^1\!{\rm d}\lambda\,\lambda
\left( \hat{\bf r}_\alpha\!-\!{\bf r}_A \right) \times
\hat{\bf B}\left[ {\bf r}_A\!+\!\lambda \left( \hat{\bf r}_\alpha
\!-\!{\bf r}_A \right) \right] ,
\qquad
\end{eqnarray}
\end{samepage}
\begin{eqnarray}
\label{2.69}
\lefteqn{
\hat{\bf f}'({\bf r},\omega) =
\hat{U} \hat{\bf f}({\bf r},\omega) \hat{U}^\dagger
}
\nonumber\\&& 
= \hat{\bf f}({\bf r},\omega)
- \frac{i}{\hbar} \sqrt{\frac{\hbar}{\pi\varepsilon_0}\, 
\varepsilon_{\rm I}({\bf r},\omega)}\, 
\frac{\omega}{c^2}
\int\! {\rm d}^3{\bf r}'\, \hat{\bf P}^{\perp}_{\rm A}({\bf r}')
\mbb{G}^{\ast}({\bf r}',{\bf r},\omega).
\qquad
\end{eqnarray}
Employing equations (\ref{2.67}) -- (\ref{2.69})
and using Eqs.~(\ref{2.22d}), (\ref{2.29}), and (\ref{A.35a}),
we can express the Hamiltonian $\hat{H}$ in Eq.~(\ref{2.51})
in terms of the new variables $\hat{\bf r}_\alpha'$ $\!=$
$\!\hat{\bf r}_\alpha$, $\hat{\bf p}_\alpha'$, and
$\hat{\bf f}'({\bf r},\omega)$ in order to
obtain the multipolar Hamiltonian
\begin{eqnarray}
\label{2.72}
\lefteqn{
\hat{H} = 
\int\! {\rm d}^3{\bf r} \int_0^\infty\! {\rm d}\omega\,
\hbar\omega\,\hat{\bf f}'{^\dagger}({\bf r},\omega)
\hat{\bf f}'({\bf r},\omega)
}
\nonumber \\ && 
+ \sum_\alpha \frac{1}{2m_{\alpha}} \bigg\{
\hat{\bf p}_{\alpha}' + 
 q_\alpha\int_0^1\!{\rm d}\lambda\,\lambda
\left( \hat{\bf r}_\alpha\!-\!{\bf r}_A \right) \times
\hat{\bf B}'\!\left[ {\bf r}_A\!+\!\lambda \left( \hat{\bf r}_\alpha
\!-\!{\bf r}_A \right) \right]
\bigg\}^2
\nonumber\\ &&
+\,\frac{1}{2}\int\! {\rm d}^3{\bf r}\,\hat{\rho}_{\rm A}({\bf r}) 
\hat{\varphi}_{\rm A}({\bf r}) 
+\frac{1}{2\varepsilon_0}\int\! {\rm d}^3{\bf r}\,
\hat{{\bf P}}^{\perp}_{\rm A}({\bf r})
\hat{{\bf P}}^{\perp}_{\rm A}({\bf r})
\nonumber\\ &&
+ \int\! {\rm d}^3{\bf r}\,\hat{\rho}_{\rm A}({\bf r})
\hat{\varphi}_{\rm M}'({\bf r})
- \int\! {\rm d}^3{\bf r}\,\hat{{\bf P}}^{\perp}_{\rm A}({\bf r})
\hat{{\bf E}}_{\rm M}'^{\perp}({\bf r}),
\end{eqnarray}
where the relations $\hat{\bf B}'$ $\!=$ $\!\hat{\bf B}$,
$\hat{\varphi}_{\rm M}'$ $\!=$ $\!\hat{\varphi}_{\rm M}$
[cf. Eq.~(\ref{A.37a})], and
\begin{equation}
\label{2.72a}
\hat{{\bf E}}_{\rm M}'^{\perp}({\bf r})
=\hat{{\bf E}}_{\rm M}^{\perp}({\bf r})
+ \frac{1}{\varepsilon_0}\hat{\bf P}_A^\perp({\bf r})
\end{equation}
are valid.

In particular when the charged particles form a neutral
atomic system ($q_{\rm A}$ $\!=$ $\!0$),
then we may write, on integrating by parts and recalling that
$\hat{\bf E}_{\rm A(M)}^\|$ $\!=$
$\!-\hat{\bf P}_{\rm A(M)}^\|/\varepsilon_0$,
\begin{equation}
\label{2.73b}
\int {\rm d}^3{\bf r}\,\hat{\rho}_{\rm A}({\bf r})
\hat{\varphi}_{\rm A}({\bf r})
= \frac{1}{\varepsilon_0}
\int {\rm d}^3{\bf r}\,
\hat{{\bf P}}_{\rm A}^\|({\bf r})\hat{{\bf P}}_{\rm A}^\|({\bf r}),
\end{equation}
and
\begin{eqnarray}
\label{2.73c}
\lefteqn{
\int {\rm d}^3{\bf r}\,\hat{\rho}_{\rm A}({\bf r})
\hat{\varphi}_{\rm M}({\bf r})
= \frac{1}{\varepsilon_0}
\int {\rm d}^3{\bf r}\,
\hat{{\bf P}}_{\rm A}^\|({\bf r})\hat{{\bf P}}_{\rm M}^\|({\bf r})
}
\nonumber\\&&
=\frac{1}{\varepsilon_0} \int {\rm d}^3{\bf r}\,
\hat{{\bf P}}_{\rm A}({\bf r})\hat{{\bf P}}_{\rm M}({\bf r})
-\frac{1}{\varepsilon_0} \int {\rm d}^3{\bf r}\,
\hat{{\bf P}}_{\rm A}^\perp({\bf r})
\hat{{\bf P}}_{\rm M}^\perp({\bf r}) .
\end{eqnarray}
Combining Eqs.~(\ref{2.72}) -- (\ref{2.73c}) and taking into account
that $\hat{\bf P}'_{\rm M}$ $\!=$ $\!\hat{\bf P}_{\rm M}$ [cf.
Eqs.~(\ref{A.35}) and (\ref{A.38})], we may rewrite
the multipolar Hamiltonian as
\begin{eqnarray}
\label{2.75}
\lefteqn{
\hat{H} = \int\! {\rm d}^3{\bf r} \int_0^\infty\! {\rm d} \omega
\,\hbar\omega\,\hat{\bf f}'^{\dagger}({\bf r},\omega)
\hat{\bf f}'({\bf r},\omega)
}
\nonumber\\&& 
+ \sum_\alpha \frac{1}{2m_{\alpha}} \bigg\{
\hat{{\bf p}}'_{\alpha} + 
q_\alpha\int_0^1\!{\rm d}\lambda\,\lambda 
\left( \hat{\bf r}_\alpha\!-\!{\bf r}_A \right) \times
\hat{\bf B}'\left[ {\bf r}_A\!+\!\lambda \left( \hat{\bf r}_\alpha
\!-\!{\bf r}_A \right) \right]
\bigg\}^2
\nonumber \\ &&
+\,\frac{1}{2\varepsilon_0}\int\! {\rm d}^3{\bf r} \,
\hat{\bf P}_{\rm A}({\bf r}) \hat{\bf P}_{\rm A}({\bf r})
+\,\frac{1}{\varepsilon_0}\int\! {\rm d}^3{\bf r} \,
\hat{\bf P}_{\rm A}({\bf r}) \hat{\bf P}'_{\rm M}({\bf r}) 
\nonumber\\&&
-\,\frac{1}{\varepsilon_0}
\int\! {\rm d}^3{\bf r}\, \hat{{\bf P}}_{\rm A}({\bf r})
\hat{\bf D}'{^\perp}({\bf r}), 
\end{eqnarray}
where
\begin{equation}
\label{2.75a}
\hat{\bf D}'^{\perp}({\bf r}) 
= \hat{\bf D}_{\rm M}'^\perp({\bf r})
=\varepsilon_0\hat{\bf E}_{\rm M}'^\perp({\bf r})
+\hat{\bf P}_{\rm M}'^\perp({\bf r}) .
\end{equation}
{F}rom Eq.~(\ref{2.75}) the interaction Hamiltonian is seen to be
\begin{eqnarray}
\label{2.76}
\lefteqn{
\hat{H}_{\rm int} =
\frac{1}{\varepsilon_0}
\int\! {\rm d}^3{\bf r}\,\hat{{\bf P}}_{\rm A}({\bf r})
\hat{{\bf P}}_{\rm M}'({\bf r})
-\frac{1}{\varepsilon_0}
\int\! {\rm d}^3{\bf r}\,\hat{{\bf P}}_{\rm A}({\bf r})
\hat{\bf D}'^{\perp}({\bf r}) 
} 
\nonumber\\&&
-\sum_\alpha \frac{q_\alpha}{2m_{\alpha}} 
\int_0^1\!{\rm d}\lambda\,\lambda\left\{ 
\left[\left( \hat{\bf r}_\alpha\!-\!{\bf r}_A \right)
\times\hat{{\bf p}}'_{\alpha}\right]
\hat{\bf B}'\!\left[ {\bf r}_A\!+\!\lambda \left( \hat{\bf r}_\alpha
\!-\!{\bf r}_A \right) \right] + \mbox{H.c.}\right\}
\nonumber \\ &&
+\sum_\alpha \frac{q_\alpha^2}{2m_{\alpha}} \bigg\{ 
\int_0^1\!{\rm d}\lambda\,\lambda 
\left( \hat{\bf r}_\alpha\!-\!{\bf r}_A \right) \times
\hat{\bf B}'\!\left[ {\bf r}_A\!+\!\lambda \left( \hat{\bf r}_\alpha
\!-\!{\bf r}_A \right) \right]
\bigg\}^2 .
\end{eqnarray}
The first term on the right-hand side in Eq.~(\ref{2.76})
is a contact term  between the medium polarization and the
polarization of the atomic system. The second term
describes the interaction of the polarization of the atomic system with
the transverse part of the overall displacement field
[cf. Eq.~(\ref{2.75a})], 
and the last two terms refer to magnetic interactions.

So far we have transformed the dynamical variables but left unchanged
the Hamiltonian. Instead the Hamiltonian can be transformed to
obtain the new one
\begin{equation}
\label{2.77}
\hat{\cal{H}} = \hat{U}^\dagger \hat{H} \hat{U}.
\end{equation}
Obviously, the new  Hamiltonian expressed in terms of
the new variables formally looks like the untransformed
minimal-coupling Hamiltonian expressed in terms of the old
variables. Hence expressing in the new Hamiltonian the new
variables in terms of the old ones, we arrive at a
multipolar-coupling Hamiltonian that
(for a neutral atomic system) looks like the 
Hamiltonian given in Eq.~(\ref{2.75}), i.e.,
\begin{samepage}
\begin{eqnarray}
\label{2.79}
\lefteqn{
\hat{\cal H} = \int\! {\rm d}^3{\bf r} \int_0^\infty\! {\rm d} \omega
\,\hbar\omega\,\hat{\bf f}^{\dagger}({\bf r},\omega)
\hat{\bf f}({\bf r},\omega)
}
\nonumber\\&& 
+ \sum_\alpha \frac{1}{2m_{\alpha}} \bigg\{
\hat{{\bf p}}_{\alpha} + 
q_\alpha\int_0^1\!{\rm d}\lambda\,\lambda 
\left( \hat{\bf r}_\alpha\!-\!{\bf r}_A \right) \times
\hat{\bf B}\left[ {\bf r}_A\!+\!\lambda \left( \hat{\bf r}_\alpha
\!-\!{\bf r}_A \right) \right]
\bigg\}^2
\nonumber \\ &&
+\,\frac{1}{2\varepsilon_0}\int\! {\rm d}^3{\bf r} \,
\hat{\bf P}_{\rm A}({\bf r}) \hat{\bf P}_{\rm A}({\bf r})
+\,\frac{1}{\varepsilon_0}\int\! {\rm d}^3{\bf r} \,
\hat{\bf P}_{\rm A}({\bf r}) \hat{\bf P}_{\rm M}({\bf r}) 
\nonumber\\&&
-\,\frac{1}{\varepsilon_0}
\int\! {\rm d}^3{\bf r}\, \hat{{\bf P}}_{\rm A}({\bf r})
\hat{\bf D}{^\perp}({\bf r}). 
\end{eqnarray}
\end{samepage}
In fact, the Hamiltonians in Eqs.~(\ref{2.75}) and (\ref{2.79})
have different meanings. Since the expectation value
of an observable associated with an operator $\hat{O}$ must
not change, the use of $\hat{\cal{H}}$ necessarily requires a
transformation of both the operator [$\hat{O}$
$\!\to$ $\!\hat{U}^\dagger\hat{O}\hat{U}$]
and the state [$\hat{\varrho}$ $\!\to$
$\!\hat{U}^\dagger\hat{\varrho}\hat{U}$, with
$\hat{\varrho}$ being the density operator], so that
\begin{equation}
\label{2.81}
i\hbar \frac{{\rm d}}{{\rm d}t} \langle \hat{O} \rangle
= {\rm Tr}\!\left\{
\hat{\varrho} \big[ \hat{O}, \hat{H} \big] \right\}
= {\rm Tr}\!\left\{
\hat{U}^\dagger \hat{\varrho} \hat{U}
\big[ \hat{U}^\dagger \hat{O} \hat{U},
\hat{\cal H}
\big] \right\} .
\end{equation}
Moreover, in Eq.~(\ref{2.75}) the atomic polarization field 
couples to the transverse component of the overall displacement
field, which also contains the transverse component of the
atomic polarization, rather then the ordinary transverse
displacement field in Eq.~(\ref{2.79}).

\setcounter{equation}{0}
\section{Input--output coupling}
\label{sec:inoutcoup}

The quantization scheme developed in Sec.~\ref{sec:quantization}
is best suited to study the input--output behaviour of optical fields
at dielectric devices, because both dispersion and absorption are
exactly included in the resulting input--output
relations, and characteristic quantities such as the
transmission and reflection coefficients of the setup are
expressed in terms of its complex refractive-index profile
\cite{Gruner96b}. 
Input--output relations are a very efficient description of the action of
macroscopic bodies on radiation. In particular, they can advantageously be
used to obtain the quantum statistics of the outgoing light
from that of the incoming light, either in terms of radiation-field
correlation functions or directly in terms of the density matrix.
In what follows we will restrict our attention to four-port
devices. The extension of the method to (higher-order)
multiport-devices is straightforward.

\subsection{Operator input--output relations}
\label{sec:inout}

\begin{figure}[ht]
 \unitlength=1cm
 \begin{center}
 \begin{picture}(6.5,4)
 \put(0,0){
 \psfig{file=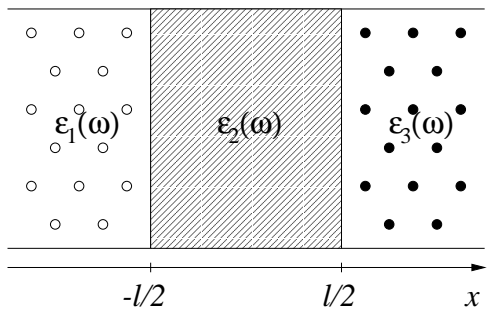,height=4cm}} 
 \end{picture}
 \end{center}
\caption{\label{fig:grunerslab}
Scheme of a three-layered structure: a dielectric plate of
thickness $l$ and permittivity $\varepsilon_2(\omega)$ is surrounded
by dielectric matter of permittivities $\varepsilon_1(\omega)$ (on the
left-hand side) and $\varepsilon_3(\omega)$ (on the right-hand side).}
\end{figure}

Let us study the problem of propagation (in the $x$-direction) of
quantized radiation through a dielectric plate that is the middle
part of a three-layered planar structure as sketched in 
Fig.~\ref{fig:grunerslab}.
The setup can be characterized by a piecewise constant permittivity
\begin{equation}
\label{ior1.1}
\varepsilon(x,\omega) = \sum_{j=1}^3 \lambda_j(x)
\varepsilon_j(\omega), \qquad \lambda_j(x) =
\left\{ \begin{array}{ll}
1 & \mbox{if \,$x_{j-1}<x<x_j$} \\
0 & \mbox{otherwise}
\end{array} \right. \,,
\end{equation}
where $\varepsilon_j(\omega)$ is the complex permittivity of the
$j$th domain, and $x_0$ $\!\to$ $\!-\infty$, $x_1$ $\!=$ $\!-l/2$,
$x_2$ $\!=$ $\!l/2$, and $x_3$ $\!\to$ $\!+$ $\!\infty$.
The Green function $G(x,x',\omega)$ satisfies the
partial differential equation (\ref{2.45}) from which it follows 
that it can be decomposed into two parts\footnote{For the
   calculation of Green tensors and a large variety of examples, 
   see \cite{Chew}.}
\begin{equation}
\label{ior1.2}
G(x,x',\omega) = \sum_{j=1}^3 \lambda_j(x) \lambda_j(x')
G_j^{\rm bulk}(x,x',\omega) 
+ R(x,x',\omega),
\end{equation}
where $G_j^{\rm bulk}(x,x',\omega)$ is the Green function of the bulk
material as given by Eq.~(\ref{2.46}) (with $\varepsilon_j$ in place
of $\varepsilon$), and the reflection term $R(x,x',\omega)$ is a
solution to the homogeneous wave equation,
\begin{equation}
\label{ior1.2a}
R(x,x',\omega) = \sum_{j=1}^3 \lambda_j(x)
\big[C_{j+}(x',\omega){\rm e}^{in_j(\omega)\omega x/c}
+C_{j-}(x',\omega){\rm e}^{-in_j(\omega)\omega x/c} 
\big],
\end{equation}
where the coefficients $C_{j\pm}(x',\omega)$ are to be determined in
such a way that continuity and differentiability at the surfaces of
discontinuity at $x$ $\!=$ $\!-l/2$ and $x$ $\!=$ $\!l/2$ are ensured.
Combining Eqs.~(\ref{2.44}) and (\ref{ior1.2}) [together with
Eq.~(\ref{ior1.2a})], the vector potential $\hat{A}(x)$ for the
$j$th domain may be represented as, similar to Eq.~(\ref{2.47}),
\begin{eqnarray}
\label{ior1.3}
\lefteqn{
\hat{A}(x)
= \int_0^\infty {\rm d}\omega\,
\Bigg\{
\sqrt{\frac{\hbar}{4\pi\varepsilon_0 c \omega n_{j\rm R}(\omega) {\cal A}}}
\frac{n_{j\rm R}(\omega)}{n_j(\omega)}
}
\nonumber \\[.5ex] && \hspace{2ex}
\times\,
\left[{\rm e}^{i n_{j\rm R}(\omega)\omega x/c} \hat{a}_{j+}(x,\omega) +
{\rm e}^{-i n_{j\rm R}(\omega)\omega x/c} \hat{a}_{j-}(x,\omega) \right]
+ \mbox{H.c.}
\Bigg\}
\qquad
\end{eqnarray}
[$x_{j-1}$ $\!\le$ $\!x$ $\!\le$ $\!x_j$], where the dependence on $x$ 
of the amplitude operators $\hat{a}_{j\pm}(x,\omega)$ is governed by
quantum Langevin equations of the type given in Eq.~(\ref{2.49a})
together with Eq.~(\ref{2.49}).

The amplitude operators of the incoming fields are given according
to Eq.~(\ref{2.48}) and satisfy the commutation relations\footnote{Note
  that the amplitude operators of the field inside the plate (domain $2$)
  and the amplitude operators of the outgoing fields are not given
  according to Eq.~(\protect\ref{2.48}) and thus do not satisfy 
  commutation relations of this type in general.} 
\begin{equation}
\label{ior1.4a}
\big[ \hat{a}_{1+}(x,\omega), \hat{a}_{1+}^\dagger(x',\omega')\big]
= {\rm e}^{-n_{1{\rm I}}(\omega) \omega|x-x'|/c}
\delta(\omega-\omega'),
\end{equation}
\begin{equation}
\label{ior1.4b}
\big[ \hat{a}_{3-}(x,\omega), \hat{a}_{3-}^\dagger(x',\omega')\big]
= {\rm e}^{-n_{3{\rm I}}(\omega) \omega |x-x'|/c}
\delta(\omega-\omega'),
\end{equation}
\begin{equation}
\label{ior1.4c}
\big[ \hat{a}_{1+}(x,\omega), \hat{a}_{3-}^\dagger(x',\omega') \big] = 0 
\end{equation}
[cf. Eq.~(\ref{2.49})]. Thus, the incoming fields from the left and
right behave like the fields in the corresponding bulk dielectrics
and may be regarded as independent variables.
Further, it can be shown that the amplitude operators
of the outgoing fields can be related to the amplitude
operators of the incoming fields and appropriately chosen
operators $\hat{g}_\pm(\omega)$ of the plate as
\begin{equation}
\label{ior1.5}
{\hat{a}_{1-}(-\frac{1}{2}l,\omega)\choose\hat{a}_{3+}(\frac{1}{2}l,\omega)}
= {\bf T}(\omega)
{\hat{a}_{1+}(-\frac{1}{2}l,\omega)\choose\hat{a}_{3-}(\frac{1}{2}l,\omega)}
+ {\bf A}(\omega)
{\hat{g}_+(\omega)\choose\hat{g}_-(\omega)},
\end{equation}
where the 2$\times$2-matrices ${\bf T}(\omega)$ and
${\bf A}(\omega)$ are the characteristic transformation
and absorption matrices of the plate expressed in terms of the
thickness and the permittivity of the plate and the permittivities
of the surrounding media, and the operators $\hat{g}_\pm(\omega)$ read  
\begin{eqnarray}
\label{ior1.5a}
\lefteqn{
\hat{g}_\pm(\omega) =
i\sqrt{\frac{\omega}{2c \lambda_\pm(l,\omega)}}
\,{\rm e}^{in_2(\omega) \omega l/(2c)}
}
\nonumber\\&&\times
\int_{-l/2}^{l/2}{\rm d}x'\,
\big[{\rm e}^{in_2(\omega) \omega x'/c}
\pm {\rm e}^{-in_2(\omega) \omega x'/c}
\big]\hat{f}(x',\omega),
\end{eqnarray}
with
\begin{equation}
\label{ior1.5b}
\lambda_\pm(l,\omega) = {\rm e}^{-n_{2{\rm I}}(\omega) \omega l/c}
\left\{
\frac{\sinh[ n_{2{\rm I}}(\omega)\omega l/c]}{n_{2{\rm I}}(\omega)}
\pm
\frac{\sin[ n_{2{\rm R}}(\omega)\omega l/c]}{n_{2{\rm R}}(\omega)}
\right\}
\end{equation}
(for details, see \cite{Gruner96b}).
It is not difficult to prove, on recalling the basic commutation
relations (\ref{2.26}) and (\ref{2.27}), that the commutation relations
\begin{equation}
\label{ior1.5c}
\big[ \hat{g}_\pm(\omega),\hat{g}^\dagger_\pm(\omega') \big]
= \delta(\omega-\omega'),
\end{equation}
\begin{equation}
\label{ior1.5d}
\big[ \hat{g}_\pm(\omega),\hat{g}^\dagger_\mp(\omega')
\big] = 0
\end{equation}
are valid.
Hence, the operators $\hat{g}_\pm(\omega)$ and
$\hat{g}^{\dagger}_\pm(\omega)$ are respectively annihilation
and creation operators of bosonic excitations associated with the
plate. Obviously, they commute with the amplitude operators of the
incoming fields $\hat{a}_{1+}(x,\omega)$ and $\hat{a}_{3-}(x,\omega)$
and are thus independent variables. It should be mentioned that
the output amplitude operators $\hat{a}_{1-}(x,\omega)$, $x$ $\!\le$
$\!-l/2$, and $\hat{a}_{3+}(x,\omega)$, $x$ $\!\ge$
$\!l/2$, can easily be obtained from $\hat{a}_{1-}(-l/2,\omega)$
and $\hat{a}_{3+}(l/2,\omega)$, respectively, by means of
the corresponding solutions of the Langevin equations (\ref{2.49a}).      

Whereas the input amplitude operators $\hat{a}_{1+}(-l/2,\omega)$, 
$\hat{a}^\dagger_{1+}(-l/2,\omega)$ and $\hat{a}_{3-}(l/2,\omega)$,
$\hat{a}^\dagger_{3-}(l/2,\omega)$ satisfy ordinary bosonic
commutation relations [see Eqs.~(\ref{ior1.4a}) -- (\ref{ior1.4c})],   
the amplitude operators of the outgoing fields,
$\hat{a}_{1-}(-l/2,\omega)$,
$\hat{a}^\dagger_{1-}(-l/2,\omega)$ and $\hat{a}_{3+}(l/2,\omega)$,
$\hat{a}^\dagger_{3+}(l/2,\omega)$ do not, if the plate
is surrounded by (absorbing) matter. When the plate is surrounded
by vacuum, then the characteristic transformation and
absorption matrices ${\bf T}(\omega)$ and ${\bf A}(\omega)$
fulfill the matrix relation\footnote{Note that a unitary
   transformation together with rescaling can be applied
   to the output amplitude operators such that the new
   operators are bosonic and thus Eq.~(\protect\ref{ior2.2}) 
   can be assumed to be valid, without loss of generality.}
\begin{equation}
\label{ior2.2}
{\bf T}(\omega){\bf T}^+(\omega) +
{\bf A}(\omega){\bf A}^+(\omega)
= {\bf I}
\end{equation}
(${\bf I}$, unit matrix), and from Eq.~(\ref{ior1.5}) it then
follows that the output amplitude operators also satisfy bosonic
commutation relations. In fact, it can be shown that
this is not only true at the very input and output ports
of the plate but in the whole half-spaces on the left
and the right. In this case both the input amplitude
operators and the output amplitude operators can be regarded
as being bosonic operators associated with ordinary
monochromatic incoming and outgoing modes, respectively,
and the matrices ${\bf T}(\omega)$ and ${\bf A}(\omega)$
read as [$n(\omega) \equiv n_2(\omega)$]
\begin{equation}
\label{ior1.5e}
T_{11}(\omega) = T_{22}(\omega) = {\rm e}^{-i\omega l/c} r(\omega)
\left[1 -t_1(\omega) {\rm e}^{2in(\omega) \omega l/c}
\vartheta(\omega) t_2(\omega) \right] ,
\end{equation}
\begin{equation}
\label{ior1.5f}
T_{12}(\omega) = T_{21}(\omega) = {\rm e}^{-i\omega l/c} 
t_1(\omega) {\rm e}^{in(\omega) \omega l/c} \vartheta(\omega)
t_2(\omega) ,
\end{equation}
\begin{eqnarray}
\label{ior1.5g}
\lefteqn{
A_{11}(\omega) = A_{21} (\omega) =
\sqrt{n_{\rm I}(\omega) n_{\rm R}(\omega)}\,
{\rm e}^{-i\omega l/(2c)} t_1(\omega) \vartheta(\omega)
}
\nonumber \\ && \hspace{20ex} \times\,
\sqrt{\lambda_+(l,\omega)}
\left[ 1- {\rm e}^{in(\omega) \omega l/c} r(\omega) \right],
\end{eqnarray}
\begin{eqnarray}
\label{ior1.5h}
\lefteqn{
A_{12}(\omega) = -A_{22}(\omega) =
\sqrt{n_{\rm I}(\omega) n_{\rm R}(\omega)}
{\rm e}^{-i\omega l/(2c)} t_1(\omega) \vartheta(\omega)
}
\nonumber \\&& \hspace{20ex}\times
\sqrt{\lambda_-(l,\omega)}
\left[ 1+ {\rm e}^{in(\omega) \omega l/c} r(\omega) \right] .
\end{eqnarray}
Here,
\begin{equation}
\label{ior1.5i}
r(\omega) = \frac{1-n(\omega)}{1+n(\omega)}
\end{equation}
and
\begin{equation}
\label{ior1.5j}
t_1(\omega) = \frac{2}{1+n(\omega)}\,,
\quad
t_2(\omega) = \frac{2n(\omega)}{1+n(\omega)}
\end{equation}
are the interface reflection and transmission coefficients, respectively, 
and the factor 
\begin{equation}
\label{ior1.5k}
\vartheta(\omega) = \left[ 1- r^2(\omega)
{\rm e}^{2in(\omega) \omega l/c} \right]^{-1} 
\end{equation}
arises from multiple reflections inside the plate.

\begin{figure}[ht]
 \unitlength=1cm
 \begin{center}
 \begin{picture}(8,4)
 \put(0,0){
 \psfig{file=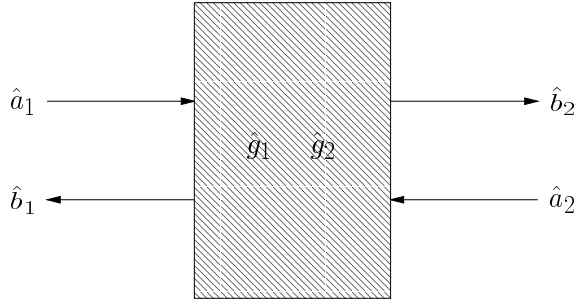,height=4cm}}
 \end{picture}
 \end{center}
\caption{\label{fig:slab} Scheme of a four-port device: Two incoming
fields [photonic operators $\protect\hat{a}_1(\omega)$
and $\protect\hat{a}_2(\omega)$] are superimposed to produce
two outgoing fields [photonic operators $\protect\hat{b}_1(\omega)$
and $\protect\hat{b}_2(\omega)$], the $\protect\hat{g}_j(\omega)$
being the operators of the relevant device excitations.}
\end{figure}

Operator input--output relations of the type given in
Eq.~(\ref{ior1.5}) are of course
valid also for more complicated four-port devices such as multilayer
plates. Obviously, the only difference consists in the actual expressions for 
the characteristic transformation and absorption matrices.
For notational reasons it will be convenient to call
the input amplitude operators $\hat{a}_i(\omega)$
[i.e., $\hat{a}_{1+}(-l/2,\omega)$ $\!\to$ $\!\hat{a}_1(\omega)$,
$\hat{a}_{3-}(l/2,\omega)$ $\!\to$ $\!\hat{a}_2(\omega)$],
the output amplitude operators $\hat{b}_i(\omega)$
[i.e., $\hat{a}_{1-}(-l/2,\omega)$ $\!\to$ $\!\hat{b}_1(\omega)$,
$\hat{a}_{3+}(l/2,\omega)$ $\!\to$ $\!\hat{b}_2(\omega)$],
the device operators $\hat{g}_i(\omega)$ [i.e.,
$\hat{g}_\pm(\omega)$ $\!\to$ $\!\hat{g}_i(\omega)$]
(see Fig.~\ref{fig:slab}), and to introduce the definitions
\begin{equation}
\label{ior2.1a}
\hat{\bf a}(\omega)
= {\hat{a}_1(\omega)\choose\hat{a}_2(\omega)},
\quad
\hat{\bf b}(\omega)
= {\hat{b}_1(\omega)\choose\hat{b}_2(\omega)},
\quad
\hat{\bf g}(\omega)
= {\hat{g}_1(\omega)\choose\hat{g}_2(\omega)}.
\end{equation}
The operator input-output relations (\ref{ior1.5}) at a four-port
device can then be written in the compact form of
\begin{equation}
\label{ior2.1}
\hat{\bf b}(\omega) = {\bf T}(\omega)  \hat{\bf a}(\omega)
+{\bf A}(\omega) \hat{\bf g}(\omega). 
\end{equation}
In what follows
we assume that the characteristic transformation and
absorption matrices ${\bf T}(\omega)$ and ${\bf A}(\omega)$,
respectively, obey the equation (\ref{ior2.2}).

\subsection{Quantum-state transformation}
\label{sec:state}

The operator input--output relations (\ref{ior2.1})
can be used to calculate various moments and correlations
of the outgoing fields in a straightforward way 
\cite{Gruner96b,Patra99}. The output operators are
expressed in terms of the input operators and the device
operators which for themselves act on the quantum state
the incoming field and the device are prepared in. 
Instead the photonic operators may be left unchanged
but the quantum state is transformed.
This equivalent procedure is suitable in view of
problems, such as the determination of the
entanglement of the outgoing fields, where 
knowledge of the quantum state of the outgoing field
as a whole is required. Hence, we are interested in a
unitary transformation that transforms
the input-state density operator $\hat{\varrho}_{\rm in}$
(i.e., the density operator of the quantum state the incoming field
and the device are prepared in) into an output-state density operator
$\hat{\varrho}_{\rm out}$ according to 
\begin{equation}
\label{ior2.2a}
\hat{\varrho}_{\rm out} = \hat{U} \hat{\varrho}_{\rm in}
\hat{U}^\dagger,
\end{equation}
from which the density operator of the state the outgoing field is
prepared in can be obtained by taking the trace with regard
to the device variables.
  
Let us shortly digress to lossless devices and assume, for a moment,
that \mbox{${\bf A}(\omega)$ $\!=$ $\!0$}. It is clear from
Eq.~(\ref{ior2.2}) that in this case the characteristic transformation
matrix ${\bf T}(\omega)$ must be a unitary matrix and thus represents
for each $\omega$ [and ${\rm det}\,{\bf T}(\omega)$ $\!=$ $\!1$] an
element of the group SU(2)
\cite{Yurke86,Prasad87,Ou87,Fearn87,Campos89,Leonhardt93}.
For absorbing devices this can surely not be the 
case. Because of the coupling to the environment, we will definitely
not be able to construct any unitary transformation
that acts on the electromagnetic field operators alone.
But we may look for one in the larger Hilbert space that comprises
both the electromagnetic field
and the device. Introducing some
auxiliary field operators $\hat{\bf h}(\omega)$ and defining
the ``four-vectors''
\begin{equation}
\label{ior2.2b}
\hat{\mbb{\alpha}}(\omega) =
{\hat{\bf a}(\omega) \choose \hat{\bf g}(\omega)}, 
\quad
\hat{\mbb{\beta}}(\omega) =
{\hat{\bf b}(\omega) \choose \hat{\bf h}(\omega)},
\end{equation}
we may extend the input--output relations (\ref{ior2.1}) to the
four-form of
\begin{equation}
\label{ior2.3}
\hat{\mbb{\beta}}(\omega) = \mbb{\Lambda}(\omega) 
\hat{\mbb{\alpha}}(\omega), 
\end{equation}
where $\mbb{\Lambda}(\omega)$ is a unitary
4$\times$4-matrix, hence $\mbb{\Lambda}(\omega)
\mbb{\Lambda}^+(\omega)$ $\!=$ $\!\mbb{I}$. After separation of
some phases from the matrices 
${\bf T}(\omega)$ and ${\bf A}(\omega)$  and inclusion of them in the
input operators, the matrix $\mbb{\Lambda}(\omega)$ can be regarded
(for each $\omega$) as an element of the group SU(4), and it can be
expressed in terms of the matrices ${\bf T}(\omega)$ and
${\bf A}(\omega)$ \cite{Knoll99} as
\begin{equation}
\label{ior2.3a}
\mbb{\Lambda}(\omega) =
\left(\!\!
\begin{array}{cc}
{\bf T}(\omega) & {\bf A}(\omega)\\[.5ex]
-{\bf S}(\omega)  {\bf C}^{-1}(\omega) 
{\bf T}(\omega) & {\bf C}(\omega) 
{\bf S}^{-1}(\omega)  {\bf A}(\omega)
\end{array}
\!\!\right), 
\end{equation}
where
\begin{equation}
\label{ior2.3b}
{\bf C}(\omega) = \sqrt{{\bf T}(\omega){\bf T}^+(\omega)}
\end{equation}
and
\begin{equation}
\label{ior2.3c}
{\bf S}(\omega) = \sqrt{{\bf A}(\omega){\bf A}^+(\omega)}
\end{equation}
are commuting positive Hermitian $2\times 2$-matrices. Note that
${\bf C}^2(\omega)$ $\!+$ $\!{\bf S}^2(\omega)$ $\!=$ ${\bf I}$.

The matrix transformation in Eq.~(\ref{ior2.3}) may be
realized also as a unitary operator transformation  
\begin{equation}
\label{neu01}
\hat{\mbb{\beta}}(\omega) = \hat{U}^\dagger
\hat{\mbb{\alpha}}(\omega)\hat{U}
= \mbb{\Lambda}(\omega)\hat{\mbb{\alpha}}(\omega), 
\end{equation} 
where
\begin{equation}
\label{neu01a}
\hat{U} = \exp\! \left\{ -i \int_0^\infty \!{\rm d}\omega \big[
\hat{\mbb{\alpha}}^\dagger(\omega) \big]^T 
\mbb{\Phi}(\omega)  \hat{\mbb{\alpha}}(\omega) \right\},
\end{equation}
with the $4\times 4$-matrix $\mbb{\Phi}(\omega)$ being defined
according to
\begin{equation}
\label{neu01b}
\exp[-i\mbb{\Phi}(\omega)] = \mbb{\Lambda}(\omega) .
\end{equation}
Obviously, $\hat{U}$ is just the unitary operator that transforms
$\hat{\varrho}_{\rm in}$ into $\hat{\varrho}_{\rm out}$ in
Eq.~(\ref{ior2.2a}). Since the input density operator
can be regarded as being an operator functional of 
$\hat{\mbb{\alpha}}(\omega)$ and $\hat{\mbb{\alpha}}^\dagger(\omega)$,
$\hat{\varrho}_{\rm in}$ $\!=$
$\!\hat{\varrho}_{\rm in}[\hat{\mbb{\alpha}}(\omega),\hat{\mbb{\alpha}}^\dagger(\omega)]$, 
from Eqs.~(\ref{ior2.2a}) and (\ref{neu01}) it follows that
the transformed density operator can be given by 
\begin{equation}
\label{neu01c}
\hat{\varrho}_{\rm out} = \hat{\varrho}_{\rm in} \big[
\mbb{\Lambda}^+(\omega)  \hat{\mbb{\alpha}}(\omega),
\mbb{\Lambda}^T(\omega)  \hat{\mbb{\alpha}}^\dagger(\omega) \big].
\end{equation}
Projecting $\hat{\varrho}_{\rm out}$ onto the Hilbert space
of the radiation field then yields the
density operator of the outgoing fields 
\begin{equation}
\label{ior2.4}
\hat{\varrho}^{({\rm F})}_{\rm out} = {\rm Tr}^{({\rm D})}\!\left\{
\hat{\varrho}_{\rm in} \big[ \mbb{\Lambda}^+(\omega)
 \hat{\mbb{\alpha}}(\omega), \mbb{\Lambda}^T(\omega)
 \hat{\mbb{\alpha}}^\dagger(\omega) \big] \right\},
\end{equation}
where ${\rm Tr}^{({\rm D})}$ means the trace with respect to the
device.\footnote{The similarity to the usual open-systems approach to
   dissipation is not 
   accidental. However, master (or related) equations, to which an
   open-systems theory would lead \cite{Giulini}, are not required here,
   because the action of the environment (i.e., the device) is explicitly
   known. For chosen quantum state the device is initially prepared in,
   all the necessary information is contained in the characteristic
   transformation and absorption matrices
   ${\bf T}(\omega)$ and ${\bf A}(\omega)$, which in turn are
   determined by the Green function of the phenomenological
   Maxwell equations of the classical problem.} 

It is often useful and illustrative to describe quantum states in terms
of phase-space functions such as the familiar $s$-parametrized
phase-space functions (see, e.g., \cite{Cahill69,Perina}).
{F}rom Eq.~(\ref{neu01c}) it follows that the $s$-parametrized
phase-space functional $P_{\rm out}[\mbb{\alpha}(\omega);s]$ that
corresponds to $\hat{\varrho}_{\rm out}$ is simply
given by\footnote{Since $\omega$ is continuous,
   $P_{\rm in}[\mbb{\alpha}(\omega);s]$ and
   $P_{\rm out}[\mbb{\alpha}(\omega);s]$  are functionals
   rather than functions.}
\begin{equation}
\label{ior2.4a}
P_{\rm out}[\mbb{\alpha}(\omega);s] =
P_{\rm in}\!\left[\mbb{\Lambda}^+(\omega)
\mbb{\alpha}(\omega);s\right],
\end{equation}
so that the phase-space functional of the outgoing radiation reads as
\begin{equation}
\label{ior2.4b}
P_{\rm out}^{({\rm F})}[\mbb{\alpha}(\omega);s]
= \int {\cal D}{\bf g} \,
P_{\rm out}[\mbb{\alpha}(\omega);s]
= \int {\cal D}{\bf g}\,
P_{\rm in}\!\left[\mbb{\Lambda}^+(\omega)
\mbb{\alpha}(\omega);s\right] ,
\end{equation}
where the functional integration (notation ${\cal D}{\bf g}$)
is taken over the continua of the complex phase-space variables
$g_1(\omega)$ and $g_2(\omega)$ of the dielectric device.
In Eq.~(\ref{ior2.4a}) we have used the fact that
application of the unitary transformation under consideration
implies preservation of operator ordering, i.e., the
annihilation and creation operators are not mixed 
by the quantum-state transformation. It should be pointed
out that this is not the case for amplifying devices \cite{Scheel00}. 

For practical purposes a description of the incoming and outgoing
radiation in terms of discrete (quasi-mono\-chro\-matic) modes is
frequently preferred to be used. For this,  we divide the
frequency axis into sufficiently small intervals\footnote{The
   matrices ${\bf T}(\omega)$ and ${\bf A}(\omega)$
   must not change substantially over an interval.}
of mid-frequencies $\omega_m$ and widths $\Delta\omega_m$
and define the discrete photonic input operators
\begin{equation}
\label{2.5100}
\hat{\mbb{\alpha}}_m = \frac{1}{\sqrt{\Delta\omega_m}}
\int_{\Delta\omega_m} \!{\rm d}\omega \,\hat{\mbb{\alpha}}(\omega),
\end{equation}
and the discrete photonic output operators $\hat{\mbb{\beta}}_m$
accordingly. Then we assign to each pair of
operators $\hat{\mbb{\alpha}}_m$ and $\hat{\mbb{\beta}}_m$ the
input--output relation (\ref{ior2.3}) with the 4$\times$4-matrix
$\mbb{\Lambda}_m=\mbb{\Lambda}(\omega_m)$, and, according to
Eq.~(\ref{neu01a}), the unitary operator $\hat{U}$ then reads as
\begin{equation}
\label{2.5200}
\hat{U} = \prod_m \hat{U}_m ,
\end{equation}
where
\begin{equation}
\label{2.5300}
\hat{U}_m 
= \exp\!\left[-i 
\big(\hat{\mbox{\boldmath $\alpha$}}_m^\dagger\big)^T
\mbox{\boldmath $\Phi$}_m
\hat{\mbox{\boldmath $\alpha$}}_m
\right]
\end{equation}
with $\mbox{\boldmath $\Phi$}_m$ $\!=$ $\!\mbox{\boldmath
$\Phi$}(\omega_m)$. In particular, the functional integral in
Eq.~(\ref{ior2.4b}) becomes an ordinary multiple integral.

\subsubsection{Examples}

Let us restrict our attention to (quasi-)monochromatic
fields, so that it is sufficient to consider
only a single frequency component $\omega_m$.  
Suppose the incoming field and the device are prepared in coherent
states\footnote{For notational convenience we
   omit the subscript $m$.}
\begin{equation}
\label{ior2.4c}
|\psi_{\rm in}\rangle = |\mbb{\gamma}\rangle = \exp\big(
\mbb{\gamma}^T  \hat{\mbb{\alpha}}^\dagger
-\mbb{\gamma}^+ \hat{\mbb{\alpha}} \big) |0\rangle,
\qquad 
\mbb{\gamma} = {{\bf c} \choose {\bf d}},
\end{equation}
with $c_j$ and $d_j$ ($j$ $\!=$ $\!1,2$), respectively,
being the coherent-state amplitudes of the input fields and the device.
Application of Eq.~(\ref{neu01c}) yields $|\psi_{\rm out}\rangle$ $\!=$
$\!|\mbb{\gamma}'\rangle$, with $\mbb{\gamma}'$ $\!=$
$\!\mbb{\Lambda}\mbb{\gamma}$ in place of
$\mbb{\gamma}$ in Eq.~(\ref{ior2.4c}), and it follows, on applying
Eq.~(\ref{ior2.4}), that the outgoing fields are prepared
in coherent states, i.e.,
\begin{equation}
\label{ior2.4d}
\hat{\varrho}^{({\rm F})}_{\rm out} =
|{\bf c}'\rangle \langle {\bf c'}|,
\qquad
{\bf c'} = {\bf T}{\bf c}+{\bf A}{\bf d}.
\end{equation}
Thus, the coherent-state amplitudes $c_1'$ and $c_2'$ of the outgoing
fields are not only determined by the characteristic transformation
matrix ${\bf T}$ but also by the absorption matrix ${\bf A}$ via the
coherent-state amplitudes of the device.

Next let us consider the case where the field in one of the
two input channels is prepared in an $n$-photon Fock state and
the field in the other input channel and the
device are left in vacuum, i.e., $|\psi_{\rm in}\rangle$ $\!=$
$\!|n000\rangle$. The Wigner function of the input state
reads\footnote{$W_{\rm in}(\mbb{\alpha})$
   $\!=$ $\!W_n(a_1)W_0(a_2)W_0(g_1)W_0(g_2)$, where
   $W_k(a)$ $\!=$ $\!(2/\pi)(-1)^k  L_k(4|a|^2){\rm e}^{-2|a|^2}$
   [$L_k(z)$, Laguerre polynomial] is the Wigner function of
   a $k$-quanta Fock state (see, e.g., \cite{Gardiner}).}
\begin{equation}
\label{ior2.4e}
W_{\rm in}(\mbb{\alpha}) = (-1)^n \left( \frac{2}{\pi}
\right)^4 L_n\!\left( 4|a_1|^2 \right)
\exp\big( -2|\mbb{\alpha}|^2\big).
\end{equation}
Applying Eqs.~(\ref{ior2.4a}) and (\ref{ior2.4b})
and integrating over the phase space of one outgoing field,
the Wigner function of the field in the $j$th output channel
is derived to be \cite{Knoll99}
\begin{equation}
\label{ior2.5}
W_{\rm out}^{({\rm F})}\left( a_j \right) =
\sum_{k=0}^n {n \choose k}
|T_{j1}|^{2k} \left( 1-|T_{j1}|^2 \right)^{n-k} W_k(a_j).
\end{equation}
Obviously, Eq.~(\ref{ior2.5}) does not only hold for the Wigner
function but for any $s$-parametrized phase-space
function, and hence the corresponding density operator 
reads in the Fock basis as
\begin{equation}
\label{ior2.6}
\hat{\varrho}^{({\rm F})}_{{\rm out}\,j} = \sum_{k=0}^n {n \choose k}
|T_{j1}|^{2k} \left( 1-|T_{j1}|^2 \right)^{n-k} |k\rangle\langle k|.
\end{equation}
Note that the quantum state the outgoing field is prepared in contains
only Fock states up to the photon number $n$ of the input Fock
state. That is a direct consequence of the action of the compact group
SU(4) which leaves the total number of quanta unchanged.
Furthermore, the Fock state with the highest photon number (i.e., the
input state) appears in the output state
with a weight $|T_{j1}|^{2n}$, that is, the probability
of finding the same number of photons after transmission through
(reflection at) an absorbing device decreases as $|T_{j1}|^{2n}$. 
In fact, for a device in the ground state, as it is the case here,
there is some classical reasoning explaining this result. One could
imagine that each of the $n$ photons ``feels'' the effect of a
transmission (reflection) coefficient smaller than unity, which for
$n$ photons amounts to the $n$th power of the transmission
(reflection) coefficient. This reasoning fails when the device is
prepared in anything else then the ground state.

Finally, let us assume that the field in one of the
two input channels is prepared in a Schr\"odinger-cat state
and the field in the other input channel and the
device are left in vacuum, i.e., 
\begin{equation}
|\psi_{\rm in}\rangle
= \frac{1}{\sqrt{N}}
\left( |\gamma\rangle +|-\gamma\rangle \right)|000\rangle,
\end{equation}
where $|\gamma\rangle$ is a coherent state, and
$N$ $\!=$ $\!2 [ 1$ $\!+$ $\!\exp(-2|\gamma|^2)]$ the proper
normalization factor. Using Eq.~(\ref{ior2.4d}) with 
\mbox{$c_1$ $\!=$ $\!\pm\gamma$}, $c_2$ $\!=$ $\!d_1$ $\!=$ $\!d_2$
$\!=$ $\!0$, 
it is not difficult to derive the density operator of the
field in the $j$th output channel. The result is
\begin{eqnarray}
\label{ior2.8}
\lefteqn{
\hat{\varrho}^{({\rm F})}_{{\rm out}\,j} = \frac{1}{N} \Big\{
|\gamma T_{j1}\rangle \langle \gamma T_{j1}|
+|-\gamma T_{j1}\rangle \langle -\gamma T_{j1}|
}
\nonumber \\ && 
+\big(
|\gamma T_{j1}\rangle \langle -\gamma T_{j1}|
+|-\gamma T_{j1}\rangle \langle \gamma T_{j1}| \big)
\exp\!\left[-2|\gamma|^2( 1-|T_{j1}|^2)\right] 
\Big\}.
\qquad
\end{eqnarray}
Whereas the two peaks decay as $|T_{j1}|^2$, the quantum
interference, in contrast, decays exponentially
as $|T_{j1}|^2\exp[-2|\gamma|^2( 1-|T_{j1}|^2)]$. The larger
the mean number of photons
$\langle\hat{n}\rangle$ $\!=$ $\!|\gamma|^2 \tanh|\gamma|^2$
becomes, the faster is the decay.
Let us consider, e.g., the transmitted Schr\"odinger cat
($j$ $\!=$ $\!2$). {F}rom Eq.~(\ref{ior2.2}) it follows that
the squares of the absolute values of the transmission
coefficient $|T_{21}|^2$, the reflection coefficient
$|T_{22}|^2$, and the absorption coefficients
$|A_{21}|^2$ and $|A_{22}|^2$  are related to each other as
$1$ $\!-$ $\!|T_{21}|^2$ $\!=$ $\!|T_{22}|^2$ $\!+$ $\!|A_{21}|^2$
$\!+$ $\!|A_{22}|^2$, so that $\exp[-2|\gamma|^2( 1-|T_{j1}|^2)]$ $\!=$
$\!\exp[-2|\gamma|^2|T_{22}|^2)$ $\!\exp[-2|\gamma|^2(|A_{21}|^2$
$\!+$ $\!|A_{22}|^2)]$. The terms $\exp(-2|\gamma|^2|T_{22}|^2)$
and $\exp[-2|\gamma|^2(|A_{21}|^2$ $\!+$ $\!|A_{22}|^2)]$
then respectively describe the decoherence associated with the
losses owing to reflection and absorption.

\subsubsection{Entanglement degradation}
\label{channel}

Quantum information processing such as quantum teleportation and quantum
cryptography \cite{Bennett93,Braunstein98,Furusawa98,Bennett84,Peres}
is essentially based on entanglement,
which can be regarded as being the nonclassical contribution to
the overall correlation between two parts of a system. In particular,
when two fields that are prepared in nonclassical states are
superimposed by a four-port device, then the two outgoing
fields are prepared in an entangled state in general.
Entanglement of a bipartite quantum state $\hat{\varrho}$ is commonly
quantified by measures which fulfill some basic requirements as
non-negativity (being zero only for separable states), invariance
under local unitary transformations of the subsystems, and
non-increase under arbitrary positive trace-preserving maps
\cite{Vedral98,Horodecki99}. 
Further, the reduced von Neumann entropy should be recovered for
pure states which in addition gives a normalization condition for
the measure. So far, the distance $E(\hat{\varrho})$ of $\hat{\varrho}$
to the set ${\cal S}$ of all separable quantum states
$\hat{\sigma}$\footnote{A quantum state $\hat{\sigma}$ of a system
  that consists of two subsystems $A$ and $B$ is called separable
  if \mbox{$\hat{\sigma}$ $\!=$ $\!\sum_i p_i \hat{\sigma}^{(A)}_i
  \otimes \hat{\sigma}^{(B)}_i$}, where $\hat{\sigma}^{(A)}_i$ and
  $\hat{\sigma}^{(B)}_i$ represent quantum states of the subsystems,
  and \mbox{$\sum_i p_i$ $\!=$ $\!1$}.}
which is measured by means of the relative entropy has been proven
to be one measure satisfying all the given conditions
\cite{Vedral98}. Thus, 
\begin{equation}
\label{ior3.2}
E(\hat{\varrho}) = \min_{\hat{\sigma}\in{\cal S}} {\rm Tr}\!\left[
\hat{\varrho} \left( \ln \hat{\varrho} -\ln \hat{\sigma} \right)
\right]. 
\end{equation}

Let us consider the entanglement degradation that is observed
when two fields that are prepared in a Bell basis
state\footnote{For the entanglement degradation of a two-mode
   squeezed vacuum, see \cite{Scheel00b}.} 
\begin{equation}
\label{ior3.1a}
|\Psi^\pm\rangle = \frac{1}{\sqrt{2}} \left(
|01\rangle \pm |10\rangle
\right)
\end{equation}
or
\begin{equation}
\label{ior3.1b}
|\Phi^\pm\rangle = \frac{1}{\sqrt{2}} \left(
|00\rangle \pm |11\rangle
\right)
\end{equation}
are transmitted through absorbing four-port devices prepared
in the ground state.\footnote{For an application to photon
   tunneling through absorbing dielectric barriers, see
   \cite{Gruner97}.}
Note that the entanglement of each of the Bell basis states
(\ref{ior3.1a}) and (\ref{ior3.1b}) is equal to $\ln 2$ (sometimes
also named $1$ bit). 
Applying Eqs.~(\ref{neu01c}) and (\ref{ior2.4}) to the input states
$|\psi_{\rm in}\rangle$ $\!=$
$\!|\Psi^\pm\rangle\otimes|0\rangle_{\rm D}$
and $|\psi_{\rm in}\rangle$ $\!=$
$\!|\Phi^\pm\rangle\otimes|0\rangle_{\rm D}$
($|0\rangle_{\rm D}$, ground state of the two devices),
we easily find that the outgoing fields are
prepared in the quantum states  
\begin{eqnarray}
\label{ior3.4a}
\lefteqn{
\hat{\varrho}_{\rm out}^{\rm (F)}\big(|\Psi^\pm\rangle)
= {\textstyle\frac{1}{2}} \left[ \left( 2-|T_1|^2-|T_2|^2 \right)
|00\rangle\langle 00| \right]
}
\nonumber \\ &&\hspace{4ex}
+{\textstyle\frac{1}{2}} \left( T_2 |01\rangle \pm T_1 |10\rangle \right)
\left( T_2^\ast \langle 01| \pm T_1^\ast \langle 10| \right),
\end{eqnarray}
\begin{eqnarray}
\label{ior3.4b}
\lefteqn{
\hat{\varrho}_{\rm out}^{\rm (F)}\big(|\Phi^\pm\rangle)
= {\textstyle\frac{1}{2}} \left[ \left( 1-|T_1|^2 \right)
\left( 1-|T_2|^2 \right) |00\rangle\langle 00|\right.
}
\nonumber \\ && 
\left. +|T_1|^2 \left( 1-|T_2|^2 \right) |10\rangle\langle 10|
+|T_2|^2 \left( 1-|T_1|^2 \right) |01\rangle\langle 01| \right]
\nonumber \\ &&
+{\textstyle\frac{1}{2}} \left( |00\rangle \pm T_1T_2 |11\rangle \right)
\left( \langle 00| \pm (T_1T_2)^\ast \langle 11| \right).
\end{eqnarray}
Here and in the following the notation $(T_l)_{11}$ $\!=$
$\!(T_l)_{22}$ $\!\equiv$ $\!R_l$ and $(T_l)_{12}$ $\!=$ 
$\!(T_l)_{21}$ $\!\equiv$ $\!T_l$ for the elements of the
characteristic transformation matrix ${\bf T}_l$ of the
$l$th four-port device is used ($l$ $\!=$ $\!1,2$) [cf.
Eqs.~(\ref{ior1.5e}) and (\ref{ior1.5f})].
The entanglement contained in the quantum states of the
outgoing fields, $\hat{\varrho}_{\rm out}^{\rm (F)}(|\Psi^\pm\rangle)$ 
and $\hat{\varrho}_{\rm out}^{\rm (F)}(|\Phi^\pm\rangle)$,
can be estimated by using the convexity property of the
relative entropy.\footnote{$E(\sum_i
   p_i \hat{\varrho}_i)$ $\!\le$ $\!\sum_i p_i E(\hat{\varrho}_i)$,
   where $\sum_i p_i$ $\!=$ $\!1$ \cite{Wehrl78}.
   The partition into separable states and a single pure state is not
   unique. Is has been proven, however, that for a pair of
   spin-$\textstyle\frac{1}{2}$ parties there exists a unique
   (sometimes called optimal) decomposition such that the weight of
   the separable state is maximized \cite{Lewenstein98}. The reduced
   von Neumann entropy of the extracted pure state  with corresponding
   minimal weight is then exactly the amount of entanglement, hence
   the inequality reduces to an equality.}
In Eqs.~(\ref{ior3.4a}) and (\ref{ior3.4b}) the output state 
is written as a sum of separable states and a single pure
state. Since separable states have zero entanglement
by definition, the entanglement of the whole state is bounded from
above by the reduced von Neumann entropy of the respective pure
states. If we assume equal transmission coefficients
of the two devices ($T_1$ $\!=$ $\!T_2$ $\!=$ $\!T$), the bounds
are given according to \cite{Scheel00}
\begin{equation}
\label{ior3.6a}
E\big[\hat{\varrho}^{\rm (F)}_{\rm out}(|\Psi^\pm\rangle)\big]
\le |T|^2 \ln 2 
\end{equation}
and
\begin{equation}
\label{ior3.6b}
E\big[\hat{\varrho}^{\rm (F)}_{\rm out}(|\Phi^\pm\rangle)\big]
\le {\textstyle\frac{1}{2}}
\left[ \left( 1+|T|^4 \right) \ln \left( 1+|T|^4 \right)
-|T|^4 \ln |T|^4 \right] \,.
\end{equation}

Suppose the two fields are transmitted through (equal) optical fibres
with perfect input coupling ($R$ $\!=$ $\!0$), so that the 
transmission coefficient $T$ may be given by
\begin{equation}
\label{ior3.7}
T = {\rm e}^{in\omega l/c}
= {\rm e}^{in_{\rm R}\omega l/c} {\rm e}^{-l/L},
\end{equation}
with $l$ and $L$ $\!=$ $\!c/(n_{\rm I}\omega)$ being 
respectively the propagation length through the fibres and
the absorption length of the fibres. Substituting of this
expression into the inequalities (\ref{ior3.6a}) and
(\ref{ior3.6b}) yields the dependence on $l$ of the
bounds of entanglement. 
\begin{figure}[h]
 \unitlength=1cm
 \begin{center}
 \begin{picture}(7.5,5)
 \put(0,0){
 \psfig{file=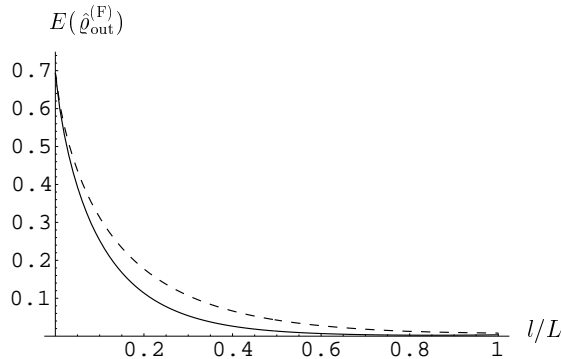,width=7.5cm}}
 \end{picture}
 \end{center}
\caption{\label{fig:vergleich}
Entanglement degradation of Bell basis states
\protect$|\Phi^\pm\rangle$ (full curve) and \protect$|\Psi^\pm\rangle$
(dashed curve) after transmission through absorbing fibres 
as a function of the fibre length.}
\end{figure}
The exact dependence on $l$ of the entanglement degradation
calculated by applying Eq.~(\ref{ior3.2}) is shown in
Fig.~\ref{fig:vergleich}. One observes that the states
$|\Phi^\pm\rangle$ decay faster than the states $|\Psi^\pm\rangle$.
Since the device has been left in the ground state, we can
again use some classical reasoning to explain this behaviour.
When the two fields are initially prepared in a state
$|\Psi^\pm\rangle$, then only a single photon (i.e., either the photon
from the first field or the photon from the second one) is effectively
subject to absorption. By contrast, the two photons
get effected simultaneously, if the initial state is
a state $|\Phi^\pm\rangle$. 

\setcounter{equation}{0}
\section{Spontaneous decay}
\label{sec:decay}

Spontaneous emission is a prime example of the action of
ground-state fluctuations on physically measurable processes. Einstein
\cite{Einstein17} already pointed out that, in order to obtain the
Planck radiation law, a process as spontaneous emission must
necessarily be included in the theory of atomic decay. Later on, the
radiation properties of an excited atom located in free space have been 
a subject of many studies. In particular, the rate of spontaneous
emission of an excited (two-level) atom in free space is given
by (the famous Einstein $A$-coefficient)
\begin{equation}
\label{se1.1}
\Gamma_0 = \frac{\omega_{\rm A}^3 d^2}{3\pi \hbar \varepsilon_0 c^3}\,,
\end{equation}
where $d$ is the absolute value of the matrix element
${\bf d}$ $\!=$ $\!\langle l| \hat{\bf d}_{\rm A} |u\rangle$
of the atomic dipole operator 
$\hat{\bf d_{\rm A}}$ [Eq.~(\ref{se1.7})] between the upper
state $|u\rangle$ and the lower
state $|l\rangle$, and $\omega_{\rm A}$ is the corresponding
atomic transition frequency.
When the atom is surrounded by (dielectric) matter, then
the ground state felt by the radiating atom is changed and
thus the rate formula (\ref{se1.1}) must be corrected\footnote{In
  the strong-coupling regime the decay becomes non-exponential
  and cannot be described by a rate (Sec.~\ref{sec:beyond}).}
\cite{Agarwal75,Wylie84,Meschede90,Barnett92,Juzeliunas94,Fichet95,%
Barnett96,Koshino96,Yeung96,Juzeliunas97,Tomas97,Scheel99a,Scheel99b,%
Fleischhauer99a,Tomas99,Fleischhauer99b}.  

\subsection{Time dependence of the atom--field system}
\label{sec:twolevel}

\subsubsection{Hamiltonian}

Let us assume that the surrounding matter can be regarded
as being a dielectric of given complex permittivity. 
Applying the minimal-coupling scheme (Sec.~\ref{section2.4.3.1}),
we may decompose the Hamiltonian [Eq.~(\ref{2.51})] of the coupled
system consisting of an atom and the medium-assisted
electromagnetic field as
\begin{equation}
\label{1.3a}
\hat{H} = \hat{H}_{\rm A} + \hat{H}_{\rm M} +\hat{H}_{\rm int},
\end{equation}
where
\begin{equation}
\label{1.3b}
\hat{H}_{\rm A} = \sum_\alpha \frac{\hat{\bf p}_\alpha^2}{2m_\alpha}
+ {\textstyle\frac{1}{2}} \int\!{\rm d}^3{\bf r} \,
\hat{\rho}_{\rm A}({\bf r}) \hat{\varphi}_{\rm A}({\bf r})
\end{equation}
is the Hamiltonian of the atom, 
\begin{equation}
\label{se1.3}
\hat{H}_{\rm M} = \int \!{\rm d}^3{\bf r} \int_0^\infty \!
{\rm d} \omega \, \hbar \omega \, \hat{\bf f}^\dagger({\bf r},\omega)
 \hat{\bf f}({\bf r},\omega) 
\end{equation}
is the Hamiltonian of the electromagnetic field and the dielectric
matter, and
\begin{equation}
\label{1.3c}
\hat{H}_{\rm int} = -\sum_\alpha \frac{q_\alpha}{m_\alpha}
\hat{\bf p}_\alpha  \hat{\bf A}({\bf r}_\alpha) +
\int\!{\rm d}^3{\bf r} \, \hat{\rho}_{\rm A}({\bf r})
\hat{\varphi}_{\rm M}({\bf r})
\end{equation}
is the interaction energy. Here we have omitted the
$\hat{\bf A}^2$ term.
In the electric-dipole approximation, the first term
on the right-hand side of Eq.~(\ref{1.3c}) simplifies to
\begin{equation}
\label{se1.7}
-\sum_\alpha \frac{q_\alpha}{m_\alpha} \hat{\bf p}_\alpha 
\hat{\bf A}({\bf r}_\alpha)
= - \frac{1}{i\hbar} \big[
\hat{\bf d}_{\rm A}, \hat{H}_{\rm A} \big] 
\hat{\bf A}({\bf r}_{\rm A}), 
\end{equation}
where
\begin{equation}
\hat{\bf d}_{\rm A}
= \sum_\alpha q_\alpha \hat{\bf r}_\alpha 
\end{equation}
is the atomic dipole operator.
Restricting our attention to a two-level system, the atomic
Hamiltonian $\hat{H}_{\rm A}$ reduces to
\begin{equation}
\label{se1.7a}
\hat{H}_{\rm A} = \hbar\omega_u |u\rangle\langle u| + \hbar \omega_l
|l\rangle\langle l|
= {\textstyle\frac{1}{2}}\hbar\omega_{\rm A}\hat{\sigma}_z
+ \mbox{const.},
\end{equation}
where $\hat{\sigma}_z$ $\!=$ $\!|u\rangle\langle u|$ $\!-$
$\!|l\rangle\langle l|$. 
Thus we may further simplify Eq.~(\ref{se1.7}) to obtain,
on applying the rotating wave approximation and
recalling Eqs.~(\ref{2.38}) and (\ref{2.38a}),
\begin{equation}
\label{se1.8}
-\sum_\alpha \frac{q_\alpha}{m_\alpha} \hat{\bf p}_\alpha  
\hat{\bf A}({\bf r}_\alpha)
= - \hat{\sigma}^\dagger
\hat{\bf E}_{\rm M}^{\perp(+)}({\bf r}_{\rm A})  {\bf d}
+\mbox{H.c.} 
\end{equation}
(${\bf d}$ real), with $\hat{\sigma}$ being defined by
$\hat{\sigma}$ $\!=$ $\!|l\rangle\langle u|$.

For a neutral atom, the atomic polarization defined by
Eq.~(\ref{2.65}) reduces in the dipole approximation to  
\begin{equation}
\label{se1.8a}
\hat{\bf P}_{\rm A}({\bf r}) = \hat{\bf d}_{\rm A}
\delta({\bf r}-{\bf r}_{\rm A}),
\end{equation}
and thus application of Eq.~(\ref{2.73c}) yields
($\hat{\bf E}_{\rm M}^\|$ $\!=$ $\!-\hat{\bf P}_{\rm M}^\|/\varepsilon_0$)
\begin{equation}
\label{se1.8b}
\int{\rm d}^3{\bf r} \, \hat{\rho}_{\rm A}({\bf r})
\hat{\varphi}_{\rm M}({\bf r})
= - \hat{\bf d}_{\rm A}  \hat{\bf E}_{\rm M}^\|({\bf r}_{\rm A}),
\end{equation}
which for a two-level atom in the rotating wave approximation reads
\begin{equation}
\label{se1.10}
\int{\rm d}^3{\bf r} \, \hat{\rho}_{\rm A}({\bf r})
\hat{\varphi}_{\rm M}({\bf r})
= - \hat{\sigma}^\dagger
\hat{\bf E}_{\rm M}^{\|(+)}({\bf r}_{\rm A})
 {\bf d} +\mbox{H.c.} .
\end{equation}

Eqs.~(\ref{se1.8}) and (\ref{se1.10}) yield
\begin{equation}
\label{se1.11}
\hat{H}_{\rm int}
= - \hat{\sigma}^\dagger
\hat{\bf E}_{\rm M}^{(+)}({\bf r}_{\rm A})  {\bf d} +\mbox{H.c.}. 
\end{equation}

\subsubsection{Heisenberg picture}

{F}rom the Hamiltonian given in Eq.~(\ref{1.3a}) together
with Eqs.~(\ref{se1.3}), (\ref{se1.7a}), and (\ref{se1.11}), 
the Heisenberg equations of motion read \cite{Scheel99b}
\begin{equation}
\label{se1.11a}
{\dot{\hat\sigma}}_z
= \frac{2i}{\hbar}\,\hat{\sigma}^\dagger
\hat{\bf E}_{\rm M}^{(+)}({\bf r}_{\rm A})
 {\bf d} +\mbox{H.c.},
\end{equation}
\begin{equation}
\label{se1.11b}
\dot{\hat\sigma}{^\dagger}
=i\omega_{\rm A}\hat{\sigma}^\dagger
+ \frac{i}{\hbar} \hat{\bf E}_{\rm M}^{(-)}({\bf r}_{\rm A})
 {\bf d}\,\hat{\sigma}_z,
\end{equation}
\begin{equation}
\label{se1.11c}
\dot{\hat{\bf f}}({\bf r},\omega)
= -i \omega
\hat{\bf f}({\bf r},\omega) +
\frac{\omega^2}{c^2}
\sqrt{\frac{\varepsilon_{\rm I}({\bf r},\omega)}{\hbar\pi\varepsilon_0}}
\,{\mbb{G}}^\ast({\bf r},{\bf r}_{\rm A},\omega){\bf d}
\, \hat{\sigma}.
\end{equation}
Substituting in the expression for the electric field strength
[Eq.~(\ref{2.32}) together with Eq.~(\ref{2.29})] for
$\hat{\bf f}({\bf r},\omega)$ the formal solution of
Eq.~(\ref{se1.11c}), we derive, on using the relation (\ref{2.22d}), 
\begin{eqnarray}
\label{se1.15}
\lefteqn{
\hat{\bf E}_{\rm M}^{(+)}({\bf r},t)
= \hat{\bf E}_{\rm M\,free}^{(+)}({\bf r},t)
}
\nonumber\\&&
+\,\frac{i}{\pi\varepsilon_0} \int_0^\infty
\!{\rm d}\omega \, \frac{\omega^2}{c^2}
\,{\rm Im}\,\mbb{G}({\bf r},{\bf r}_{\rm A},\omega)
{\bf d}
\int_{t'}^t \!{\rm d}\tau \, {\rm e}^{-i\omega(t-\tau)}
\hat{\sigma}(\tau). 
\end{eqnarray}
Substitution of this expression into Eqs.~(\ref{se1.11a}) and
(\ref{se1.11b}) then yields a system of integro-differential
equations for the atomic quantities.

In the Markov approximation the integro-differential equations
reduce to Langevin-type differential equations. It is assumed
that [after performing the $\omega$-integration in Eq.~(\ref{se1.15})]
the time integral effectively runs over a small correlation
time interval $\tau_{\rm c}$. As long as we require that
$t$ $\!-$ $\!t'$ $\!\gg$ $\!\tau_{\rm c}$, we may extend the lower
limit of the $\tau$-integral to minus infinity with little error.
Further we require that $\tau_{\rm c}$ be small on a time
scale on which the atomic system is changed owing to the coupling
to the (medium-assisted) electromagnetic field. In this case,
in the $\tau$-integral in Eq.~(\ref{se1.15}) the slowly varying
atomic quantity $\hat{\sigma}(\tau){\rm e}^{i\omega_{\rm A}\tau}$
can be taken at time $t$ and put in front of the integral, thus
\begin{equation}
\label{se1.15a}
\hat{\bf E}_{\rm M}^{(+)}({\bf r},t)
= \hat{\bf E}_{\rm M\,free}^{(+)}({\bf r},t)
+\hat{\sigma}(t)\,
\frac{i}{\pi\varepsilon_0} \int_0^\infty
\!{\rm d}\omega \,
\frac{\omega^2}{c^2}
\,{\rm Im}\,\mbb{G}({\bf r},{\bf r}_{\rm A},\omega)
{\bf d}
\,\zeta(\omega_{\rm A}-\omega)
\end{equation}
[$\zeta(x)$ $\!=$ $\pi\delta(x)$ $\!+$ $\!i{\cal P}x^{-1}$].
Substitution of this expression into Eqs.~(\ref{se1.11a}) and
(\ref{se1.11b}) yields
\begin{equation}
\label{se1.15b}
{\dot{\hat\sigma}}_z
= -\Gamma (1+\hat{\sigma}_z)
+\Big[\frac{2i}{\hbar}\,\hat{\sigma}^\dagger
\hat{\bf E}_{\rm M\,free}^{(+)}({\bf r}_{\rm A})
 {\bf d} +\mbox{H.c.}\Big],
\end{equation}
\begin{equation}
\label{se1.15c}
\dot{\hat\sigma}{^\dagger}
= \big[ i(\omega_{\rm A}-\delta\omega) -{\textstyle\frac{1}{2}} \Gamma \big]
\hat{\sigma}^\dagger
+\frac{i}{\hbar} \hat{\bf E}_{\rm M\,free}^{(-)}({\bf r}_{\rm A})
 {\bf d}\,\hat{\sigma}_z,
\end{equation}
where
\begin{equation}
\label{se1.20}
\Gamma = \frac{2\omega_{\rm A}^2 d_i d_{j}}{\hbar\varepsilon_0 c^2}
\,{\rm Im}\, G_{ij}({\bf r}_{\rm A},{\bf r}_{\rm A},\omega_{\rm A})
\end{equation}
is the rate of spontaneous decay of the upper state and
\begin{equation}
\label{se1.20a}
\delta\omega = \frac{d_i d_j}{\hbar\pi\varepsilon_0}\,\,
{\cal P}\!\int_0^\infty \!{\rm d}\omega \,
\frac{\omega^2}{c^2}
\frac{{\rm Im}\,G_{ij}({\bf r}_{\rm A},{\bf r}_{\rm A},\omega)}
{\omega-\omega_{\rm A}}
\end{equation}
is the contribution to the Lamb shift.\footnote{For the
    the overall (vacuum) Lamb shift, see, e.g., \cite{Milonni}.} 
Note that from Eq.~(\ref{dft}) it follows that 
Eq.~(\ref{se1.20}) can be given in the equivalent form of
\begin{equation}
\label{se1.20b}
\Gamma = \frac{2\pi}{\hbar^2}\,d_i d_j
\int_0^\infty\!{\rm d}\omega\,
\langle 0|\underline{\hat{E}}_i({\bf r}_{\rm A},\omega)
\underline{\hat{E}}_{j}^\dagger({\bf r}_{\rm A},\omega_{\rm A}) |0 \rangle, 
\end{equation}
which exactly corresponds to Fermi's Golden Rule (see, e.g., \cite{Loudon}).

\subsubsection{Schr\"odinger picture}

The above given equations of motion do not only apply to
the spontaneous emission but are also suitable for the
study of the evolution of a two-level atom driven by an external
medium-assisted electromagnetic field. If the atom is initially
prepared in the upper state and there is no driving field,
then the use of the wave equation
\begin{equation}
\label{se1.20c}
i\hbar \frac{{\rm d}}{{\rm d}t}|\Psi\rangle = \hat{H}|\Psi\rangle
\end{equation} 
may be more appropriate for the study of the motion of the
coupled atom--field system \cite{Ho00}. According to
the Hamiltonian in Eq.~(\ref{1.3a}) together
with Eqs.~(\ref{se1.3}), (\ref{se1.7a}), and (\ref{se1.11}), 
the state vector $|\Psi(t)\rangle$ can be expanded as
\begin{eqnarray}
\label{e15}
\lefteqn{
|\Psi(t)\rangle = C_{u}(t) 
{\rm e}^{-i \omega_u t}|u\rangle\otimes|0\rangle
}
\nonumber\\&&\hspace{4ex}
+\int\!{\rm d}^3{\bf r} \int_0^\infty {\rm d}\omega\, 
C_{li}({\bf r},\omega,t)
{\rm e}^{-i (\omega+\omega_l)t} |l\rangle\otimes|i,{\bf r},\omega\rangle,
\end{eqnarray}
where $|0\rangle$ is the vacuum state of the fundamental fields
$\hat{f}_i({\bf r},\omega)$,  and $|i,{\bf r},\omega\rangle$ 
is the state, where one of them is excited in a single-quantum
Fock state. It is not difficult to prove that the  
probability amplitudes $C_{u}$ and $C_{li}$ satisfy
the differential equations
\begin{eqnarray}
\label{e16}
\lefteqn{
\dot{C}_u(t) =
-\frac{d_j}{\sqrt{\pi\epsilon_0\hbar}}           
\int_0^\infty {\rm d}\omega \,\frac{\omega^2}{c^2} 
\int {\rm d}^3{\bf r}\,\Big[
\sqrt{\epsilon_{\rm I}({\bf r},\omega)}
}
\nonumber \\&&\hspace{10ex} \times\,
G_{ji}({\bf r}_{\rm A},{\bf r},\omega) \,
C_{li}({\bf r},\omega,t) {\rm e}^{-i(\omega-\omega_{\rm A})t}\Big] ,
\end{eqnarray}
\begin{equation}
\label{e17}
\dot{C}_{li}({\bf r},\omega,t) =
\frac{d_j}{\sqrt{\pi\epsilon_0\hbar}}\,
\frac{\omega^2}{c^2}\,\sqrt{\epsilon_{\rm I}({\bf r},\omega)}
G_{ji}^\ast({\bf r}_{\rm A},{\bf r},\omega)\,
C_{u}(t) {\rm e}^{i(\omega-\omega_{\rm A})t} . 
\end{equation}
We now substitute the result of formal integration of Eq.~(\ref{e17})
[$C_{li}({\bf r},\omega,0)$ $\!=$ $\!0$] into Eq.~(\ref{e16}).
Making use of the relationship (\ref{2.22d}), 
we obtain the integro-differential equation
\begin{equation}
\label{e18}
\dot{C}_u(t) =\int_0^t {\rm d}t'\, K(t-t') C_{u}(t'),
\end{equation}
with the kernel function
\begin{equation}
\label{e19}
K(t-t') =
- \frac{d_i d_j }{\hbar\pi\epsilon_0}
\int_0^\infty {\rm d}\omega \,
\frac{\omega^2}{c^2}\, 
{\rm Im}\,G_{ij}({\bf r}_{\rm A},{\bf r}_{\rm A},\omega)
{\rm e}^{-i(\omega-\omega_{\rm A})(t-t')}.
\end{equation}
Taking the time integral of both sides of Eq.~(\ref{e18}), 
we easily derive, on changing the order of integrations
on the right-hand side,
\begin{equation}
\label{se1.21}
C_{u}(t) =\int_0^t {\rm d}t'\, \bar{K}(t-t') C_{u}(t') + 1
\end{equation}
[$C_{u}(0)$ $=$ $\!1$], where 
\begin{equation}
\label{se1.22}
\bar{K}(t-t') =
\frac{d_i d_j}{\hbar\pi\epsilon_0}
\int_0^\infty {\rm d}\omega\,\frac{\omega^2}{c^2}\,
\frac{{\rm Im}\,G_{ij}({\bf r}_{\rm A},{\bf r}_{\rm A},\omega)}
{i(\omega-\omega_{\rm A})}
\left[{\rm e}^{-i(\omega-\omega_{\rm A})(t-t')}-1 \right].
\end{equation}
Note that in the Markov approximation the kernel in
Eq.~(\ref{se1.22}) simply becomes
\begin{equation}
K(t-t') = -\textstyle\frac{1}{2} \Gamma+i\delta\omega,
\end{equation}
where $\Gamma$ and $\delta\omega$ are respectively given by
Eqs.~(\ref{se1.20}) and (\ref{se1.20a}).

The equation (\ref{se1.21}) is a well-known Volterra integral equation
of the second kind. 
It is worth noting that the integro-differential equation (\ref{e18})
and the equivalent integral equation (\ref{se1.21}) apply to 
the spontaneous decay of an atom in the presence of an 
arbitrary configuration of dispersing and absorbing dielectric 
bodies. All the matter parameters that are relevant for  
the atomic evolution are contained, via the 
Green tensor, in the kernel functions (\ref{e19}) and
(\ref{se1.22}). In particular when absorption is
disregarded and the permittivity is regarded as being 
a real frequency-independent quantity (which of course
can change with space), then the formalism yields the results 
of standard mode decomposition (see, e.g. \cite{Cook87,Feng89,Ujihara99}).

\subsection{Atom near a dielectric surface}
\label{sec:planar}

As a first example, let us consider the spontaneous decay of an
excited atom near an absorbing planar dielectric surface.
To be more specific, the distance $z$ to the surface of the atom is
assumed to be small compared to the atomic transition wavelength.
For real permittivity, this configuration has been studied
extensively in connection with Casimir and
van der Waals forces (see, e.g., \cite{Meschede90,Fichet95} and
references cited therein). Further, it can be regarded
as being the basic configuration in scanning near-field optical
microscopy (see, e.g., \cite{Henkel98}) and related applications.

In order to calculate the decay rate $\Gamma$, Eq.~(\ref{se1.20}),
the imaginary part of the Green tensor of two (infinite) half-spaces with
a common interface \cite{Maradudin75,Mills75} is required.
When the atom's half-space is the vacuum, then the Green tensor
in this half-space can be decomposed into the vacuum Green
tensor and a reflection term, i.e., 
\begin{equation}
\label{se2.0}
G_{kl}({\bf r},{\bf r}',\omega) =
G_{kl}^{\rm V}({\bf r},{\bf r}',\omega) +
R_{kl}({\bf r},{\bf r}',\omega),
\end{equation}
where the vacuum Green tensor $G_{kl}^{\rm V}({\bf r},{\bf r}',\omega)$ is 
obtained from Eq.~(\ref{se2.1}) for $q$ $\!=$ $\!\omega/c$. {F}rom
Eq.~(\ref{se2.2a}) it is seen that ${\rm Im}\,G_{kl}^{{\rm V}\|}
({\bf r},{\bf r}',\omega)$ $\!=$ $\!0$, and thus we obtain,
on applying Eq.~(\ref{se2.2c}),
\begin{equation}
\label{se2.2d}
{\rm Im}\,G_{kl}^{\rm V}({\bf r},{\bf r},\omega)
= \frac{\omega}{6\pi c}\,\delta_{kl}.
\end{equation}
Combining Eqs.~(\ref{se1.20}), (\ref{se2.0}), and (\ref{se2.2d}),
we obtain
\begin{equation}
\label{se2.2d1}
\Gamma = \Gamma_0 + \frac{2\omega_{\rm A}^2d_kd_l}{\hbar\varepsilon_0c^2}
\,{\rm Im}\,R_{kl}({\bf r}_{\rm A},{\bf r}_{\rm A},\omega_{\rm A}),
\end{equation}
where $\Gamma_0$ is given by Eq.~(\ref{se1.1})

In the coincidence limit, the reflection term
$R_{kl}({\bf r},{\bf r},\omega)$ can be given in the form
\cite{Ho98,Tomas95}
\begin{eqnarray}
\label{se2.2e}
\lefteqn{
R_{xx}(z,z,\omega)
= R_{yy}(z,z,\omega)
}
\nonumber\\&&
= -\frac{i}{8\pi q^2} \int_0^\infty
\!{\rm d}k \,k\beta\, {\rm e}^{2i\beta z} r^p(k)
+\frac{i}{8\pi} \int_0^\infty \!{\rm d}k \,
\frac{k{\rm e}^{2i\beta z}}{\beta} \,r^s(k) ,
\end{eqnarray}
\begin{equation}
\label{se2.2f}
R_{zz}(z,z,\omega) = \frac{i}{4\pi q^2} \int_0^\infty \!{\rm d}k \, k^3
\frac{{\rm e}^{2i\beta z}}{\beta} \,r^p(k), 
\end{equation}
where $r^p(k)$ and $r^s(k)$ are the usual 
Fresnel reflection coefficients for $p$- and $s$-polarized
waves ($q$ $\!=$ $\!\omega/c$, $\beta$ $\!=$ $\!\sqrt{q^2-k^2}$;
the origin of the co-ordinates is on the interface). 
When the distance $z$ of the atom from the surface is small compared
to the wavelength, $qz$ $\!\ll$ $\!1$, then
the integrals in Eqs.(\ref{se2.2e}) and (\ref{se2.2f}) can be evaluated
asymptotically to obtain
\begin{equation}
\label{se2.2g}
R_{zz}=\frac{1}{16\pi q^2z^3}\frac{n^2-1}{n^2+1}
+\frac{1}{8\pi z}\frac{(n-1)^2}{n(n+1)}
+ \frac{iq}{12\pi}\frac{(n-1)(2n-1)}{n(n+1)} + O(z), 
\end{equation}
\begin{equation}
\label{se2.2h}
R_{xx}=R_{yy}={\textstyle\frac{1}{2}}R_{zz}
-\frac{1}{16\pi z}\frac{n^2-1}{n^2+1}
-\frac{iq}{3\pi}\frac{n-1}{n+1} + O(z).  
\end{equation}
[$\varepsilon$ $\!=$ $\!\varepsilon(\omega)$,
$n$ $\!=$ $\!\sqrt{\varepsilon(\omega)}$].
Inserting Eqs.~(\ref{se2.2g}) and (\ref{se2.2h})
into Eq.~(\ref{se2.2d1}) yields (\mbox{$\omega$ $\!=$ $\!\omega_{\rm A}$})
\cite{Scheel99c}
\begin{equation}
\label{se2.3}
\Gamma = \Gamma_0 \,\frac{3}{8}
\left( 1+\frac{d_z^2}{d^2} \right)
\left(\frac{c}{\omega_{\rm A} z}\right)^3
\frac{\varepsilon_{\rm I}(\omega_{\rm A})}{|\varepsilon(\omega_{\rm A})+1|^2}
+ O(z^{-1}) .
\end{equation}
It is worth noting that the leading term $\sim z^{-3}$ in
Eq.~(\ref{se2.3}) exactly agrees with the result of the
involved microscopic approach in \cite{Yeung96}.
It typically occurs for absorbing media and reflects
the possibility of nonradiative decay, if the excited
atom is sufficiently near the medium.

\subsection{Real-cavity model of spontaneous decay
in a dielectric medium} 
\label{sec:realcavity}

Let us consider the situation when an excited atom is placed
inside an empty micro-sphere (embedded in an otherwise homogeneous
dielectric) with a radius much smaller than the wavelength of the atomic
transition. This might serve as a model for spontaneous decay of a
(guest) atom in an absorbing dielectric.\footnote{Such a model applies
   as long as the atom cannot ``resolve'' the atomic structure of the
   surrounding matter.}
{F}rom simple arguments
based on the change of the mode density it is suggested that the
spontaneous emission rate inside a nonabsorbing dielectric should
be modified according to \mbox{$\Gamma$ $\!=$ $\!n\Gamma_0$},
where $n$ is the (real) refractive index
of the medium and $\Gamma_0$ is given by Eq.~(\ref{se1.1})
\cite{Dexter56,diBartolo,Yariv,Nienhuis76}.
Here it is assumed that the local field the atom interacts with is exactly
the same as the electromagnetic field in the continuous medium. In
reality, the atom is located in a small region of free space, and the
local field is different from the field in the continuous medium which 
leads to the introduction of the so-called local-field correction
factor $\xi$, thus giving \mbox{$\Gamma$ $\! =$ $\!n\xi \Gamma_0$}.
Different models have been used to compute $\xi$. Here we want to
concentrate on the (Glauber-Lewenstein) real-cavity model,
which for real $n$ leads to 
\cite{Glauber91}
\begin{equation}
\label{se3.3}
\xi  = \left( \frac{3n^2}{2n^2+1} \right)^2. 
\end{equation}
Experiments suggest that this model is a good candidate for
describing the decay of substitutional guest atoms
different from the constituents of the dielectric
\cite{Rikken95,Schuurmans98,deVries98}. 

Now let us turn to the question what will appear
for absorbing media. According to Eq.~(\ref{se1.20}),
the rate of spontaneous decay is proportional to the
imaginary part of the Green tensor
in the coincidence limit (i.e., the two spatial arguments are
to be taken at the position of the atom). In the real-cavity model
the Green tensor that enters the rate formula
is the Green tensor for the system disturbed by a small
free-space inhomogeneity. Obviously, it can again
be given in a form like in Eq.~(\ref{se2.0}), where
for a sphere (with the centre as origin of co-ordinates)
the reflection term $\mbb{R}({\bf r},{\bf r}',\omega)$
reads \cite{Li94}
\begin{eqnarray}
\label{se3.8}
\lefteqn{
\mbb{R}({\bf r},{\bf r}',\omega)
= \frac{i\omega}{4\pi c} \sum_{e,o} \sum_{n=1}^\infty
\sum_{m=0}^n \bigg\{ \frac{2n+1}{n(n+1)} \frac{(n-m)!}{(n+m)!}
\left( 2-\delta_{0m} \right)
}
\nonumber \\ && \hspace{10ex}\times 
\bigg[ C_n^M(\omega)
{\bf M}_{{e \atop o}nm}\!\left({\bf r},\frac{\omega}{c}\right)
\otimes
{\bf M}_{{e \atop o}nm}\!\left({\bf r}',\frac{\omega}{c}\right)
\nonumber \\ &&\hspace{15ex}
+ C_n^N(\omega)
{\bf N}_{{e \atop o}nm}\!\left({\bf r},\frac{\omega}{c}\right)
\otimes
{\bf N}_{{e \atop o}nm}\!\left({\bf r}',\frac{\omega}{c}\right)
\bigg] \bigg\},
\qquad
\end{eqnarray}
where
${\bf M}_{{e \atop o}nm}({\bf r},k)$ and
${\bf N}_{{e \atop o}nm}({\bf r},k)$ are the ($e$ven and $o$dd) vector 
Debye potentials defined by
\begin{equation}
\label{se3.9}
{\bf M}_{{e \atop o}nm}({\bf r},k) = \mbb{\nabla} \times \left[
\psi_{{e \atop o}nm}({\bf r},k)\, {\bf r} \right] \,,
\end{equation}
\begin{equation}
\label{se3.10}
{\bf N}_{{e \atop o}nm}({\bf r},k) = \frac{1}{k} \mbb{\nabla} \times
\mbb{\nabla} \times \left[ \psi_{{e \atop o}nm}({\bf r},k) \,{\bf r}
\right], 
\end{equation}
with
\begin{equation}
\label{se3.10a}
\psi_{{e \atop o}nm}({\bf r},k) = j_n(kr) P_n^m(\cos\Theta)
{\cos\choose\sin} (m\phi) 
\end{equation}
[$j_n(kr)$, spherical Bessel function of the first kind; 
$P_n^m(\cos\Theta)$, associated Legendre polynomial;
for the rather lengthy expressions of the generalized
reflection coefficients $C_n^{M(N)}$, see \cite{Li94}].

In particular in the coincidence limit ${\bf r}$ $\!\to$
${\bf r}'$ $\!\to$ $\!0$ 
only the TM-wave vector Debye potentials
${\bf N}_{{e \atop o}10}({\bf r},k)$ and
${\bf N}_{{e \atop o}11}({\bf r},k)$ contribute to
$R_{kl}({\bf r},{\bf r}',\omega)$ and we find that
\begin{equation}
\label{se3.11}
\left.R_{kl}({\bf r},{\bf r},\omega)\right|_{{\bf r}=0}
= \frac{i\omega}{6\pi c}\, C_1^N(\omega) \delta_{kl},
\end{equation}
where the generalized reflection coefficient $C_1^N(\omega)$ reads 
\begin{eqnarray}
\label{se3.12}
\lefteqn{
C_1^N(\omega) =
}
\nonumber \\ &&
\frac{\left[i+\varrho(n+1)-i\varrho^2n-\varrho^3n^2/(n+1) \right]
{\rm e}^{i\varrho}}{\sin \varrho-\varrho(\cos \varrho+in\sin \varrho)
+i\varrho^2n\cos \varrho
-\varrho^3(\cos \varrho-in\sin \varrho) n^2/(n^2-1)} \nonumber \\
\end{eqnarray}
[$n$ $\!=$ $\!\sqrt{\varepsilon(\omega)}$ and $\varrho$ $\!=$
$R\omega/c$, with $R$ being the cavity radius]. 
Hence, when the  atom is located at the centre of
the sphere, we may insert Eq.~(\ref{se3.11}) into
Eq.~(\ref{se2.2d1}) to obtain (for $\omega$ $\!=$ $\!\omega_{\rm A}$)
the decay rate in the
form of\footnote{Note that when in the coincidence
   limit (at the position of the atom) the Green tensor 
   has the form of Eq.~(\ref{se2.0}), then the
   rate of the spontaneous decay has the form of
   Eq.~(\ref{se2.2d1}). Since the vacuum Green tensor
   $G_{kl}^{\rm V}({\bf r},{\bf r}',\omega)$
   has no longitudinal imaginary part and the scattering
   component $R_{kl}({\bf r},{\bf r}',\omega)$ is
   purely transverse, the longitudinal electromagnetic
   field does not contribute to the decay rate.}
\cite{Scheel99b}    
\begin{equation}
\label{se3.13}
\Gamma = \Gamma_0 \left[ 1+{\rm Re}\, C_1^N(\omega_{\rm A}) \right] .
\end{equation}

As long as the surrounding medium can be treated as a continuum 
this result is exact and valid for all admissible cavity sizes,
without restriction to transition frequencies far from
medium resonances.
If the cavity radius is small compared to the wavelength
of the atomic transition (but large enough for the continuum
approach), we may expand the generalized reflection
coefficient $C_1^N(\omega_{\rm A})$, Eq.~(\ref{se3.12}),
in powers of \mbox{$\varrho$ $\!=$ $\!R\omega_{\rm A}/c$} to obtain
\begin{eqnarray}
\label{se3.14}
\lefteqn{
\Gamma = \Gamma_0 \bigg\{
\frac{9\varepsilon_{\rm I}(\omega_{\rm A})}
{|2\varepsilon(\omega_{\rm A})+1|^2}
\left( \frac{c}{\omega_{\rm A} R} \right)^3
}
\nonumber \\[.5ex] &&
+\,\frac{9\varepsilon_{\rm I}
(\omega_{\rm A})\left[28|\varepsilon(\omega_{\rm A})|^2 
+16\varepsilon_{\rm R}(\omega_{\rm A})+1
\right]}{5|2\varepsilon(\omega_{\rm A})+1|^4} 
\left( \frac{c}{\omega_{\rm A} R} \right)
\nonumber \\[.5ex] &&
+\frac{9n_{\rm R}(\omega_{\rm A})}{|2\varepsilon(\omega_{\rm A})+1|^4}
\left[ 4|\varepsilon(\omega_{\rm A})|^4
+4 \varepsilon_{\rm R}(\omega_{\rm A}) 
|\varepsilon(\omega_{\rm A})|^2  
+\varepsilon_{\rm R}^2(\omega_{\rm A})
-\varepsilon_{\rm I}^2(\omega_{\rm A}) \right] 
\nonumber \\[.5ex] &&
-\frac{9n_{\rm
I}(\omega_{\rm A})\varepsilon_{\rm I}(\omega_{\rm A})}
{|2\varepsilon(\omega_{\rm A})+1|^4}  
\left[ 4|\varepsilon(\omega_{\rm A})|^2
+2\varepsilon_{\rm R}(\omega_{\rm A})
\right] \bigg\} +O\!\left( \frac{R\omega_{\rm A}}{c} \right) \,.
\end{eqnarray}
For $\varepsilon_{\rm I}(\omega_{\rm A})$ $\!=$ $\!0$, i.e, when 
absorption is fully disregarded,
in Eq.~(\ref{se3.14}) only the third term survives and reproduces
exactly the Glauber-Lewenstein local-field correction factor
(\ref{se3.3}). For an absorbing medium terms proportional to
$R^{-3}$ and $R^{-1}$ are observed, which can give rise
to a strong dependence of the decay rate on the cavity radius.
In particular, the leading term proportional to $R^{-3}$ 
can be regarded as corresponding to nonradiative decay
via dipole-dipole energy transfer from the atom to the
medium.

\begin{figure}[ht]
 \unitlength=1cm
\begin{minipage}{0.5\textwidth}
\begin{center}
 \begin{picture}(7,4.5)
\psfig{file=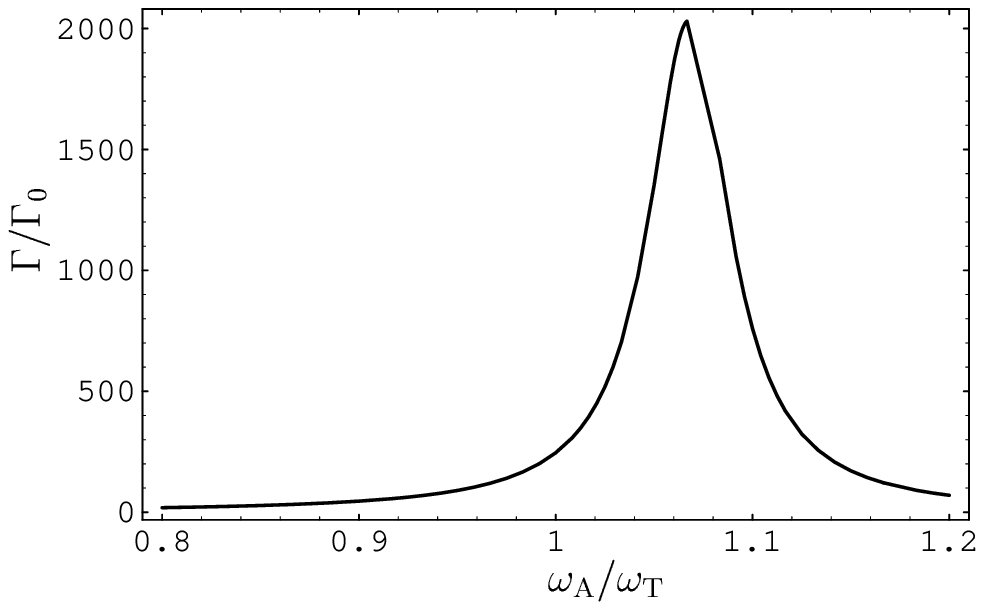,width=7cm} 
 \end{picture}
\end{center} 
\end{minipage}
\hfill
\begin{minipage}{0.5\textwidth}
\begin{center}
 \begin{picture}(7,4.5)
\psfig{file=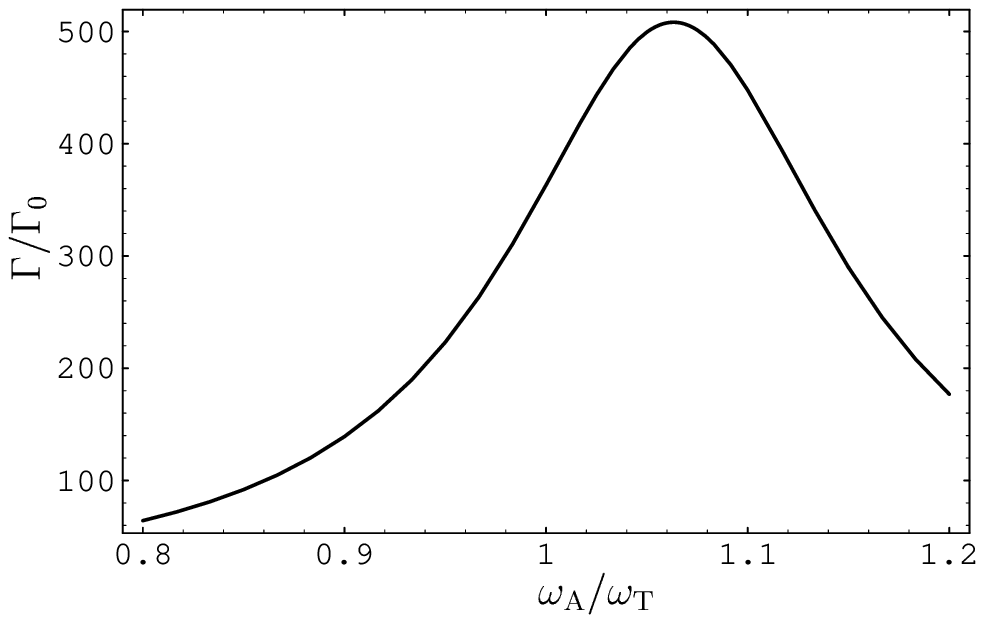,width=7cm} 
 \end{picture}
\end{center} 
\end{minipage}
\caption{\label{fig:se}
The spontaneous decay rate $\Gamma$, Eq.~(\protect\ref{se3.13}),
for the real-cavity model of an atom in an absorbing dielectric
is shown as a function of the atomic transition frequency $\omega_{\rm A}$
near a medium resonance for the model permittivity (\protect\ref{se3.15})
(\mbox{$\omega_{\rm P}$ $\!=$ $\!0.46\,\omega_{\rm T}$}, $\gamma$ $\!=$
$\!0.05\,\omega_{\rm T}$ [left figure] and $\gamma$ $\!=$
$\!0.2\,\omega_{\rm T}$
[right figure]) and $R$ $\!=$ $\!0.02\,\lambda_{\rm A}$.
}
\end{figure}

Examples of the dependence of rate of spontaneous decay on the atomic
transition frequency are plotted in Fig.~\ref{fig:se} for a
model permittivity of Lorentz type 
\begin{equation}
\label{se3.15}
\varepsilon(\omega) =
1+\frac{\omega_{\rm P}^2}{\omega_{\rm T}^2-\omega^2-i\gamma\omega } \,.
\end{equation}
It shows a band gap between the (transverse) resonance 
frequency $\omega_{\rm T}$ and the longitudinal frequency
$\omega_{\rm L}$ $\!=$ 
$\!\sqrt{\omega_{\rm T}^2+\omega_{\rm P}^2}$. One observes, that
precisely in the 
gap region the spontaneous decay strongly increases due to
non-radiative decay channels which become important in that frequency
range.

\subsection{Cavity QED}
\label{sec:beyond}

If the radius of the microcavity is not small compared with the wavelength 
of the atomic transition, the cavity can act as a resonator, and it is
well known that the spontaneous decay of an excited atom can be
strongly modified when it is placed in
a microresonator (see, e.g., \cite{Berman,Hinds}). There are typically 
two qualitatively different regimes: the weak coupling regime and the 
strong-coupling regime. In the weak coupling regime the Markov
approximation applies and a monotonous exponential decay is observed,
the decay rate being enhanced or reduced compared
to the free-space value depending on whether the atomic transition
frequency fits a cavity resonance or not. The strong-coupling regime,
in contrast, 
is characterized by reversible Rabi oscillations where the energy of the
initially excited atom is periodically exchanged between the atom
and the radiation field. This usually requires that the emission
is in resonance with a high-quality cavity mode. Recent progress
in constructing certain types of microresonators such as microspheres
has rendered it possible to approach the ultimate quality level
determined by intrinsic material losses \cite{Gorodetsky96}.

\begin{figure}[ht]
 \unitlength=1cm
 \begin{center}
 \begin{picture}(4,4)
 \put(0,0){
 \psfig{file=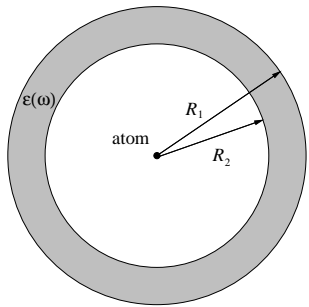,height=4cm}}
 \end{picture}
 \end{center}
\caption{\label{fig:slabho} Scheme of the spherical microresonator.}
\end{figure}

Let us consider the spontaneous decay of an excited atom placed inside
a spherical three-layer structure (Fig.~\ref{fig:slabho}). The outer layer
($r$ $\!>$ $\!R_1$) and the inner layer
\mbox{($0$ $\!\le$ $\!r$ $\!<$ $\!R_2$)} 
are vacuum, whereas the middle layer 
($R_2$ $\!\le$ $\!r$ $\!\le$ $\!R_1$), which plays
the role of the resonator wall, is a dispersive and absorbing dielectric.
In particular when a Lorentz-type dielectric is assumed whose permittivity is
of the form (\ref{se3.15}), the wall would be perfectly reflecting
in the band-gap zone, i.e., $\omega_{\rm T}$ $\!<$ $\!\omega_{\rm A}$ 
$\!<$ $\!\omega_{\rm L}$, provided that absorption could be disregarded 
($\gamma$ $\!\to$ $\!0$). The Green tensor $\mbb{G}({\bf r},{\bf r}',\omega)$
for a spherical three-layer geometry can be found in \cite{Li94}.

\subsubsection{Weak coupling}

For ${\bf r}$ and
${\bf r}'$ inside the inner (vacuum) sphere, the Green tensor looks
like the one for the two-layered spherical geometry considered
in Sec.~\ref{sec:realcavity}. It
has again the form given in Eq.~(\ref{se2.0}) together with
Eqs.~(\ref{se3.8}) -- (\ref{se3.10a}), but with different  
reflection coefficients $C_n^{M(N)}$ (see \cite{Li94}).
Hence, when the atom is located at the centre of the inner sphere, then 
the rate of spontaneous decay can again be given in the form of
Eq.~(\ref{se3.13}).  
Obviously, for a sufficiently thick cavity wall, 
$\exp[-in_{\rm I}(\omega_{\rm A})(R_1-R_2)\omega/c]$ $\!\ll$ $\!1$,
the generalized reflection coefficient $C_1^N$ in Eq.~(\ref{se3.13})
reduces to that one for the two-layered configuration
[Eq.~(\ref{se3.12}) with $R$ $\!=$ $\!R_2$], and
thus for a true microresonator, $R_2\omega_{\rm A}/c$ $\!\gg$ $\!1$, 
the decay rate becomes \cite{Ho00} 
\begin{eqnarray}
\label{se4.2}
\lefteqn{
\Gamma \simeq \Gamma_0 \,{\rm Re}\! \left[
\frac{n(\omega_{\rm A})-i\tan(R_2\omega_{\rm A}/c)}
{1-in(\omega_{\rm A})\tan(R_2\omega_{\rm A}/c)} 
\right]
}
\nonumber \\ &&
= \Gamma_0 \,\frac{n_{\rm R}(\omega_{\rm A})
[1+\tan^2(R_2\omega_{\rm A}/c)]}
{[1+n_{\rm I}(\omega_{\rm A})\tan(R_2\omega_{\rm A}/c)]^2
+n_{\rm R}^2(\omega_{\rm A}) \tan^2(R_2\omega_{\rm A}/c)} \,.
\qquad
\end{eqnarray}
\begin{figure}[ht]
 \unitlength=1cm
 \begin{center}
 \begin{picture}(7.5,5)
\put(0,0){
\psfig{file=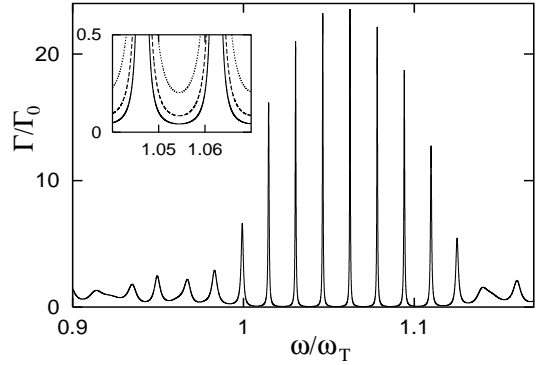,width=7.5cm}}
 \end{picture}
 \end{center}
\caption{\label{fig:ho1} The spontaneous decay rate
\protect$\Gamma/\Gamma_0$, Eq.~(\protect\ref{se3.13}),
of an atom in the microresonator in Fig.~\protect\ref{fig:slabho}
is shown for $R_2$ $\!=$ $\!30\lambda_{\rm T}$,
$R_1$ $\!-$ $\!R_2$ $\!=$ $\!\lambda_{\rm T}$, $\omega_{\rm P}$ $\!=$ 
$\!0.5\omega_{\rm T}$, and $\gamma$ $\!=$ $\!10^{-2}\omega_{\rm T}$.
The curves in 
the inset correspond to $\gamma$ $\!=$ $\!10^{-2}\omega_{\rm T}$ (solid
line), $\gamma$ $\!=$ $\!2\times 10^{-2}\omega_{\rm T}$ (dashed line), and
\mbox{$\gamma$ $\!=$ $\!5\times 10^{-2}\omega_{\rm T}$ (dotted line)}. }
\end{figure}
{F}rom Fig.~\ref{fig:ho1} it is seen that
[for the model permittivity (\ref{se3.15})]
the rate of spontaneous decay sensitively depends on
the transition frequency. Narrow-band enhancement of spontaneous decay
($\Gamma/\Gamma_0$ $\!>$ $\!1$) alternates with broadband inhibition
($\Gamma/\Gamma_0$ $\!<$ $\!1$). The frequencies where the maxima of
enhancement are observed correspond to the resonance frequencies
of the cavity. Within the band gap the heights and widths of the
frequency intervals in which spontaneous decay is feasible are
essentially determined by the material losses. Outside the
band-gap zone the change of the decay rate is less pronounced,
because of the relatively large input--output coupling, the
(small) material losses being of secondary importance. 

\subsubsection{Strong coupling}

The strength of the coupling between the atom and the electromagnetic
field increases when the frequency of the atomic transition frequency
approaches a cavity resonance frequency. 
In order to gain insight into the solution of the integral 
equation (\ref{se1.21}) for the strong-coupling regime, 
let us consider the limiting case of only
a single cavity resonance (frequency $\omega_{\rm C}$)
being involved in the atom--field interaction.
In this case we may approximate, on using Eqs.~(\ref{se1.20}) and
(\ref{se3.13}), the kernel function (\ref{se1.22}) by
\begin{eqnarray}
\label{se4.3}
\lefteqn{
\bar{K}(t-t') \simeq -\frac{\Gamma_{\rm C}(\delta\omega_{\rm C})^2}{2\pi}
{\rm e}^{-i(\omega_{\rm C}-\omega_{\rm A})(t-t')}
\int_{-\infty}^\infty \!{\rm d}\omega \,
\frac{{\rm e}^{-i(\omega-\omega_{\rm C})(t-t')}}{(\omega-\omega_{\rm C})^2
+\left(\delta\omega_{\rm C} \right)^2}
}
\nonumber \\&&\hspace{5ex}
= -{\textstyle\frac{1}{2}} \Gamma_{\rm C} \delta\omega_{\rm C}
{\rm e}^{-i(\omega_{\rm C}-\omega_{\rm A})(t-t')}
{\rm e}^{-\delta\omega_{\rm C}|t-t'|},
\hspace*{20ex}
\end{eqnarray}
and thus the integral equation (\ref{se1.21}) corresponds to
the differential equation \cite{Ho00}
\begin{equation}
\label{se4.4}
\ddot{C}_u(t) +\left[i(\omega_{\rm C}-\omega_{\rm A})
+\delta\omega_{\rm C} \right]
\dot{C}_u(t) +{\textstyle\frac{1}{2}}\Gamma_{\rm C}
\delta\omega_{\rm C} C_u(t) = 0. 
\end{equation}
where $\Gamma_{\rm C}$ is the decay rate at the cavity resonance
[i.e., $\omega_{\rm A}$ $\!=$ $\!\omega_{\rm C}$ in the
rate formulas (\ref{se3.13}) and (\ref{se4.2}], and
$\delta\omega_{\rm C}$ is the width of the cavity resonance.
For small absorption, $\gamma$ $\!\ll$
$\!\omega_{\rm T},\,\omega_{\rm P},\,\omega_{\rm P}^2/\omega_{\rm T}$,
the resonance lines in the
band-gap zone can be regarded as being Lorentzians, and
\begin{equation}
\label{se4.4a}
\delta\omega_{\rm C} = \frac{c\Gamma_0}{R_2\Gamma_{\rm C}}\,.
\end{equation}
Equation (\ref{se4.4}) reveals that (under the assumptions made)
the upper-state probability amplitude of the atom obeys the equation
of motion for a damped harmonic oscillator.
In the strong-coupling regime, where $\omega_{\rm A}$ $\!=$
$\!\omega_{\rm C}$ and $\Gamma_{\rm C}$ $\!\gg$
$\!\delta\omega_{\rm C}$, damped Rabi oscillations are
observed:\footnote{Note that in the opposite case where
   $\Gamma_{\rm C}$ $\!\ll$ $\!\delta\omega_{\rm C}$ the
   solution of Eq.~(\protect\ref{se4.4}) is $|C_u(t)|^2$
   $\!=$ $\!{\rm e}^{-\Gamma_{\rm C}t}$ for \mbox{$\omega_{\rm A}$ $\!=$
   $\!\omega_{\rm C}$}.}
\begin{equation}
\label{se4.5}
|C_u(t)|^2 = {\rm e}^{-\delta\omega_{\rm C} t} \cos^2(\Omega t/2),
\end{equation}
where the Rabi frequency reads
\begin{equation}
\label{se4.5a}
\Omega = \sqrt{2\Gamma_{\rm C} \delta\omega_{\rm C}}\,.
\end{equation}
\begin{figure}[ht]
 \unitlength=1cm
 \begin{center}
 \begin{picture}(7.5,5)
 \put(0,0){
 \psfig{file=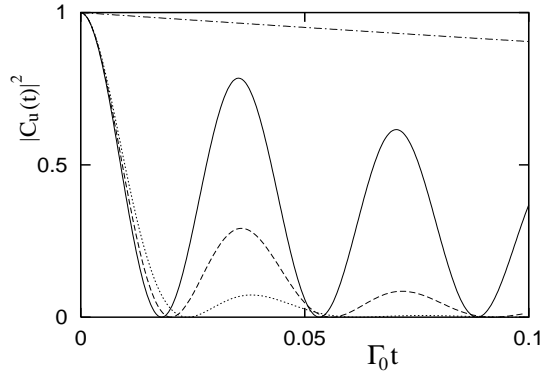,width=7.5cm}}
 \end{picture}
 \end{center}
\caption{\label{fig:ho2} The temporal evolution of the occupation
probability \protect$|C_u(t)|^2$ of the upper atomic state is shown
for $R_2$ $\!=$ $\!30\lambda_{\rm T}$, $R_1$ $\!-$ $\!R_2$ $\!=$
$\!\lambda_{\rm T}$, 
$\omega_{\rm P}$ $\!=$ $\!0.5\omega_{\rm T}$, $\omega_{\rm A}$
$\!=$ $\!1.046448\omega_{\rm T}$,
$\Gamma_0\lambda_{\rm T}/(2c)$ $\!=$ $\!10^{-6}$,
and $\gamma$ $\!=$ $\!10^{-4}\omega_{\rm T}$ (solid line),
$\gamma$ $\!=$ $\!5\times 10^{-4}\omega_{\rm T}$ (dashed line),
$\gamma$ $\!=$ $\!10^{-3}\omega_{\rm T}$ (dotted line).
For comparison, the exponential decay in free space is shown
(dashed-dotted line).}
\end{figure}
Typical examples for the time evolution of
the upper-state occupation probability are shown in
Fig.~\ref{fig:ho2}. The curves are the exact (numerical) solutions
of the integral equation (\ref{se1.21}) [together with the kernel
function (\ref{se1.22})] for the model permittivity (\ref{se3.15})
The figure shows that with increasing value of the
intrinsic damping constant $\gamma$ of the material of the cavity
wall the Rabi oscillations become less pronounced. Physically, larger
$\gamma$ means larger absorption probability of the emitted photon by
the cavity wall and thus reduced probability of atom-field energy
interchange.

\setcounter{equation}{0}
\section{Extensions to other media}
\label{sec:extensions}

So far we have focused our attention to quantum electrodynamics in 
absorbing, isotropic, non-magnetic, linear media. 
The quantization scheme outlined in Sec.~\ref{sec:quantization} can,
of course, also be generalized to other media.
The starting point is the basic formula (\ref{2.29}) relating the
operator of the electric field strength to the dynamical
variables of the system composed of the electromagnetic field
and the medium. Recalling Eqs.~(\ref{2.18}) and (\ref{2.25}),
we may write Eq.~(\ref{2.29}) as 
\begin{equation}
\label{2.29aa}
\underline{\hat{\bf E}}({\bf r},\omega)
= i\omega\mu_0 \int\! {\rm d}^3{\bf r}' \,
\mbb{G}({\bf r},{\bf r}',\omega)
 \underline{\hat{\bf j}}_{\rm N}({\bf r}',\omega),
\end{equation}
where
\begin{equation}
\label{2.29bb}
\underline{\hat{\bf j}}_{\rm N}({\bf r},\omega)=
\omega\sqrt{\frac{\hbar\varepsilon_0}{\pi}\,
\varepsilon_{\rm I}({\bf r},\omega)}\, \hat{\bf f}({\bf r},\omega). 
\end{equation}
In what follows we briefly comment on possible extensions. 

\subsection{Amplifying media}
\label{sec:extensions1}

Let us consider a dielectric which in some space and 
frequency regions is amplifying, but in other ones absorbing.
Amplification can be described by a complex permittivity 
$\varepsilon({\bf r},\omega)$
which exhibits the familiar properties as in the case of absorption,
most importantly, it still fulfills the Kramers--Kronig relations
(\ref{2.4.9}) and (\ref{2.12}), 
except that the imaginary part is negative,
$\varepsilon_{\rm I}({\bf r},\omega)$ $\!<$ $\!0$.
It can be shown that the quantization scheme in
Sec.~\ref{sec:quantization} also applies to amplifying media
if Eq.~(\ref{2.29bb}) is modified according to \cite{Scheel98}
\begin{equation}
\label{ext1.2}
\underline{\hat{\bf j}}_{\rm N}({\bf r},\omega)
= \omega \sqrt{\frac{\hbar\varepsilon_0}{\pi} 
|\varepsilon_{\rm I}({\bf r},\omega)|} \,
\left[\Theta(\varepsilon_{\rm I})
\hat{\bf f}({\bf r},\omega) + 
\Theta(-\varepsilon_{\rm I})
\hat{\bf f}^\dagger({\bf r},\omega)
\right]
\end{equation}
[$\Theta(x)$, Heaviside step function],
which reflects the well-known fact that amplification requires the
roles of the noise annihilation and creation operators to be exchanged
\cite{Caves82,Jeffers93,Jeffers96,Matloob97,Artoni98}.
Substitution of this expression into Eq.~(\ref{2.29aa})
yields the representation of the ($\omega$-components of the) operator
of the electric field strength in terms of the fundamental fields,
which replaces Eq.~(\ref{2.29}). All the other formulas in
Sec.~\ref{sec:quantization} remain valid.
In particular, the proof of the commutation relations (\ref{2.35})
and (\ref{2.36}) can be given in the same way as in Appendix \ref{sectionB}
for absorbing media. 

\subsection{Anisotropic media}
\label{sec:extensions2}

In anisotropic dielectrics the scalar $\varepsilon({\bf r},\omega)$ has to be
replaced by a tensor ${\mbb{\varepsilon}}({\bf r},\omega)$, which 
is symmetric for reciprocal media (see \cite{Chew}),
\begin{equation}
\label{ext2.1}
\varepsilon_{ij}({\bf r},\omega) = \varepsilon_{ji}({\bf r},\omega).
\end{equation}
Thus the Green tensor has to be determined from the
partial differential equation   
\begin{equation}
\label{ext2.2aa}
\mbb{\nabla}\times  
\mbb{\nabla}\times
\mbb{G}({\bf r},{\bf r}',\omega) 
-\frac{\omega^2}{c^2} \mbb{\varepsilon}({\bf r},\omega) 
\mbb{G}({\bf r},{\bf r}',\omega)
= \mbb{\delta}({\bf r}-{\bf r}'),
\end{equation}
which replaces Eq.~(\ref{2.22}). The solution for a uniaxial
bulk dielectric is given in Eq.~(\ref{ext2.2bb}).
The symmetry property (\ref{ext2.1}) ensures that the reciprocity relation
(\ref{2.22c}) is valid, and the generalization of Eq.~(\ref{2.22d}) reads
\begin{equation}
\label{ext2.5}
\int \!{\rm d}^3{\bf s}\, \frac{\omega^2}{c^2}\,
\varepsilon_{{\rm I}kl}({\bf s},\omega) G_{ik}({\bf s},{\bf r},\omega)
G_{jl}^\ast({\bf s},{\bf r}',\omega) = {\rm Im}\,G_{ij}({\bf r},{\bf r}',\omega). 
\end{equation}

The extension of Eq.~(\ref{2.29bb}) to anisotropic, absorbing
media then reads\footnote{Using the results of
   Sec.~\ref{sec:extensions1}, it is not difficult to include
   in the scheme also amplification.}
\begin{equation}
\label{2.30bb}
\underline{\hat{\bf j}}_{\rm N}({\bf r},\omega)=
\omega\sqrt{\frac{\hbar\varepsilon_0}{\pi}}\,
{\mbb{\varepsilon}}^{1/2}_{\rm I}
({\bf r},\omega)  \hat{\bf f}({\bf r},\omega), 
\end{equation}
where
\begin{equation}
\label{2.31bb}
\mbb{\varepsilon}^{1/2}_{\rm I}({\bf r},\omega) =
\mbb{O}({\bf r},\omega)
\mbb{\varepsilon}'^{1/2}_{\rm I}({\bf r},\omega)
{\mbb{O}}^{-1}({\bf r},\omega).   
\end{equation} 
Here the orthogonal matrix  $\mbb{O}({\bf r},\omega)$
transforms ${\mbb{\varepsilon}}_{\rm I}({\bf r},\omega)$ into its
(positive) diagonal form
${\mbb{\varepsilon}}'_{\rm I}({\bf r},\omega)$,\footnote{Note that,
   depending on the type of anisotropy in the dielectric, the direction
   of the optical axes might be frequency dependent \cite{Roemer}.}
\begin{equation}
\label{2.32bb}
{\mbb{\varepsilon}}'_{\rm I}({\bf r},\omega) =
{\mbb{O}}^{-1}({\bf r},\omega)
\mbb{\varepsilon}_{\rm I}({\bf r},\omega)
\mbb{O}({\bf r},\omega). 
\end{equation}
Substitution of Eq.~(\ref{2.30bb}) into Eq.~(\ref{2.29aa})
yields the ($\omega$-components of the) operator of the electric
field strength expressed
in terms of the fundamental fields, which replaces Eq.~(\ref{2.29}).
The further procedure is same as outlined in Sec.~\ref{sec:quantization},
and it can again been shown that the commutation
relations (\ref{2.35}) and (\ref{2.36}) are fulfilled. 


\subsection{Magnetic media}
\label{sec:extensions3}

Let us now include a possible magnetic response of the matter
in the quantization scheme, restricting our attention to
isotropic media, so that the susceptibility $\mu({\bf r},\omega)$
is a scalar. We  start from the constitutive relations, written in a
non-standard form, which preserves the
interpretation of the fields ${\bf E}$ and ${\bf B}$ as the fundamental
electromagnetic fields, i.e.,
\begin{equation}
\label{ext3.1}
\underline{\hat{\bf D}}({\bf r},\omega) = \varepsilon_0
\underline{\hat{\bf E}}({\bf r},\omega) +
\underline{\hat{\bf P}}({\bf r},\omega) ,
\end{equation}
\begin{equation}
\label{ext3.1a}
\underline{\hat{\bf H}}({\bf r},\omega) = \kappa_0
\underline{\hat{\bf B}}({\bf r},\omega) -
\underline{\hat{\bf M}}({\bf r},\omega) .
\end{equation}
Here, the polarization is given according to Eq.~(\ref{2.4.7a}),
\begin{equation}
\label{ext3.2}
\underline{\hat{\bf P}}({\bf r},\omega) = \varepsilon_0
\left[ \varepsilon({\bf r},\omega) -1 \right]
\underline{\hat{\bf E}}({\bf r},\omega) +
\underline{\hat{\bf P}}_{\rm N}({\bf r},\omega),
\end{equation}
and the magnetization reads 
\begin{equation}
\label{ext3.3}
\underline{\hat{\bf M}}({\bf r},\omega) = \kappa_0
\left[ 1-\kappa({\bf r},\omega) \right]
\underline{\hat{\bf B}}({\bf r},\omega) +
\underline{\hat{\bf M}}_{\rm N}({\bf r},\omega),
\end{equation}
where $\kappa({\bf r},\omega)$ $\!=$ $\!\mu^{-1}({\bf r},\omega)$
and $\kappa_0$ $\!=$ $\!\mu_0^{-1}$. It is not difficult to prove that
the Green tensor now obeys the equation
\begin{equation}
\label{ext3.6}
\left[ \mbb{\nabla}\times \kappa({\bf r},\omega) \mbb{\nabla}\times \;
-\frac{\omega^2}{c^2} \varepsilon({\bf r},\omega) \right]
\mbb{G}({\bf r},{\bf s},\omega) = \mbb{\delta}({\bf r}-{\bf s}). 
\end{equation}
Subsequently, the integral relation (\ref{2.22d}) has to be
replaced by
\begin{eqnarray}
\label{ext3.9}
\lefteqn{
\int\!{\rm d}^3{\bf s}\, \kappa_{\rm I}({\bf s},\omega)
\partial_n^s G_{ki}({\bf s},{\bf r},\omega)
\left[ \partial_k^s G_{nj}^\ast({\bf s},{\bf r}',\omega)
-\partial_n^s G_{kj}^\ast({\bf s},{\bf r}',\omega) \right]
}
\nonumber \\ && 
+\int\!{\rm d}^3{\bf s} \,
\frac{\omega^2}{c^2}\,\varepsilon_{\rm I}({\bf s},\omega)
G_{in}({\bf s},{\bf r},\omega) G_{jn}^\ast({\bf s},{\bf r}',\omega)
= {\rm Im}\,G_{ij}({\bf r},{\bf r}',\omega) ,
\qquad
\end{eqnarray}
which can be proved correct in analogy to the derivation
given in Appendix \ref{sectionA} for non-magnetic matter.

The contribution to $\underline{\hat{\bf j}}_{\rm N}({\bf r},\omega)$ of
the ``magnetization'' changes Eq.~(\ref{2.29bb})
to\footnote{Note that Eq.(\protect\ref{2.18})
   changes to $\underline{\bf j}_{\rm N}({\bf r},\omega)$
   $\!=$ $\!-i\omega\underline{\bf P}_{\rm N}({\bf r},\omega)$ $\!+$ 
   $\!\mbb{\nabla}\times \underline{\,\mbb{\cal M}}_{\rm N}({\bf r},\omega)$.}
\begin{equation}
\label{extext3.8aaa}
\underline{\hat{\bf j}}_{\rm N}({\bf r},\omega) = \omega
\sqrt{\frac{\hbar\varepsilon_0}{\pi} \varepsilon_{\rm I}({\bf r},\omega)}
\,\hat{{\bf f}}_e({\bf r},\omega)  
+ \mbb{\nabla}\times
\sqrt{\frac{\hbar\kappa_0}{\pi}|\kappa_{\rm I}({\bf r},\omega)|} 
\,\hat{{\bf f}}_m({\bf r},\omega),
\end{equation}
with $\hat{{\bf f}}_e({\bf r},\omega)$ and
$\hat{{\bf f}}_m({\bf r},\omega)$ being the bosonic
vector fields associated with the electric and magnetic
properties respectively.
Here we have assumed that the medium is absorbing, so that
$\mu_{\rm I}({\bf r},\omega)$ $\!>$ $\!0$ \cite{LandauVIII} and thus
$\kappa_{\rm I}({\bf r},\omega)$ $\!=$ 
$\!-\mu_{\rm I}({\bf r},\omega)/|\mu({\bf r},\omega)|^2$ $\!<$ $\!0$.

Substituting for $\underline{\hat{\bf j}}_{\rm N}({\bf r}',\omega)$ in
Eq.~(\ref{2.29aa}) the result of Eq.~(\ref{extext3.8aaa}),
we obtain the ($\omega$-components of the)  operator of the electric
field strength $\underline{\hat{\bf E}}({\bf r},\omega)$ expressed
in terms of the dynamical variables of the system.
The fields $\underline{\hat{\bf B}}({\bf r},\omega)$
and $\underline{\hat{\bf D}}({\bf r},\omega)$ can then be
constructed in the same way as in Sec.~\ref{sec:quantization},
and by means of Eq.~(\ref{ext3.1a}) the field
$\underline{\hat{\bf H}} ({\bf r},\omega)$ can be obtained accordingly. 
As before, the commutation relations (\ref{2.35}) and (\ref{2.36})
are fulfilled. 

\subsection{Nonlinear media}
\label{sec:extensions4}

In general, the linear response of a dielectric medium to the
electromagnetic field is only the first term in the series expansion
of the polarization in powers of the electric field strength, and in
strong fields nonlinear interactions play an important role. For
example, the propagating of intense light in Kerr-type nonlinear
media can lead to the formation of soliton-like pulses, the
dispersion of the group velocity being compensated by the Kerr
nonlinearity \cite{Hasegawa,Akhmanov,Taylor}.
There have been a number of approaches to quantize the electromagnetic
field in nonlinear dielectrics
\cite{Hillery84,Abram91,Joneckis93,Duan97}, but mostly absorption is
not taken into account in a consistent way.

A possible way to treat absorption is to include in the Hamiltonian
(\ref{2.28}) an appropriately chosen nonlinear term that
may be thought of as being an integral over a nonlinear function of the
medium polarization field ${\bf P}({\bf r})$, and thus the extended
Hamiltonian is given by
\begin{equation}
\label{ext4.1}
\hat{H} = \int\!{\rm d}^3{\bf r}
\int_0^\infty\!{\rm d}\omega \, \hbar\omega \,
\hat{\bf f}^\dagger({\bf r},\omega) \hat{\bf f}({\bf r},\omega) + 
\hat{H}_{\rm NL}
\end{equation}
with
\begin{equation}
\label{ext4.1a}
\hat{H}_{\rm NL} = \int\!{\rm d}^3{\bf r}
\,h_{\rm NL}[\hat{\bf P}({\bf r})], 
\end{equation}
where
\begin{equation}
\label{ext4.2}
\hat{\bf P}({\bf r})
= \int {\rm d}\omega\,
\underline{\hat{\bf P}}({\bf r},\omega)
+ \mbox{H.c.}
\end{equation}
with $\underline{\hat{\bf P}}({\bf r},\omega)$ being defined
according to Eq.~(\ref{2.4.7a}).

The Heisenberg equation of motion of $\hat{\bf f}({\bf r},\omega)$  
then reads 
\begin{equation}
\label{ext4.2a}
i\hbar \dot{\hat{\bf f}}({\bf r},\omega) =
\big[ \hat{\bf f}({\bf r},\omega) ,\hat{H} \big] =
\hbar\omega \,\hat{\bf f}({\bf r},\omega) +
\big[ \hat{\bf f}({\bf r},\omega) ,\hat{H}_{\rm NL} \big],
\end{equation}
which can be rewritten as $\hat{\omega} \hat{\bf f}({\bf r},\omega)$
$\!=$ $\!\omega\hat{\bf f}({\bf r},\omega)$, where
\begin{equation}
\label{ext4.3}
\hat{\omega} = i \partial_t +\hbar^{-1}\hat{H}^\times_{\rm NL},
\end{equation}
with $\hat{H}^\times_{\rm NL} \hat{O}$ $\!\equiv$
$\![ \hat{H}_{\rm NL},\hat{O}]$.
Hence the frequency-integrated (operator version of) Eq.~(\ref{2.20})
changes to \cite{Schmidt97}
\begin{equation}
\label{ext4.5}
\mbb{\nabla}\times  
\mbb{\nabla}\times
\hat{\bf E}({\bf r}) 
-  \hat{K}({\bf r})\, \hat{\bf E}({\bf r}) 
= \int {\rm d}\omega\, i\omega \mu_0\,
\underline{\hat{\bf j}}_{\rm N}({\bf r},\omega)
+ \mbox{H.c.},
\end{equation}
with $\hat{K}$ being the superoperator 
\begin{equation}
\label{ext4.3aaa}
\hat{K}({\bf r}) = 
\frac{1}{c^2} \left( 
i\partial_t + \hbar^{-1} \hat{H}^\times_{\rm NL} \right)^2
\varepsilon \big({\bf r}, i\partial_t+\hbar^{-1}
\hat{H}^\times_{\rm NL}\big) .
\end{equation}
A similar equation holds for the (transverse) vector potential,
from which the $\hat{\bf B}$-field can be obtained. Obviously,
the approach ensures that the commutation relations (\ref{2.35}) and
(\ref{2.36}) are satisfied. 

For a linear medium, the superoperator $\hat{K}$ simply describes
the effects of dispersion and absorption, but for a nonlinear medium it
introduces via $\hat{H}^\times_{\rm NL}$ additional nonlinear coupling
terms. An application of 
Eq.~(\ref{ext4.5}) to one-dimensional propagation of light
in a Kerr-type nonlinear medium is considered in \cite{Schmidt97},
and a simplified version has been used to study
soliton propagation in absorbing optical fibres \cite{Schmidt99}.

\vspace{2ex}
\noindent
{\bf Acknowledgements}
We are grateful to Ho Trung Dung,
Tom\'{a}\v{s} Opatrn\'{y}, Eduard Schmidt,
and Adriaan Tip for valuable discussions and comments.
This work was supported by
the Deutsche Forschungsgemeinschaft.



\setcounter{section}{0}
\setcounter{equation}{0}
\renewcommand{\thesection}{\Alph{section}}
\renewcommand{\theequation}{\Alph{section}.\arabic{equation}}
\section{The Green tensor} 
\label{sectionA}

\subsection{Basic relations}

In Eq.~(\ref{2.22}),
the Green tensor $\mbb{G}({\bf r},{\bf r}',\omega)$
as a function of ${\bf r}$ and ${\bf r}'$  
can be regarded as being the matrix
elements in the position basis of a (tensor-valued) Green
operator $\hat{\mbb{G}}$ $\!=$ $\!\hat{\mbb{G}}(\omega)$
in an abstract 1-particle Hilbert space,  
\begin{equation}
\label{A.01}
\mbb{G}({\bf r},{\bf r}',\omega) = 
\langle {\bf r}|\hat{\mbb{G}}|{\bf r}'\rangle.
\end{equation}
The matrix elements of the position operator $\hat{\bf r}$
are given by
\begin{equation}
\label{A.02}
\langle {\bf r}|\hat{{\bf r}}|{\bf r}'\rangle = {\bf r}
\delta({\bf r}-{\bf r}'),
\end{equation}
and the matrix elements of the associated momentum
operator $\hat{\bf p}$,
\begin{equation}
\label{A.03a}
\left[ \hat{x}_i,\hat{p}_j\right]
= i\delta_{ij},
\end{equation}
read
\begin{equation}
\label{A.03}
\langle {\bf r}|\hat{{\bf p}}|{\bf r}'\rangle = 
\frac{1}{i}\,\mbb{\nabla} \delta({\bf r}-{\bf r}').
\end{equation}
Let $\hat{\mbb{H}}$ $\!=$ $\!\hat{\mbb{H}}(\omega)$
be the tensor-valued operator
\begin{equation}
\label{A.04aaa}
\hat{\mbb{H}}  = 
i\hat{{\bf p}}\times  
i\hat{{\bf p}}\times
- \hat{q}^2\hat{\mbb{I}}
= \hat{\bf p}^2\hat{\mbb{I}}-\hat{\bf p}\otimes\hat{\bf p}
- \hat{q}^2\hat{\mbb{I}}, 
\end{equation}
where
\begin{equation}
\label{A.04aaaa}
\hat{q}^2 = \frac{\omega^2}{c^2}\,\varepsilon(\hat{{\bf r}},\omega),
\end{equation}
and $\hat{\mbb{I}}$ is the unit operator,
\begin{equation}
\label{A.04aaaaa}
\langle{\bf r}|\hat{\mbb{I}}|{\bf r}'\rangle
= \mbb{\delta}({\bf r}-{\bf r}').
\end{equation}
Equation (\ref{2.22}) corresponds to the operator equation
\begin{equation}
\label{A.05}
\hat{\mbb{H}} \hat{\mbb{G}} = \hat{\mbb{I}},
\end{equation}
as can be easily seen. Writing down Eq.~(\ref{A.05})
in the position basis,
\begin{equation}
\label{A.06}
\int\!{\rm d}^3{\bf s}\,
\langle {\bf r}|\hat{\mbb{H}} 
|{\bf s}\rangle  \langle {\bf s}|
\hat{\mbb{G}}|{\bf r}'\rangle = 
\langle {\bf r}|\hat{\mbb{I}}|{\bf r}'\rangle , 
\end{equation}   
and using Eqs.~(\ref{A.02}) and (\ref{A.03}),  
we are just left with Eq.~(\ref{2.22}).

Now it is not difficult to prove Eq.~(\ref{2.22c}).
{F}rom Eq.~(\ref{A.05}) it follows that the equation 
\begin{equation}
\label{A.08}
\hat{\mbb{G}} = \hat{\mbb{H}}{^{-1}}
\end{equation}
is valid, and thus we find,  
after multiplying it from the right by $\hat{\mbb{H}}$, 
\begin{equation}
\label{A.09}
\hat{\mbb{G}} \hat{\mbb{H}} = \hat{\mbb{I}},
\end{equation}
which in the position basis reads
\begin{equation}
\label{A.10}
\int\!{\rm d}^3{\bf s}\,
\langle {\bf r}|\hat{\mbb{G}} 
|{\bf s}\rangle  \langle {\bf s}|
\hat{\mbb{H}}|{\bf r}'\rangle = 
\langle {\bf r}|\hat{\mbb{I}}|{\bf r}'\rangle . 
\end{equation}
Using again Eqs.~(\ref{A.02}) and (\ref{A.03}), after some
straightforward calculation we derive     
\begin{equation}
\label{A.11}
\left[ \left(\partial^{r'}_j \partial^{r'}_k - \delta_{jk} 
\Delta^{r'}\right)- 
\delta_{jk} \frac{\omega^2}{c^2} \varepsilon({\bf r}',\omega) \right]
G_{ik}({\bf r}, {\bf r}',\omega) 
= \delta_{ji} \delta({\bf r}'-{\bf r}). 
\end{equation}
Comparing Eq.~(\ref{A.11}) with Eq.~(\ref{2.22a}), we
immediately see the Green tensor indeed obeys Eq.~(\ref{2.22c}). 

In order to prove Eq.~(\ref{2.22d}),
we introduce operators $\hat{\mbb{O}}{^\ddagger}$ defined by
\begin{equation}
\label{A.12}
\big(\hat{O}{^\ddagger}\big)_{ij}
= \big(\hat{O}_{ji}\big){^\dagger} \equiv \hat{O}_{ji}^\dagger.
\end{equation}
{F}rom Eq.~(\ref{A.05}) it then follows that
\begin{equation}
\label{A.13}
\hat{\mbb{G}}{^\ddagger}\hat{\mbb{H}}{^\ddagger} 
= \hat{\mbb{I}}.
\end{equation}   
Multiplying Eq.~(\ref{A.05}) from the left by
$\hat{\mbb{G}}{}^{\ddagger}$ and Eq.~(\ref{A.13}) from the
right by $\hat{\mbb{G}}$ and subtracting the resulting
equations from each other, we find 
\begin{equation}
\label{A.14}
\hat{\mbb{G}}{^\ddagger} 
\big(\hat{\mbb{H}} - \hat{\mbb{H}}{^\ddagger}\big) 
\hat{\mbb{G}} = 
\hat{\mbb{G}}{^\ddagger} - \hat{\mbb{G}},
\end{equation}  
which in Cartesian components reads 
\begin{equation}
\label{A.15}
\hat{G}_{mi}^\dagger 
\big(\hat{H}_{mn} - \hat{H}_{nm}^\dagger\big) 
\hat{G}_{nj} = 
\hat{G}_{ji}^\dagger - \hat{G}_{ij}.
\end{equation}
{F}rom Eq.~(\ref{A.04aaa}) it is easily seen that
\begin{equation}
\label{A.15a}
\hat{H}_{mn} - \hat{H}_{nm}^\dagger = 
\delta_{mn}\big(\hat{q}^{2\dagger}- \hat{q}^2\big).
\end{equation}
Representing Eq.~(\ref{A.15}) [together with Eq.~(\ref{A.15a})]
in the position basis and recalling the relations
\begin{equation}
\label{A.15b}
\langle {\bf r}|\hat{G}_{ji}^\dagger |{\bf r}'\rangle
= \langle {\bf r}'|\hat{G}_{ji} |{\bf r}\rangle^\ast =
G_{ji}^\ast({\bf r}',{\bf r},\omega) = 
G_{ij}^\ast({\bf r}, {\bf r}',\omega),
\end{equation}
we eventually arrive at Eq.~(\ref{2.22d}).

\subsubsection{Asymptotic behaviour}

The Green tensor $G_{ij}({\bf r},{\bf r}',\omega)$ is 
holomorphic in the upper half-plane of complex $\omega$, because of the
holomorphic behaviour of $\varepsilon({\bf r},\omega)$.
In order to study the behaviour of
$G_{ij}({\bf r},{\bf r}',\omega)$ for
$|\omega|$ $\!\to$ $\!\infty$ and $|\omega|$ $\!\to$ $\!0$,
we first introduce the tensor-valued 
projection operators 
\begin{equation}
\label{A.16}
\hat{\mbb{I}}{^\perp} = \hat{\mbb{I}} -
\hat{\mbb{I}}{^\|},
\quad
\hat{\mbb{I}}{^\|} = 
\frac{\hat{{\bf p}}\otimes\hat{{\bf p}}}{\hat{{\bf p}}^2}
\end{equation}
and decompose $\hat{\mbb{H}}$, Eq.~(\ref{A.04aaa}), as 
\begin{equation}
\label{A.17}
\hat{\mbb{H}} = 
\big(\hat{\bf p}{^2} - \hat{q}{^2}\big) \hat{\mbb{I}}{^\perp}
- \hat{q}{^2} \hat{\mbb{I}}{^\|}.
\end{equation}   
Applying the Feshbach formula \cite{Newton}, we then
may decompose the Green tensor operator 
$\hat{\mbb{G}}$ $\!=$ $\!\hat{\mbb{H}}{^{-1}}$,
Eq.~(\ref{A.08}), as
\begin{eqnarray}
\label{A.18}
\lefteqn{
\hat{\mbb{G}}= \hat{\mbb{H}}{^{-1}} 
= \hat{\mbb{I}}{^\|}
(\hat{\mbb{I}}{^\|} \hat{\mbb{H}} 
\hat{\mbb{I}}{^\|})^{-1} \hat{\mbb{I}}{^\|}
}  
\nonumber\\[.5ex] && \hspace{-2ex}
+ \big[\hat{\mbb{I}}{^\perp}\!-\!
\hat{\mbb{I}}{^\|}
(\hat{\mbb{I}}{^\|} \hat{\mbb{H}} 
\hat{\mbb{I}}{^\|}){^{-1}}\hat{\mbb{I}}{^\|}  
\hat{\mbb{H}\!}\hat{\mbb{I}}{^\perp}\big]
\hat{\mbb{K}}
\big[\hat{\mbb{I}}{^\perp}\!-\!
\hat{\mbb{I}}{^\perp}\hat{\mbb{H}}\hat{\mbb{I}}{^\|}
(\hat{\mbb{I}}{^\|}\hat{\mbb{H}} 
\hat{\mbb{I}}{^\|}){^{-1}}
\hat{\mbb{I}}{^\|}  
\big],\qquad
\end{eqnarray}   
where  
\begin{equation}
\label{A.19}
\hat{\mbb{K}} =
\big[
\hat{\mbb{I}}{^\perp}\hat{\mbb{H}}\hat{\mbb{I}}{^\perp}
- \hat{\mbb{I}}{^\perp}\hat{\mbb{H}}\hat{\mbb{I}}{^\|}
(\hat{\mbb{I}}{^\|}\hat{\mbb{H}} 
\hat{\mbb{I}}{^\|}){^{-1}}\hat{\mbb{I}}{^\|}  
\hat{\mbb{H}}\hat{\mbb{I}}{^\perp}
\big]^{-1}.
\end{equation}
It is not difficult to prove that
\begin{equation}
\label{A.20}
\hat{\mbb{I}}{^\|} \hat{\mbb{H}} 
\hat{\mbb{I}}{^\|} = -  \hat{\mbb{I}}{^\|}\hat{q}{^2}  
\hat{\mbb{I}}{^\|}, \quad    
\hat{\mbb{I}}{^\|} \hat{\mbb{H}} 
\hat{\mbb{I}}{^\perp} = -  \hat{\mbb{I}}{^\|}\hat{q}{^2}  
\hat{\mbb{I}}{^\perp},    
\end{equation}   
\begin{equation}
\label{A.21}
\hat{\mbb{I}}{^\perp} \hat{\mbb{H}} 
\hat{\mbb{I}}{^\|} = -  \hat{\mbb{I}}{^\perp}\hat{q}{^2}  
\hat{\mbb{I}}{^\|}, \quad    
\hat{\mbb{I}}{^\perp} \hat{\mbb{H}} 
\hat{\mbb{I}}{^\perp} =  \hat{\mbb{I}}{^\perp}
\big(\hat{{\bf p}}{^2}-\hat{q}{^2}\big)  
\hat{\mbb{I}}{^\perp}.    
\end{equation}
Combining Eqs.~(\ref{A.18}) -- (\ref{A.21}) and
using Eq.~(\ref{A.04aaaa}), we obtain for
$\hat{\mbb{G}}$ the expression
\begin{eqnarray}
\label{A.22}
\lefteqn{
\hat{\mbb{G}} 
= -\frac{c^2}{\omega^2}\,
\hat{\mbb{I}}{^\|}
(\hat{\mbb{I}}{^\|}\hat{\varepsilon}  
\hat{\mbb{I}}{^\|})^{-1} \hat{\mbb{I}}{^\|}
}  
\nonumber\\[.5ex] && 
+ \big[ \hat{\mbb{I}}{^\perp} - \hat{\mbb{I}}{^\|}
(\hat{\mbb{I}}{^\|} \hat{\varepsilon} 
\hat{\mbb{I}}{^\|})^{-1}  \hat{\mbb{I}}{^\|}  
\hat{\varepsilon} \hat{\mbb{I}}{^\perp}\big]
\hat{\mbb{K}}
\big[\hat{\mbb{I}}{^\perp}-
\hat{\mbb{I}}{^\perp}\hat{\varepsilon}\hat{\mbb{I}}{^\|}
(\hat{\mbb{I}}{^\|} \hat{\varepsilon} 
\hat{\mbb{I}}{^\|})^{-1} \hat{\mbb{I}}{^\|}  
\big],
\end{eqnarray}   
where
\begin{equation}
\label{A.23}
\hat{\mbb{K}} = 
\left[
\hat{\mbb{I}}{^\perp}\left(\hat{{\bf p}}^2
-\frac{\omega^2}{c^2}\,\hat{\varepsilon}\right) 
\hat{\mbb{I}}{^\perp}
+ \frac{\omega^2}{c^2}\,
\hat{\mbb{I}}{^\perp}\hat{\varepsilon}\hat{\mbb{I}}{^\|}
(\hat{\mbb{I}}{^\|} \hat{\varepsilon} 
\hat{\mbb{I}}{^\|})^{-1} \hat{\mbb{I}}{^\|}  
\hat{\varepsilon}\hat{\mbb{I}}{^\perp}\right]^{-1}
\end{equation}
[$\hat{\varepsilon}$ $\!=$
$\!\varepsilon(\hat{{\bf r}},\omega)$].   

Now the desired limiting processes can be performed easily.
For \mbox{$|\omega|$ $\!\to$ $\!0$} we find
\mbox{[$\hat{\varepsilon}^{(0)}$ $\!=$ $\!\varepsilon^{(0)}(\hat{\bf r})$
$\!=$ $\!\varepsilon(\hat{\bf r},\omega=0)$]} 
\begin{equation}
\label{A.24}
\lim_{|\omega|\to 0}\frac{\omega^2}{c^2}\,
\hat{\mbb{G}} = - \hat{\mbb{I}}{^\|}
(\hat{\mbb{I}}{^\|}\hat{\varepsilon}^{(0)}  
\hat{\mbb{I}}{^\|})^{-1} \hat{\mbb{I}}{^\|}
\end{equation}
and
\begin{equation}
\label{A.24a}
\lim_{|\omega|\to 0}\hat{q}{^2}\hat{\mbb{G}} 
= - \hat{\varepsilon}^{(0)}\hat{\mbb{I}}{^\|}
(\hat{\mbb{I}}{^\|}\hat{\varepsilon}^{(0)}  
\hat{\mbb{I}}{^\|})^{-1} \hat{\mbb{I}}{^\|}.
\end{equation}   
For $|\omega|$ $\!\to$ $\!\infty$ we
arrive at, on recalling that
$\hat{\mbb{I}}{^\|}\hat{\mbb{I}}{^\perp}$
$\!=$ $\!\hat{\mbb{I}}{^\perp}\hat{\mbb{I}}{^\|}$ $\!=$
$\!0$ and $\varepsilon({\bf r},\omega)$ $\!\to$ $\!1$ if
$|\omega|$ $\!\to$ $\!\infty$,
\begin{equation}
\label{A.25a}
\lim_{|\omega|\to \infty}\frac{\omega^2}{c^2}\,
\hat{\mbb{G}}
= \lim_{|\omega|\to \infty}\hat{q}{^2}
\hat{\mbb{G}} = - \hat{\mbb{I}} .
\end{equation}

Obviously, the first term on the right hand of
Eq.~(\ref{A.22}) is the singular part of $\hat{\mbb{G}}$
for $|\omega|$ $\!\to$ $\!0$. Performing in that term
the Taylor expansion
\begin{equation}
\label{A.26}
\varepsilon(\hat{{\bf r}},\omega) = 
\varepsilon^{(0)}(\hat{{\bf r}})
+ \omega\,\varepsilon^{(1)}(\hat{{\bf r}}) + \ldots,
\end{equation}
where $\varepsilon^{(0)}(\hat{\bf r})$ is real and
$\varepsilon^{(1)}(\hat{\bf r})$ is imaginary [see
Eq.~(\ref{2.12a})], we find that
\begin{equation}
\label{A.27}
{\rm Re}\,G_{ij}({\bf r}, {\bf r}',\omega)\sim \omega^{-2}
\quad
(|\omega| \to 0)
\end{equation}
and
\begin{equation}
\label{A.27a}
{\rm Im}\,G_{ij}({\bf r}, {\bf r}',\omega)\sim \omega^{-1}
\quad
(|\omega| \to 0).
\end{equation}

\subsubsection{Isotropic bulk material}

For a bulk material of given permittivity $\varepsilon(\omega)$
the Green tensor reads
($\mbb{\rho}$ $\!=$ ${\bf r}$ $\!-$ $\!{\bf r}'$)
\begin{equation}
\label{se2.1}
\mbb{G}({\bf r},{\bf r}',\omega)
= \left[\mbb{\nabla}^r\otimes\mbb{\nabla}^r + \mbb{I}q^2(\omega) \right]
\frac{{\rm e}^{iq(\omega)\rho}}
{4\pi q^2(\omega)\rho}\,,
\end{equation}
where
\begin{equation}
\label{se2.1a}
q(\omega) = \sqrt{\varepsilon(\omega)} \,\frac{\omega}{c} =
\left[ n_{\rm R}(\omega) +in_{\rm I}(\omega) \right]
\,\frac{\omega}{c}\,.
\end{equation}
It can be split up into a longitudinal and
a transverse part according to
\begin{equation}
\label{se2.2a}
\mbb{G}^\|({\bf r},{\bf r}',\omega)
= -\frac{1}{4\pi q^2} \left[
\frac{4\pi}{3} \delta(\mbb{\rho}) \mbb{I} +\left( \mbb{I}
-\frac{3\mbb{\rho}\otimes\mbb{\rho}}{\rho^2} \right) \frac{1}{\rho^3} \right]
\end{equation}
and
\begin{eqnarray}
\label{se2.2b}
\lefteqn{
\mbb{G}^\perp({\bf r},{\bf r}',\omega)
= \frac{1}{4\pi q^2} \bigg\{
\left( \mbb{I} -\frac{3\mbb{\rho}\otimes\mbb{\rho}}{\rho^2} \right)
\frac{1}{\rho^3}
}
\nonumber \\ &&\hspace{-3ex} 
+\,q^3 \bigg[\! \left(\! \frac{1}{q\rho}\! +\!\frac{i}{(q\rho)^2}
\!-\!\frac{1}{(q\rho)^3}\! \right) \mbb{I}
\!-\!\left(\! \frac{1}{q\rho}
\!+\!\frac{3i}{(q\rho)^2}\! -\!\frac{3}{(q\rho)^3}\! \right)
\frac{\mbb{\rho}\otimes\mbb{\rho}}{\rho^2}\! \bigg] {\rm e}^{iq\rho}\! \bigg\}.
\qquad
\end{eqnarray}
In particular, from Eq.~(\ref{se2.2b}) it follows that
\begin{equation}
\label{se2.2c}
{\rm Im}\,\mbb{G}^\perp({\bf r},{\bf r},\omega)
= \lim_{{\bf r}'\to{\bf r}}
{\rm Im}\,\mbb{G}^\perp({\bf r},{\bf r}',\omega)
= \frac{\omega}{6\pi c}\,n_{\rm R}(\omega) \mbb{I}.
\end{equation}

\subsubsection{Uniaxial bulk material}

Following \cite{Weiglhofer90}, the Green tensor for
a homogeneous uniaxial dielectric can be given in the form of
\begin{samepage}
\begin{eqnarray}
\label{ext2.2bb}
\lefteqn{
\mbb{G}({\bf r},{\bf r}',\omega) = q_t^{-2}(\omega)
\left[
\mbb{\nabla}^r\otimes\mbb{\nabla}^r
+ q_t^2(\omega)\varepsilon_c
\mbb{\varepsilon}^{-1} 
\right] \frac{{\rm e}^{iq_t(\omega)\rho_e}}{4\pi \rho_e}
}
\nonumber \\ &&\hspace{-2ex}
-\left[ \frac{\varepsilon_c}{\varepsilon_t}
\frac{{\rm e}^{iq_t(\omega)\rho_e}}{4\pi \rho_e} -
\frac{{\rm e}^{iq_t(\omega)\rho}}{4\pi\rho} \right]
\frac{
(\mbb{\rho}\times {\bf c}) \otimes
(\mbb{\rho}\times {\bf c})
}{(\mbb{\rho}\times {\bf c})^2}
\nonumber\\ &&\hspace{-2ex}
-\left[ \frac{{\rm e}^{iq_t(\omega)\rho_e}\!-\!{\rm e}^{iq_t(\omega)\rho}}{4\pi
iq_t(\omega)} \right] \left[
\frac{ \mbb{I}- {\bf c}\otimes{\bf c}}{(\mbb{\rho}\times {\bf c})^2}
-\frac{2 (\mbb{\rho}\times {\bf c}) \otimes
(\mbb{\rho}\times {\bf c})}{(\mbb{\rho}\times {\bf c})^4} \right],
\qquad 
\end{eqnarray}
\end{samepage}
[$\rho_e^2$ $\!=$
$\!\varepsilon_c \mbb{\rho}\mbb{\varepsilon}^{-1}\mbb{\rho}$,
$q_t$ $\!=$ $\!\sqrt{\varepsilon_t}\omega/c$].
Here, the vector ${\bf c}$ is the unit vector parallel
to the crystallographic
axis of the medium, $\varepsilon_c$ $\!=$ $\!\varepsilon_c(\omega)$
the (complex) permittivity
in that direction, and $\varepsilon_t$ $\!=$ $\!\varepsilon_t(\omega)$
the (complex) permittivity
transverse to it. Note  that for $\varepsilon_c$ $\!=$
$\!\varepsilon_t$ $\!=$ $\!\varepsilon$
[and $\mbb{\varepsilon}^{-1}$ $\!=$
$\!\varepsilon^{-1} \mbb{I}$]
the Green tensor (\ref{se2.1}) for isotropic bulk material
is recovered.

\setcounter{equation}{0}
\section{Commutation relations} 
\label{sectionB}

Let us first consider the commutation relations of the
electric field and the vector potential.
Using Eqs.~(\ref{2.32}) and (\ref{2.38}) together with
Eqs.~(\ref{2.29}) and (\ref{2.38a}), recalling the commutation relations
(\ref{2.26}) and (\ref{2.27}), and applying the integral
relation (\ref{2.22d}), after some algebra we derive    
\begin{equation}
\label{A.28}
\left[ \hat{E}_{k}({\bf r}),\hat{E}_{l}({\bf r'})\right] = 0, 
\end{equation}
\begin{equation}
\label{A.29}
\left[ \hat{A}_{k}({\bf r}),\hat{A}_{l}({\bf r'})\right] = 0, 
\end{equation}
\begin{equation}
\label{A.30}
\left[\varepsilon_0
\hat{E}_{k}({\bf r}),\hat{A}_{l}({\bf r'})\right] =
\int {\rm d}^3{\bf s}\,\bigg[
\frac{2i\hbar}{\pi} 
\int_0^\infty\!{\rm d}\omega \, \frac{\omega}{c^2}\, 
{\rm Im}\,G_{km}({\bf r}, {\bf s},\omega) \bigg]
\delta^{\perp}_{ml}({\bf s}-{\bf r}').
\end{equation}
In Eq.~(\ref{A.30}) the
$\omega$-integral can be performed by applying the rule
\begin{equation}
\label{A.30a}
\int_{0}^\infty\!{\rm d}\omega\,\ldots
=\lim_{\epsilon\to 0} \int_{\epsilon}^\infty\! {\rm d}\omega\,\ldots.
\end{equation}
Note that, according to Eq.~(\ref{A.27a}),
${\rm Im}\,G_{il}({\bf r}, {\bf r}',\omega)$
behaves like $\omega^{-1}$  as $\omega$ approaches zero.
Thus we may transform, on recalling Eq.~(\ref{2.22b}),
the $\omega$-integral into a principal-part (${\cal P}$)
integral, so that Eq.(\ref{A.30}) reads
\begin{equation}
\label{A.31}
\left[\varepsilon_0
\hat{E}_{k}({\bf r}),\hat{A}_{l}({\bf r'})\right]
= \int {\rm d}^3{\bf s} \,\bigg[
\hbar\frac{\cal P}{\pi} 
\int_{-\infty}^\infty\!\frac{{\rm d}\omega}{\omega} \,
\frac{\omega^2}{c^2}\,G_{km}({\bf r}, {\bf s},\omega)
\bigg]\delta^{\perp}_{ml}({\bf s}-{\bf r}').
\end{equation}

The evaluation of the $\omega$-integral in Eq.~(\ref{A.31}) 
can be performed by means of contour-integral techniques.
Since $G_{km}({\bf r}, {\bf s},\omega)$
is a holomorphic function of $\omega$ in the upper
complex frequency half-plane
with the asymptotic behaviour according to Eq.~(\ref{A.25a}),
the $\omega$-integrals can be calculated by 
contour integration along
an infinitely small half-circle $|\omega|$ $\!=$ $\!\rho$,
\mbox{$\rho$ $\!\to$ $\!0$}, and
an infinitely large  half-circle
$|\omega|$ $\!=$ $\!R$, $R$ $\!\to$ $\!\infty$,  
in the upper complex half-plane
\begin{equation}
\label{A.34}
{\cal P}\int_{-\infty}^\infty\!{\rm d}\omega\,\ldots  = 
\lim_{\rho \to 0}\int_{
|\omega|=\rho\atop\omega_{\rm I}>0}
\! {\rm d}\omega\,\ldots
-\lim_{R \to \infty}\int_{
|\omega|=R\atop\omega_{\rm I}>0 }
\! {\rm d}\omega \,\ldots.
\end{equation}
The definition of $\hat{\mbb{I}}{^{\perp(\|)}}$
as given in Eq.~(\ref{A.16}) implies that
\begin{equation}
\label{A.34a}
\langle {\bf r}|\hat{\mbb{I}}{^{\perp(\|)}}| {\bf r}'\rangle
= \mbb{\delta}^{\perp(\|)}({\bf r}-{\bf r}'),  
\end{equation}
and thus we may write, on applying Eq.~(\ref{A.01}), 
\begin{equation}
\label{A.34b}
\int \!{\rm d}^3{\bf s}\,\bigg[\,
\int_{\cal C}\frac{{\rm d}\omega}{\omega} \,
\frac{\omega^2}{c^2}\,G_{km}({\bf r}, {\bf s},\omega)
\bigg]\delta^{\perp}_{ml}({\bf s}-{\bf r}')
= \int_{\cal C} \frac{{\rm d}\omega}{\omega}
\,\frac{\omega^2}{c^2}\
\langle {\bf r}|\hat{\mbb{G}}(\omega)\hat{\mbb{I}}{^\perp}
|{\bf r}'\rangle.
\end{equation}
Note that
\begin{equation}
\label{deltalong}
\mbb{\delta}^\|({\bf r}) =
-\mbb{\nabla}\otimes\mbb{\nabla}\,\frac{1}{4\pi|{\bf r}|}
\end{equation}
and
\begin{equation}
\label{deltatrans}
\mbb{\delta}^\perp({\bf r}) = \mbb{\delta}({\bf r})
-\mbb{\delta}^\|({\bf r}) .
\end{equation}
{F}rom Eq.~(\ref{A.24}) it follows that
\begin{equation}
\label{A.34c}
\lim_{|\omega|\to 0}\frac{\omega^2}{c^2}
\hat{\mbb{G}}\hat{\mbb{I}}{^\perp} 
= - \hat{\mbb{I}}{^\|}
(\hat{\mbb{I}}{^\|}\hat{\varepsilon}^{(0)}  
\hat{\mbb{I}}{^\|})^{-1}
\hat{\mbb{I}}{^\|}\hat{\mbb{I}}{^\perp} = 0
\end{equation}
($\hat{\mbb{I}}{^\|}\hat{\mbb{I}}{^\perp}$ $\!=$ $\!0$)
and therefore the integral over the  small half-circle vanishes.   
Finally, from Eq.~(\ref{A.25a}) we see that
\begin{equation}
\label{A.34d}
\lim_{|\omega|\to \infty}\frac{\omega^2}{c^2}\,
\hat{\mbb{G}}  \hat{\mbb{I}}{^\perp}
= - \hat{\mbb{I}}{^\perp}.
\end{equation}
Hence,
\begin{samepage}
\begin{eqnarray}
\label{A.35a}
\lefteqn{
{\cal P}\!\int_{-\infty}^\infty \frac{{\rm d}\omega}{\omega}
\,\frac{\omega^2}{c^2}\,
\langle {\bf r}|\hat{\mbb{G}}(\omega)\hat{\mbb{I}}{^\perp}
|{\bf r}'\rangle
}
\nonumber\\&&
= \,2i\!\int_0^\infty {\rm d}\omega
\,\frac{\omega}{c^2}\,{\rm Im}\,
\langle {\bf r}|\hat{\mbb{G}}(\omega)\hat{\mbb{I}}{^\perp}
|{\bf r}'\rangle
= i\pi \mbb{\delta}^\perp({\bf r}-{\bf r}'),
\end{eqnarray}
\end{samepage}
and the sought commutator reads
\begin{equation}
\label{A.35}
\left[\varepsilon_0
\hat{E}_{k}({\bf r}),\hat{A}_{l}({\bf r'})\right] =  
i\hbar \delta^{\perp}_{kl}({\bf r}-{\bf r}').
\end{equation}
Since the corresponding commutation relations for the displacement
field [Eq.~(\ref{2.34}) together with Eqs.~(\ref{2.31}) and
(\ref{2.29})] can derived in a quite similar way, we renounce the
derivation here and immediately give the result 
\begin{equation}
\label{A.32}
\left[\hat{D}_{k}({\bf r}),\hat{D}_{l}({\bf r'})\right] = 0, 
\end{equation}
\begin{equation}
\label{A.38}
\left[
\hat{D}_{k}({\bf r}),\hat{A}_{l}({\bf r'})\right] 
=i\hbar \delta^{\perp}_{kl}({\bf r}-{\bf r}').
\end{equation}
This is the result we have expected, because the polarization
$\hat{\bf P}({\bf r})$ $\!=$ 
$\!\hat{\bf D}({\bf r})$ $\!-$ $\!\varepsilon_0
\hat{\bf E}({\bf r})$ is related to the degrees of
freedom of the matter and it should
therefore commute with the radiation field operators. 

Recalling the relations $\mbb{\nabla}\times\hat{\bf A}$ 
$\!=$ $\!\hat{\bf B}$, $\hat{\bf \Pi}$  
$\!=$ $\!-\varepsilon_0\hat{\bf E}^\perp$, 
$\!-\mbb{\nabla}\hat{\varphi}$ $\!=$ $\!\hat{\bf E}{^\|}$,
and $\hat{\bf P}$ $\!=$ 
$\!\hat{\bf D}$ $\!-$ $\!\varepsilon_0 \hat{\bf E}$
and using the commutation relations (\ref{A.28}), (\ref{A.29}),
(\ref{A.35}), (\ref{A.32}), and (\ref{A.38}),
we can then derive further commutation relations such as 
\begin{equation}
\label{A.36}
\left[\varepsilon_0\hat{E}_{k}({\bf r}),\hat{B}_{l}({\bf r}')
\right] = -i \hbar \,\epsilon_{klm} \,\partial^r_m
\delta({\bf r}-{\bf r}'),  
\end{equation}
\begin{equation}
\label{A.37}
\left[\hat{A}_{k}({\bf r}),\hat{\Pi}_l({\bf r}')\right] = 
i \hbar \,\delta^{\perp}_{kl}({\bf r}-{\bf r}'), 
\end{equation}
\begin{equation}
\label{A.37a}
\left[\hat{\varphi}({\bf r}),\hat{\varphi}({\bf r'})\right] =
\left[\hat{\varphi}({\bf r}),\hat{A}_{k}({\bf r'})\right] =  
\left[ \hat{\varphi}({\bf r}),\hat{E}_{k}({\bf r'})\right]= 0, 
\end{equation}
\begin{equation}
\label{A.39}
\left[\hat{\varphi}({\bf r}),\hat{D}_{k}({\bf r'})\right] = 0. 
\end{equation}
It should be pointed out that the commutation relations given
above are valid for both the medium-assisted electromagnetic
field considered in Sec.~\ref{sec:quantization} and the
total electromagnetic field considered in
Sec.~\ref{sec:source}. Needless to say that quantities
of the medium-assisted electromagnetic field and quantities
of the additional charged particles commute.    

\setcounter{equation}{0}
\section{Equations of motion} 
\label{sectionC}

In order to prove that the quantization scheme yields
the correct Maxwell equations and the Newtonian
equations of motion of the charged particles, let us consider,
e.g. the Maxwell equation (\ref{2.57}). Using the
minimal-coupling Hamiltonian (\ref{2.51}), we find that
\begin{eqnarray}
\label{A.40a}
\lefteqn{
\dot{\hat{\bf D}}({\bf r})  = 
 \frac{1}{i \hbar} \big[\hat{\bf D}({\bf r}),\hat{H}\big]
= \frac{1}{i \hbar}
\int\!{\rm d}^3{\bf r}'
\int_0^\infty\!{\rm d}\omega \,\hbar\omega \,
\big[\hat{\bf D}({\bf r}),
\hat{\bf f}^\dagger({\bf r}',\omega) 
\hat{\bf f}({\bf r}',\omega)
\big]
}
\nonumber \\ && \hspace{-2ex}
+\, \frac{1}{i \hbar}
\sum_\alpha
\frac{1}{2m_\alpha}
\left[\hat{\bf D}({\bf r}),
\big[\hat{\bf p}_\alpha  
- q_\alpha \hat{\bf A}(\hat{\bf r}_\alpha)
\big]^2
\right]
\nonumber \\ && \hspace{-2ex}
+\frac{1}{2 i \hbar}
\!\int\!\!{\rm d}^3{\bf r}' \,
\big[\hat{\bf D}({\bf r}),
\hat{\rho}_{\rm A}({\bf r}') \hat{\varphi}_{\rm A}({\bf r}')
\big]
\!+\frac{1}{i \hbar}
\!\int\!\!{\rm d}^3{\bf r}' \,
\big[\hat{\bf D}({\bf r}),
\hat{\rho}_{\rm A}({\bf r}') \hat{\varphi}_{\rm M}({\bf r}')
\big].
\qquad
\end{eqnarray}
Recalling the definition (\ref{2.61}) of the displacement
field and applying the commutation relation (\ref{A.39}),
we may simplify Eq.~(\ref{A.40a}) to
\begin{eqnarray}
\label{A.40}
\lefteqn{
\dot{\hat{\bf D}}({\bf r})  
= \frac{1}{i \hbar}
\int\!{\rm d}^3{\bf r}'
\int_0^\infty\!{\rm d}\omega \,\hbar\omega \,
\big[\hat{\bf D}_{\rm M}({\bf r}),
\hat{\bf f}^\dagger({\bf r}',\omega) 
\hat{\bf f}({\bf r}',\omega)
\big]
}
\nonumber \\ &&\hspace{6ex} 
+\,\frac{1}{i \hbar} 
\sum_\alpha \frac{1}{2m_\alpha}
\left[\hat{\bf D}_{\rm M}({\bf r}),
\big[
\hat{\bf p}_\alpha  
- q_\alpha \hat{\bf A}(\hat{\bf r}_\alpha)
\big]^2
\right]
\nonumber \\ &&\hspace{6ex}
-\,\frac{\varepsilon_0}{i \hbar}  
\sum_\alpha \frac{1}{2m_\alpha}
\left[\mbb{\nabla}\hat{\varphi}_{\rm A}({\bf r}),
\big[
\hat{\bf p}_\alpha 
- q_\alpha \hat{\bf A}(\hat{\bf r}_\alpha)
\big]^2
\right].
\end{eqnarray}
Using Eqs.~(\ref{2.33}) and (\ref{2.34}) together with
Eqs.~(\ref{2.30}), (\ref{2.31}), and (\ref{2.29}) and
recalling the basic commutation relations (\ref{2.26}) and (\ref{2.27}), 
we find that the first of the three commutators in
Eq.~(\ref{A.40}) gives
\begin{equation}
\label{A.41}
\frac{1}{i\hbar}
\int\!{\rm d}^3{\bf r}'
\int_0^\infty\!{\rm d}\omega \,\hbar\omega \,
\big[\hat{\bf D}_{\rm M}({\bf r}),
\hat{\bf f}^\dagger({\bf r}',\omega) 
\hat{\bf f}({\bf r}',\omega)
\big]
= \frac{1}{\mu_0}\mbb{\nabla}\times \hat{\bf B}({\bf r}) \,. 
\end{equation}
In order to calculate the second commutator, we use the commutation
relation (\ref{A.38}) and obtain
\begin{equation}
\label{A.42}
\frac{1}{i\hbar}
\sum_\alpha \frac{1}{2m_\alpha}
\left[\hat{\bf D}_{\rm M}({\bf r}),
\big[
\hat{\bf p}_\alpha  
- q_\alpha \hat{\bf A}(\hat{\bf r}_\alpha)
\big]^2
\right]
= - \hat{\bf j}^\perp_{\rm A}({\bf r}).
\end{equation}
Finally, recalling the definitions of $\hat{\varphi}_{\rm A}({\bf r})$
[Eq.~(\ref{2.52})] and $\hat{\rho}_{\rm A}({\bf r})$ [Eq.~(\ref{2.53})]
and using the particle-operator commutation relations, we derive
\begin{equation}
\label{A.43}
- \frac{\varepsilon_0}{i\hbar}
\sum_\alpha \frac{1}{2m_\alpha}
\left[\mbb{\nabla}\hat{\varphi}_{\rm A}({\bf r}),
\big[
\hat{\bf p}_\alpha 
- q_\alpha \hat{\bf A}(\hat{\bf r}_\alpha)
\big]^2
\right]
= - \hat{\bf j}^\|_{\rm A}({\bf r}).
\end{equation}
In Eqs.~(\ref{A.42}) and (\ref{A.43}), the transverse and 
longitudinal current densities  $\hat{\bf j}^\perp_{\rm A}({\bf r})$ and 
$\hat{\bf j}^\|_{\rm A}({\bf r})$, respectively, are defined by 
\begin{equation}
\label{A.44}
\hat{\bf j}^{\perp (\|)}_{\rm A}({\bf r}) = 
{\textstyle\frac{1}{2}} \sum_\alpha q_\alpha
\dot{\hat{\bf r}}_\alpha 
{\mbb{\delta}}^{\perp (\|)}({\bf r}-\hat{\bf r}_\alpha)
+ \mbox{H.c.},
\end{equation}
where the velocity operator $\dot{\hat{\bf r}}_\alpha$
of the $\alpha$th particle is
\begin{equation}
\label{A.45}
\dot{\hat{\bf r}}_\alpha = 
\frac{1}{i \hbar} \big[\hat{\bf r}_\alpha\,,\hat{H}\big] =
\frac{1}{m_\alpha} \big[ \hat{\bf p}_\alpha 
- q_\alpha \hat{\bf A}(\hat{\bf r}_\alpha) \big].
\end{equation}
Substituting Eqs.~(\ref{A.41}), (\ref{A.42}), and (\ref{A.43}) back
into Eq.~(\ref{A.40}), we just arrive at the Maxwell equation
(\ref{2.57}). The Maxwell equation (\ref{2.55}) 
and the Newtonian equations of motion (\ref{2.58}) and (\ref{2.59})
can be derived in a quite similar way.

\end{document}